\documentclass{article}
\RequirePackage[authoryear]{natbib}
\RequirePackage[colorlinks,citecolor=blue,urlcolor=blue]{hyperref}
\RequirePackage{hypernat}
\RequirePackage{graphicx}
\RequirePackage{amsmath,amsthm,amssymb}
\RequirePackage{etoolbox}
\RequirePackage{numprint}
\npthousandsep{,}
\RequirePackage[utf8]{inputenc}
\RequirePackage{booktabs}
\RequirePackage{array}
\RequirePackage{multirow}
\RequirePackage{footnote}
\RequirePackage{Sweave}
\RequirePackage{xr}
\def\spacingset#1{\renewcommand{\baselinestretch}{#1}\small\normalsize} \spacingset{1}

\newtheorem{theorem}{Theorem}[section]
\newtheorem{corollary}{Corollary}[theorem]

\newtheorem{remark}[theorem]{Remark}
\newtheorem{definition}[theorem]{Definition}

\newcommand{\Hf}{H\kern-0.3em f}                                %
\newcommand{\Ha}{H\kern-0.28em a}                               %
\newcommand{\convind}{\buildrel{\cal D}\over\longrightarrow}    %
\newcommand{\convas}{\buildrel{a.s.}\over\longrightarrow}       %
\renewcommand{\P}{{\mathbb{P}}}
\newcommand{\E}{{\mathbb{E}}}
\newcommand{\var}{{\mathrm{var}}}
\newcommand{\cov}{{\mathrm{cov}}}

\newcommand{\arxiv}[1]{arXiv: \href{https://arxiv.org/abs/#1}{#1}}
\newcommand{\tck}{'} 
\newcounter{Stbls}

\newif\ifmnscpt  
\newif\ifsuppl
\newif\ifnboth
\newif\ifblind
\newif\ifunblind

\mnscpttrue
\supplfalse
\nbothtrue
\blindfalse
\unblindtrue

\ifmnscpt
\ifnboth
\fi 
\fi 

\ifsuppl
\ifnboth
\fi 
\fi 

\begin{document}

\ifsuppl 
\ifnboth
\title{Supplementary Material Accompanying ``Average- and $\lambda$- powers under BH-FDR}
\author{~~~}
\maketitle
\fi 
\fi 

\ifmnscpt 
\ifunblind
{
  \title{Average Power and $\lambda$-power in Multiple Testing Scenarios when the Benjamini-Hochberg 
    False Discovery Rate Procedure is Used\footnote{\arxiv{1801.03989}}}
  \author{Grant Izmirlian\thanks{This article is a U.S. Government work and is in the public domain in the U.S.A.}\hspace{.2cm}\\
    Biometry Research Group, Div. of Cancer Prev., NCI\\
    9609 Medical Center Dr; Bethesda, MD 20892-9789\\
    \href{mailto:izmirlig@mail.nih.gov}{izmirlig@mail.nih.gov}}
  \maketitle
} 
\fi 
\ifblind
{
 \title{Average Power and $\lambda$-power in Multiple Testing Scenarios when the Benjamini-Hochberg 
    False Discovery Rate Procedure is Used}
  \author{~~~}
  \maketitle
} 
\fi 

\bigskip
\begin{abstract}
We discuss several approaches to defining power in studies designed around the Benjamini-Hochberg (BH) false discovery
rate (FDR) procedure. We focus primarily on the \textit{average power} and the $\lambda$-\textit{power}, which are the
expected true positive fraction and the probability that the true positive fraction exceeds $\lambda$, respectively.
We prove results concerning strong consistency and asymptotic normality for the positive call fraction (PCF), the true
positive fraction (TPF) and false discovery fraction (FDF). Convergence of their corresponding expected values, including a
convergence result for the average power, follow as a corollaries. After reviewing what is known about convergence in
distribution of the errors of the plugin procedure, \cite{GenoveseC:2004}, we prove central limit theorems for fully
empirical versions of the PCF, TPF, and FDF, using a result for stopped stochastic processes. The central limit theorem
(CLT) for the TPF is used to obtain an approximate expression for the $\lambda$-power, while the CLT for the FDF is used
to introduce an approximate procedure for determining a suitably small nominal FDR that results in a speicified bound on
the FDF with stipulated high probability. The paper also contains the results of a large simulation study covering a
fairly substantial portion of the space of possible inputs encountered in application of the results in the design of a
biomarker study, a micro-array experiment and a GWAS study.
\end{abstract}

\thispagestyle{empty}
\noindent%
{\it Keywords:} MSC-2008 Primary: 62L12; MSC-2008 Secondary: 62N022; Multiple testing; false discovery rate; average power; k power; LLN; CLT\\
\ifunblind
\arxiv{1801.03989}
\fi 
\vfill

\newpage
\spacingset{1.45} 

\pagestyle{plain}
\setcounter{page}{1}

\fi 


\ifmnscpt 
\section{Introduction}
%
%
%
%
The explosion of available high-throughput technological pipelines in the biological and medical sciences over the past
20 years has opened up many new avenues of research that were previously unthinkable. The need to understand the role of
power in the era of ``omics'' studies cannot be overstated.  As a case in point, consider the promise and pitfalls of
RNA expression micro-arrays. One of the take away themes is that new technologies such as this one enjoy initial
exuberance and early victories \citep{AlizadehA:2000}, followed by calls for caution from epidemiologists and
statisticians \citep{IonnidisJPA:2005, BaggerlyKA:2009}. Some of the most constructive things gleaned from this journey,
in hindsight, have been a thorough re-evaluation of what should constitute the ``bar for science''. More and more
researchers are starting to realize that some of the blame for lack of reproducibility is owed to lack of power
\citep{IonnidisJPA:2005}. Central to these renewed calls for scientific vetting has been the concept of multiple
testing. Very early in the history of ``omics'' researchers realized that correction for multiple testing should be done
and that the most commonly used method, Bonferroni correction, was by far too conservative for tens of thousands of
simultaneous tests. Somewhat prophetically, half a decade before, \cite{BenjaminiY:1995} and colleagues introduced a new
testing paradigm, that, in contrast to the Bonferroni procedure which controls the probability that one or more type I
errors are committed, instead controls the proportion of false discoveries among the tests called significant.  By now,
use of the Benjamini-Hochberg (BH) false discovery rate (FDR) procedure for making statistical significance calls in
multiple testing scenarios is widespread. 

\section{Definitions and Notation}
Before describing the model of the data distribution, we need the following definition.
\begin{definition}
  A family of pdf's, $\{f_{\nu}:\nu \geq 0\}$, has the \textit{monotone likelihood ratio} property if and only if  
  \begin{equation}
    \nu' > \nu \mathrm{~implies~that~} f_{\nu'}/f_{\nu} \mathrm{~is~monotone~increasing}. 
  \end{equation}
  A family of CDF's has the monotone likelihood ratio property if its members are absolutely continuous and the 
  corresponding family of pdf's has the property. 
\end{definition}
Now, consider $m$ simultaneous tests of hypotheses $i=1,2,\ldots,m$, each a test whether a location parameter is 0, ($H_{0,i}$)
or non-zero ($H_{A,i}$). We start by supposing that an expected proportion, $0 < r < 1$, of the test statistics are
distributed about non-zero location parameters. The test statistic distributions are modeled according to a mixture model,
first introduced by \cite{StoreyJD:2002} and others \cite{BaldiP:2001, IbrahimJG:2002}.  First, let $\{\xi_i\}_{i=1}^{m}$
be an i.i.d. Bernouli $\{0,1\}$ sequence, with success probability, $r$. Denote the binomially distributed sum, 
$M_m = \sum_{i=1}^m \xi_i$, which is the number of test statistics belonging to the non-zero location parameter population. 
For a sample of size $n$ replicate outcomes resulting in $m$ simultaneous tests, let $X_{i,n}$ be the $i^{th}$ test statistic. 
We assume that conditional upon $\xi_i$, that its CDF is of the form: 
\begin{equation}
F_{X_{i,n}|\xi_i} = (1-\xi_i) F_{0, n} + \xi_i F_{A,n}.
\end{equation}
In the above, $F_{0,n}$ is the common distribution of all null distributed tests, and $F_{A,n}$ is the common distribution
of all non-null distributed tests. We assume that $F_{0,n}$ is the ``minimal'' element of a class of CDF's, 
$\mathbb{F}_n = \{F_{\nu, n} : \nu \geq 0\}$, satisfying the monotone likelihood ratio principle and that 
$F_{A,n} = \sum_{\ell=1}^h s_{\ell} F_{\nu_{\ell}, n}$ is a finite mixture of elements $F_{\nu_{\ell}, n}\in \mathbb{F}_n$, 
including the possibility that the mixing proportions are degenerate at a single distribution. Note that since we will 
only consider only two-sided tests, our scope is solely focused on non-negative test statistics, $X_{i,n}$, since they 
represent the the absolute value of some intermediate quantity. We consider $X_{i,n}$ to be non-negative in the remainder 
of the paper. Let $\bar F$ denote the complementary CDF (cCDF), so that $\bar F_{0,n}(x) = \P\{ X_{i,n} > x \mid \xi_i=0\}$
and $\bar F_{A,n} = \P\{ X_{i,n} > x \mid \xi_i = 1\}$. We will for nearly the entire paper make the assumption of
independent hypothesis tests. While the Benjamini-Hochberg false discovery rate procedure is not imune to departures
from the assumption of independent tests, the effect of departures from independence (i) do not affect the expected
false discovery fraction or expected true positive fraction (defined precisely below) and (ii) discrepencies which do 
result from such departures are seen in the distribution of the FDF and TPF and are caused by a reduced effective number 
of simultaneous tests. For these reasons, results obtained under the assumption of independent hypothesis tests are still 
of great utility. We return to this discussion in the final section.

\subsection{Mixed distribution of the P-values}
For $i=1,2,\ldots,m$, let $P_i = \bar F_{0,n}(X_{i,n})$ denote the two-sided nominal p-values corresponding to the test
statistics, $X_{i,n}$, and let $P^m_{(i)}$ denote their order statistics. Notice that the nominal p-values, 
$P_i, i=1,2,\ldots,m$ are i.i.d. having mixed CDF $G(u) = \P\{ P_i \leq u \}$, 
\begin{equation}
  G(u) = (1-r) u + r \bar F_{A,n}(\bar F_{0,n}^{-1}(u))\,.\label{eqn:Gdefnd}
\end{equation}
As we shall see in the proofs of Theorems \ref{thm:Jom_convas}, \ref{thm:SoM_convas}, and \ref{thm:CLT}, below,
the requirement that the family $\{F_{\nu, n}: \nu \geq 0 \}$ satisfies the monotone likelihood ratio
principle guarantees that $G$ is concave. Next in the original unsorted list of nominal p-values, 
$\{P_i: i=1,2,\ldots,m\}$, let $\{P_{1,i}:i=1,2,\ldots, M_m\}$ be the subset of nominal p-values corresponding to test 
statistics from the non-central population, in the order that they occur in the original unsorted list, e.g., if 
$N_{m,i} = \min\{j:i=\sum_{\ell=1}^j \xi_{\ell}\}$ counts the number of non-centrally located statistics among the first $i$ in the
original unsorted list, then $P_{1, i} = P_{_{N_{m,i}}}$.  

\subsection{Numbers of significant calls, true postives and false positives}
The Benjamini and Hochberg (B-H) false discovery rate (FDR) procedure \citep{BenjaminiY:1995} provides a simultaneous test
of all $m$ null hypotheses, that controls for multiplicity in a less conservative way than Bonferroni adjustment by
changing the paradigm. Instead of controlling the probability that one or more null hypotheses is erroneously rejected, it
controls the expected proportion of null hypotheses rejected that were true, or equivalently, the posterior probability
that a test statistic has null location parameter given it was called significant. The algorithm is implemented by
specifying a tolerable false discovery rate, $f$, and then finding the largest row number, $i$, for which the
corresponding order statistic, $P^m_{(i)}$, is less than $i f/m$. The total number of test statistics in the rejection
region, $J_m$, is given by the following expression:
\begin{definition}
  \label{def:J}
\begin{equation}
J_m = \max\left\{ i : P^m_{(i)} \leq \frac{i f}{m} \right\}\label{eqn:Jdef}\,.\nonumber
\end{equation}
\end{definition}
We will refer to $J_m$ as the number of \textit{positive calls} or \textit{discoveries} which is consistent with the 
terminology of Benjamini and Hochberg, and we call the ratio  $J_m/m$ the \textit{positive call fraction}. Notice that 
expression \ref{eqn:Jdef} has the following alternate form:
\begin{equation}
J_m = \sum_{i=1}^m I\left\{ P_i \leq m^{-1}  J_m \,f\right\}\label{eqn:JmsumExch}
\end{equation}

The number of positive calls partitions into \emph{true positve calls} and \emph{false positive calls}
\begin{definition}
  \label{def:S}
Let $S_m$ denote the \emph{number of true positive calls}:
\begin{equation}
  S_m = \sum_{i=1}^m \xi_i \, I\left( P_i \leq m^{-1} J_m f \right)\,.\label{eqn:SmsumExch}
\end{equation}
\end{definition}
\begin{definition}
  \label{def:T}
Let $T_m$ denote the \emph{number of false positive calls}:
  \begin{equation}
  T_m = \sum_{i=1}^m (1-\xi_i) \, I\left( P_i \leq m^{-1} J_m f \right)\,.\label{eqn:TmsumExch}
\end{equation}
\end{definition}
There are several possible choices of normalizers for $S_m$ and $T_m$, depending upon the popultion value being 
estimated.  Because power in the single hypothesis test scenario is the probability of rejection conditional
upon the alternative distribution, it is natural to normalize by the number of non-null distributed statistics, $M_m$:
\begin{definition}
We define the true positive fraction as the ratio $S_m/M_m$.
\end{definition}
Results concerning the false discovery rate will follow as corollaries to our other results. For this reason, we
normalize the number of false positive calls, $T_m$, by the number of positive callse, $J_m$:
\begin{definition}
We define the false discovery fraction as the ratio $T_m/J_m$.
\end{definition}
In general we will use the term \textit{fraction} for a ratio that is a random quantity and \textit{rate} for its
expectation.

Table \ref{tbl:SensSpec} shows rows partitioning the test statistics into those that are non-null distributed, 
and those that are null distributed, numbering $M_m$ and $m - M_m$, respectively, and columns 
partitioning the results of hypothesis testing into the positives and negatives calls, numbering $J_m$, and $m-J_m$,
respectively. 

\subsection{False discovery rate}
Let $f_0 = (1-r) f$. \cite{BenjaminiY:1995} showed in their original paper that their procedure 
controls the expected false discovery fraction, which they called the false discovery rate:
\begin{equation}
\E\left[\frac{T_m}{J_m}\right] = f_0 \leq f \label{eqn:BHFDR}
\end{equation}
In keeping with pervasive terminology, the phrase ``false discovery rate'' is applied to both the expected false discovery
fraction, $f_0=\E\left[J_m^{-1} T_m \right]$, and in addition, the nominal value, $f$ which is used to set the threshold
on the p-value scale. We will use the symbol $\mathrm{BHFDR}(f)$ to denote the BH-FDR procedure at nominal false discovery
rate, $\mathrm{FDR}=f$.  In this paper, whenever a random variable occurs in the denominator, we tacitly define the
indeterminate 0/0 to be 0, which has the effect that all such ratios are defined jointly with the event that the
denominator is non-zero.
\vspace{0.5truein}
\begin{table}[b]
\centering
\begin{tabular}{lrrr}
  \toprule
  &rej $H_{0,i}$&acc $H_{0,i}$&row Total\\
  \cmidrule(r){1-1}\cmidrule(lr){2-2}\cmidrule(lr){3-3}\cmidrule(l){4-4}
$H_{0,i}$ is FALSE & $S_m$ & $M_m - S_m$ & $M_m$ \\ 
  $H_{0,i}$ is TRUE & $T_m$ & $(m-M_m)-T_m$ & $m - M_m$ \\ 
   \cmidrule(r){1-1}\cmidrule(lr){2-2}\cmidrule(lr){3-3}\cmidrule(l){4-4}
col Total & $J_m$ & $m-J_m$ & $m$ \\ 
   \bottomrule
\end{tabular}
\caption{Counts of true positives, false positives, false negatives and true negatives.} 
\label{tbl:SensSpec}
\end{table}
\section{The distribution of $S_m$ and notions of power in multiple testing scenarios}
In the single hypothesis test situation, $m=1$, we consider probabilities of rejection given $H_0$ is true or false.
Under the setup introduced here, when $m=1$, the BH-FDR, $f$ becomes the type I error probability, and the power as it is
usually defined for a single hypothesis test, is the conditional expectation of $S_1$ given that $\xi_1=1$. In the case of
multiple tests, $S_m$ is distributed over values from zero to as high as $m$ so that naturally there are a multitude of
avenues for conceptualizing the power. Consider first, that had we been using the Bonferroni procedure for multiple tests
adjustment to thresh-hold the test statistics arriving at $J_m$ positives and $S_m$ true positives, the distribution of
$S_m$ would have been binomial with common success probability equal to the per-test power. The fact that the distribution
of $S_m$ is not binomial when the BH-FDR criterion is used is what makes discussion of power more difficult. However, the
common thread is that any discussion of power in the multiple testing scenario must be based upon some summary of the
distribution of $S_m$, e.g. a right tail or a moment.

\subsection{Various definitions of power in multiple testing scenarios}
One of the first approaches was to use the probability that $S_m$ is non-zero: $\P\left\{S_m > 0\right\}$. 
\cite{LeeMLT:2002} used a Poisson approximation to derive a closed-form expression for the probability to observe
one or more true positives.  This kind of power, the family-wise power, is arguably not a meaningful target of
optimization for experiments built around a large number of simultaneous tests, especially when there are typically
complex underlying hypotheses relying on positive calls for a sizable portion of those tests for which the
alternative is true. For example, consider that in a micro-array experiment in which there will be downstream
pathways analysis, we would start by assuming that there are around 3\% or more of the $m$ tests for which the 
alternate hypothesis is true, and hope to make significant calls at an FDR of 15\% for at least 80\% of these non-null 
distributed statistics, so as to have a thresholded list of roughly $1600$ genes to send to an analysis of pathways.

\subsection{The Average Power and $\lambda$-power}\hfil\break
\subsection*{Average Power}
In the BH-FDR procedure for multiple testing, the role of the type I error played in the single testing scenario is 
assumed by the FDR which is an expected proportion. Therefore, it is natural in the multiple testing scenario to 
consider a power that is also defined as an expected proportion. One interpretation of power in the setting of
multiple testing that falls along this line of reasoning is the ``average power''.
\begin{definition}
The average power is the expected true positive fraction, i.e. the expected proportion of all non-null distributed 
statistics that are declared significant by the BH FDR procedure.
\begin{equation}
    \pi_{_{\mathrm{av},m}} = \E\left[\frac{S_m}{M_m}\right] \label{eqn:avgpwr}
\end{equation}
\end{definition}
Notice that here the dependence upon $m$ is made explicit, so that the average power depends upon the number of
simultaneous tests in addition to quantities named above. \cite{GlueckD:2008} provided an explicit formula for the
average power in a finite number, $m$, of simultaneous test, but its complexity grows as the factorial of the number of
simultaneous tests, and this is clearly intractable in the realm of micro-array studies and GWAS where there are tens of
thousands or even a million simultaneous tests in question.\hfil\break

\subsection{The plug-in estimate of the average power}
Thinking heuristically for the moment, if $m$ is very large as will be the case in many ``omics'' scenarios, then
the positive call fraction $J_m/m$ could be considered very close to a limiting value, $\gamma$, if such a limiting
value existed either in probability or almost surely. Continuing along heuristic lines then, we could replace the
positive call fraction, $J_m/m$, with the limiting constant, $\gamma$, in the right hand side of expression 
\ref{eqn:JmsumExch}, as well as in expressions \ref{eqn:SmsumExch} and \ref{eqn:TmsumExch}, which define $S_m$ and 
$T_m$. If such a treatment were legitimate, the resulting analysis would be extremely easy as $J_m, S_m$ and $T_m$ 
would be sums of i.i.d's and the usual L.L.N. holds, with limits given by the expected value of a single increment 
in the corresponding sum. This gives rise to the plug-in estimate of the average power,
\begin{equation}
  \pi_{\mathrm{pi}} = \P\{ P_i < \gamma f | \xi_i = 1\}  = \bar F_{A,n}(\bar F_{0,n}^{-1}(\gamma f))
\end{equation}
Several authors have discussed this plug-in estimate of average power, for example \cite{GenoveseC:2004,
  StoreyJD:2002}.  Independently, \cite{JungSH:2005} and \cite{LiuP:2007} discuss sample size and power in the
setting of multiple testing based upon the BH FDR procedure. Without actually ever calling it the average power,
they derive an expression very close to the above plugin estimate. Their derivation starts with the posterior
probability that a statistic was drawn from the null-distributed population, given that it was called
significant. Bayes theorem is used to express this in terms of the prior, $1-r$ and $r$, and conditional, $\bar
F_{0,n}$ and $\bar F_{A,n}$. They mistakenly equate this to the nominal false discovery rate, $f$, when in actuality
it is the observed false discovery rate, $f_0 = (1-r) f$. Not withstanding, their methodology is valid because the
resulting power is, as we shall see below, the average power at the ``oracle'' threshold (\cite{GenoveseC:2004}) on
the p-value scale , $\gamma f/(1-r)$. This is the largest cut-off that is still valid at the nominal false discovery
rate, $f$. Around the same time, \cite{SunWG:2007}, discussed a generalization of of the FDR procedure based upon
the local FDR, and showed via decision theoretic techniques, that the false non-discovery rate, a quantity related
to the average power, has better performance characteristics than the FDR under many circumstances.  The paper also
provided a very good survey of the then currently available results. While the ramifications of that work are great,
the results provided here have merit in that the results and methodology supplied in the form of second order
asymptotics for the TDF and the FDF are entirely new and have important ramifications in of themselves.

\subsection*{The $\lambda$-Power}
Use of the average power in designing studies or deriving operating characteristics of them makes sense only when the
width of the distribution of $S_m/M_m$ is very narrow. To have more definitive control over the true positive fraction,
some authors have introduced the ``K-power''. This was originally introduced in a model where the number of non-null
distributed tests was fixed and was defined as the probability that the number of true positives exceeded a given
integral threshold, $k$.  Under the current setup in which the number of non-null distributed tests is a binomial random
variable this no longer makes sense. We introduce instead the $\lambda$-power, which is the probability that the true
positive fraction, $S_m/M_m$ exceeds a given threshold, $\lambda \in (0, 1)$:
\begin{definition}
We define the $\lambda$-power:
\begin{equation}
\pi_{_{S/M}}(\lambda) = \P\left\{\frac{S_m}{M_m} \geq \lambda\right\}.\label{eqn:Lpower}
\end{equation}
We will also use the term ``$\lambda_{k}$-power'' to denote $\pi_{_{S/M}}(k/100)$, the $\lambda$-power at threshold $k/100$.
The associated quantile function is denoted:
\begin{equation}
\lambda_{_{S/M}}(\pi) = \pi_{_{S/M}}^{-1}(\pi).\label{eqn:leq}
\end{equation}
\end{definition}
As mentioned above, the $\lambda$-power becomes especially meaningful in experiments for which there are a small to intermediate 
number of simultaneous tests and for which the distribution of the TPF, $S_m/M_m$, becomes non-negligibly dispersed.  We
prove a CLT for the true positive fraction which we use to approximate the $\lambda$-power. The accuracy of this approximation 
will be investigated in a simulation study.

\begin{remark}
Because the distribution of the TPF is nearly symetric for even relatively small values of $m>50$, the mean and median
nearly coincide. Thus 
\begin{equation}
\pi_{_{S/M}}(\lambda) \approx 1/2 \mathrm{~when~} \lambda=\pi_{_\mathrm{pi}} \label{eqn:LpwrEqHlf}
\end{equation}
Because the $\lambda$-power takes the values $1$ when $\lambda=0$ and $0$ when $\lambda=1$ and is continuous by assumption,
there exists a quantile, $\lambda_{eq} = \pi_{_{S/M}}^{-1}(\pi_{_\mathrm{pi}})$, at which the $\lambda$ power equals the average power:
\begin{equation}
\pi_{_{S/M}}(\lambda_{eq}) = \pi_{_\mathrm{pi}}\,. \label{eqn:Lambda_eq}
\end{equation}
Because the $\lambda$-power is a cCDF it is a non-increasing function of $\lambda$,
\begin{equation}
\pi_{_{S/M}}(\lambda) < \pi_{_\mathrm{pi}} \mathrm{~for~} \lambda > \lambda_{eq} \mathrm{~and~} \pi_{_{S/M}}(\lambda) > \pi_{_\mathrm{pi}} \mathrm{~for~} \lambda < \lambda_{eq} 
\label{eqn:LPwr_AvgPwr}
\end{equation}
\end{remark}

\subsection*{Bounding the FDF}
We conclude the section on various notions of power with a brief diversion. The fact that the TPF may be non-negligibly
dispersed at small to intermediate values of $m$ leads to concern that the FDF distribution is similarly dispersed at
these small to intermediate values of $m$. This concern is addressed in one of the CLT results and in the simulation
study. We introduce some notation for its cCDF and quantile function.
\begin{definition}
At BH-FDR $f$, denote the FDF tail probability: 
\begin{equation}
\pi_{_{T/J}}(\lambda) = \P\left\{\frac{T_m}{J_m} \geq \lambda\right\}\,.\label{eqn:FDFcCDF}
\end{equation}
Denote its quantile function:
\begin{equation}
\lambda_{_{T/J}}(p) = \pi_{_{T/J}}^{-1}(p)\,.\label{eqn:lToJp}
\end{equation}
\end{definition}
At small and intermediate values of $m$, the value of $\lambda_{_{T/J}}$ required to bound the FDF by $f_0$ with probability
bounded by $f_0$ can be as much as 100\% larger than the FDR. As remarked above, this will be discussed further in the 
context of our CLT results and simulation studies below.

The remainder of the paper proceeds according to the following plan. Section 4 is a presentation of the main theoretical
results, and this is done two subsections. In subsection 4.1, almost sure limits of the positive call fraction, true
positive fraction and false discovery fraction, as the number of simultaneous tests tends to infinity, are shown to
exist and are fully characterized. Convergence of the corresponding expectations, the true positive rate or average
power, and false positive rate, follow as a corollaries. Subsection 4.2 contains central limit theorems (CLT's) for the
positive call fraction, true positive fraction and false discovery fraction.  We also provide a lower bound for the
average power at a finite number, $m$, of simultaneous tests. We show how these CLT results can be used to approximate
the $\lambda$-power allowing tighter control over the TPF in power and sample size calculations, as well as how the
approximate distirbution of the FDF can be used to tighten down control over the FDF at both the design and analysis
stage. Section 5 is devoted to a simulation study, in which we study the regions of the parameter space that are typical
to small biomarker studies, micro-array studies and GWAS studies. We also focus on characteristics of the distribution
of the FDF as the number of simultaneous tests grows. Weak consistency of $J_m/m, S_m/M_m$ and $T_m/J_m$ to $\gamma$, 
$\pi_{\mathrm{pi}}$ and $(1-r)f$ was proved in \cite{StoreyJD:2002} and \cite{GenoveseC:2002}. The paper \cite{GenoveseC:2004} is 
a study of consistency and convergence in distribution of the paths of the plug-in estimator, considered a stochastic
process in the p-value plugin criterion, $t\in(0,1)$. The strong consistency results and weak convergence results for
the observed positive call fraction, true positive fraction and false discovery fraction presented here are new. Note that
almost sure convergence of the positive call fraction is necessary for almost sure convergence of the true positive and 
false discovery fractions.c
\section{Theoretical Results}
\subsection{Law of Large Numbers}\hfil\break
\subsection*{LLN for Positive Call Fraction, $J_m/m$}
\begin{theorem}
  \label{thm:Jom_convas}
If the family $\{F_{\nu,n}:\nu \geq 0\}$ is absolutely continuous and has the monotone likelihood
ratio property, then 
\begin{equation}
\lim_{m\rightarrow\infty} m^{-1}\,J_m = \sup \{ u: u = G(u f) \} \equiv \gamma \mathrm{~almost~surely,}
\end{equation}
\end{theorem}
Proofs of this and all other results are contained in the accompanying supplemental material. 

\begin{remark}
  When the family $\{F_{\nu,n}: \nu \geq 0\}$ has the monotone likelihood ratio property, $\gamma$ will be the unique 
  non-zero solution of $G(u f) = u$. 
\end{remark}
Once the almost sure convergence of $J_m/m$ is established we can apply a result of \cite{TaylorRL:1985} 
to establish convergence results for the true positive fraction. 
\subsection*{LLN for the True Positive Fraction, $S_m/M_m$}
\begin{theorem}
  \label{thm:SoM_convas}
Under the conditions of theorem \ref{thm:Jom_convas},
\begin{eqnarray}
\lim_{m\rightarrow\infty} m^{-1}\,S_m &=& \P\{P_{i} \leq \gamma f\,, \xi_i=1\} = r\,\bar F_{A,n}(\bar F_{0,n}^{-1}(\gamma f)) 
                                     \mathrm{~a.s.}\,,\label{eqn:Som_convas}\\
&&\nonumber\\
\lim_{m\rightarrow\infty} M_m^{-1}\,S_m &=& \P\{P_{i} \leq \gamma f \mid \xi_i=1\} = \bar F_{A,n}(\bar F_{0,n}^{-1}(\gamma f))
                                      \equiv \pi_{_\mathrm{pi}}\mathrm{~a.s.}\label{eqn:SoM_convas}\\
\mathrm{and}&&\nonumber\\
\lim_{m\rightarrow\infty} \pi_{_{\mathrm{av},m}} &=& \lim_{m\rightarrow\infty}\E\left[M_m^{-1}\,S_m\right] =\pi_{_\mathrm{pi}}\label{eqn:E_SoM_conv}
\end{eqnarray}
\end{theorem}

Corresponding convergence results for the false discovery fraction and its expected value follow as a corollary.

\subsection*{LLN for the False Discovery Fraction, $T_m/J_m$}
\begin{corollary}
  \label{cor:FDF_convas}
Under the conditions of theorem \ref{thm:Jom_convas},
\begin{eqnarray}
\lim_{m\rightarrow\infty} m^{-1}\,T_m &=& \P\{P_{i} \leq \gamma f\,, \xi_i=0\} = (1-r)\,\gamma f 
                                     \mathrm{~a.s.}\,,\label{eqn:Tom_convas}\\
&&\nonumber\\
\lim_{m\rightarrow\infty} J_m^{-1}\,T_m &=& \P\{\xi_i = 0 \mid P_{i} \leq \gamma f \} = (1-r) f 
                                      \mathrm{~a.s.}\label{eqn:FDF_convas}\\
\mathrm{and}&&\nonumber\\
\lim_{m\rightarrow\infty} \E\left[J_m^{-1}\,T_m\right] &=& (1-r) f
\end{eqnarray}
\end{corollary}

\begin{remark}
Because $T_m = J_m - S_m$ then by Theorem \ref{cor:FDF_convas} and its corollary \ref{thm:SoM_convas}, we obtain the
identity $(1-r) f = 1 - r \pi_{_\mathrm{pi}}/\gamma$, which can be rearranged to obtain an expression for the limiting
positive call fraction:
\begin{equation} 
  \gamma = \frac{r \pi_{_\mathrm{pi}}}{1-f_0}.\label{eqn:gamma_expr}
\end{equation}
\end{remark}

\begin{remark}
  \label{rmk:oracle}
If the nominal false discovery rate, $f$, is replaced by the inflated value, $f/(1-r)$, resulting in the 
$\mathrm{BHFDR}(f/(1-r))$ procedure, note that the FDR is still controlled at the nominal value, $f$, since in this 
case, $\mathbb{E}[ J - S / J ] = f$ due to cancellation.  This threshold, $\gamma\,f/(1-r)$, on the p-value scale, has
been called the oracle threshold by some authors, \cite{GenoveseC:2004}, because it is the criterion resulting in the 
largest power which is still valid for a given FDR, $f$. Call this the oracle average power, $\pi_o$. The actual 
difference only begins to get appreciable as $r$ increases in size. In practice, as we will see in our simulation 
study, $r$ must be as large as 50\% or more before this has a dramatic effect on the average power. Keep in mind that 
in practice when analyzing a given dataset, this increased power is only attainable at the stage of estimation if 
a reasonably good estimate of $r$ is possible. The fact that this is very problematic has also been a topic of 
much discussion.
\end{remark}

\noindent Next, if we replace $\gamma$ in the definition of the IST average power \ref{eqn:SoM_convas} by the expression
\ref{eqn:gamma_expr}, we arrive at a new equation which gives an implicit definition for the IST average power.
\begin{corollary}
  \label{cor:pi1Alt}
Under the conditions of theorem \ref{thm:Jom_convas},
  \begin{equation}
\pi_{_\mathrm{pi}} = \bar F_{A,n}\left(\bar F_{0,n}^{-1}\left((1-f_0)^{-1} r \pi_{_\mathrm{pi}} f\right)\right)\,.
\end{equation}
\end{corollary}

\noindent The almost sure convergence results given in Theorems \ref{thm:Jom_convas}, \ref{cor:FDF_convas} and 
\ref{thm:SoM_convas} above each have corresponding central limit results which we state now. The first, a CLT for 
the centered and $\sqrt{m}$-scaled positive call fraction, $J_m/m$, is needed in the proof of the second 
and third results, CLT's for centered and scaled versions of the false discovery fraction and the true positive 
fraction.

\subsection{CLTs for the PCF, FDF, and TPF; Lower Bound for Average Power}
\begin{theorem}
  \label{thm:CLT}
Under the conditions of theorem \ref{thm:Jom_convas},
  \begin{eqnarray}
  \sqrt{m} \Big(m^{-1} J_m \kern-0.75em &-&\kern-0.75em \gamma \Big) \convind N(0, \tau^2)\,.\label{eqn:CLT_Jortm}\\
  &&\nonumber\\
  \sqrt{m} \Big(J_m^{-1} T_m \kern-0.75em &-&\kern-0.75em f_0 \Big) \convind N(0, \alpha^2)\,.\label{eqn:CLT_rtmFDF}\\
  &&\nonumber\\
  \sqrt{m} \Big(M_m^{-1} S_m\kern-0.75em &-&\kern-0.75em \pi_{_\mathrm{pi}}\Big)\convind N(0,\sigma^2)\,.\label{eqn:CLT_rtmSoM}
  \end{eqnarray}
\end{theorem}
The proof, which uses results on convergence of stopped stochastic processes, is constructive in nature producing fully
characterized limiting distributions yielding asymptotic variance formulae. We reiterate the practical
implications of these results.  

\subsection*{Approximating the $\lambda$-power via the CLT for the TPF}
Currently, multiple testing experiments are designed using the average power, $\pi_{_\mathrm{pi}}$, which is the mean of
the distribution of $S_m/M_m$. In cases in which the width of this distribution is non-negligible, e.g. $m<1000$
simultaneous tests or so, we recommend using the $\lambda$-power instead of the average power. As we defined above, in
equation \ref{eqn:Lpower}, the $\lambda$-power is the probability that the TPF exceeds a given $\lambda$. We will see in 
our simulation study that in the ranges of the parameter space investigated, this CLT approximation is quite good and can
be used to approximate the $\lambda$-power:
\begin{equation}
 \pi_{_{S/M}}(\lambda) = \P\{ S_m/M_m \geq \lambda\} \approx \Phi(\sqrt{m}/\sigma (\pi_{_\mathrm{pi}} - \lambda))\,,\label{eqn:lpwrCLT}
\end{equation}
where $\sigma$ above is the square-root of the asymptotic variance, $\sigma^2$, given in formula \ref{eqn:s2} in the proof 
of theorem \ref{thm:CLT}. 

\subsection*{Enhanced control of the FDF via its CLT}
As defined above in expression \ref{eqn:lToJp} and the text leading up to it, $\lambda_{_{T/J}}(p) = \pi_{_{T/J}}^{-1}(p)$ is the
quantile of the FDF distribution at upper tail probability, $p$. The CLT for the FDF can be used to approximate it:
\begin{equation}
\lambda_{_{T/J}}(p) \approx  f_0 + \alpha/\sqrt{m}  \Phi^{-1}(1 - p)\,.
\end{equation}
Here, $f_0 = (1-r) f$ as above, $\alpha$ is the square root of the asymptotic variance, $\alpha^2$ given in formula \ref{eqn:a2}
in the proof of \ref{thm:CLT}, and $\Phi^{-1}$ is the standard normal quantile function. This can be used in several different
ways to bound the FDF with specified probability. Three possibilities are as follows. First, as a kind of loss function on 
lack of control inherent in the use of the $\mathrm{BHFDR}(f)$ procedure, we could determine how large a threshhold is required 
so that the FDF is bounded by $\lambda$ except for a tail probability of $f_0$
\begin{equation}
\lambda_{_{T/J}}(f_0) =  f_0 + \alpha/\sqrt{m}  \Phi^{-1}(1 - f_0)
\end{equation}
A second way would to find the solution $f\tck$ to the following equation.
\begin{equation}
f_0 =  (1-r) f\tck + \alpha/\sqrt{m}  \Phi^{-1}(1 - (1-r) f_0)\,.\label{eqn:fpr}
\end{equation}
This would produce a reduced FDR, $f\tck < f$, at which the $\mathrm{BHFDR}(f\tck)$ procedure would result in a FDF, $T_m/J_m$
of no more than $f_0$ with probability $1-f_0$. The solution is 
\begin{equation}
f\tck = f - \alpha/(\sqrt{m}(1-r))  \Phi^{-1}(1 - (1-r) f_0)
\end{equation}
A third way to do this, and the most conservative of the three, would be to determine the value of a reduced FDR, $f\tck$, at which 
the $\mathrm{BHFDR}(f\tck)$ procedure would result in a FDF no more than $f_0$ with probability $1-(1-r) f\tck$, by solving the 
following equation numerically: 
\begin{equation}
f_0 =  (1-r) f\tck + \alpha/\sqrt{m}  \Phi^{-1}(1 - (1-r) f\tck)\,.\label{eqn:fprime}
\end{equation}

\begin{remark}
The farther apart $f\tck$ is from $f_0$ is an indication of the dispersion of the distribution of $T_m/J_m$. 
\end{remark}
  
\begin{remark}
The procedure summarized in equation \ref{eqn:fprime}, for finding a reduced FDR, $f\tck$, at which
$\mathrm{BHFDR}(f\tck)$ would produce an FDF of no more than $f_0$ with probability $1-(1-r)f\tck$, can also be used at the
analysis phase. Note that expression \ref{eqn:a2} in the proof of \ref{thm:CLT} for the asymptotic variance, $\alpha^2$,
of $T_m/J_m$ depends only upon $f_0=(1-r)f$ and $\gamma$. Thus we can replace $f_0$ with $f$ and estimate $\gamma$
from the data using the plug-in estimate, $J_m/m$. This has important ramifications for the setting of small to intermediate 
number of simultaneous tests, $m\leq 1000$.
\end{remark}

\subsection*{Lower Bound for finite simultaneous tests average power}
As we will see in the simulation study which follows, the IST average power is in fact extremely close to 
the finite simultaneous tests (FST) average power for the broad ranges of the parameters studied. Nevertheless, it is 
still useful to have bounds for the FST average power. 

\begin{theorem}
\label{thm:LowerBdd}
The FST average power, $\pi_{_{\mathrm{av},m}}$, is bounded below by the following quantity, $\pi^{L}_{\mathrm{av},m}$, given below. 
\begin{eqnarray}
\pi_{_{\mathrm{av},m}} &\geq& \sum_{{\ell}=1}^{m} \binom{m}{{\ell}} r^{\ell} (1-r)^{m-{\ell}} \frac{1}{{\ell}} \sum_{s=1}^{{\ell}} 
             \bar B_{{\ell}-s+1,s}\left(\bar F_{\nu,n}\left(1-\bar F_{0,n}^{-1}\left(\frac{s f}{m}\right)\right)\right) \nonumber\\
    &\equiv& \pi^{L}_{\mathrm{av},m} \nonumber
\end{eqnarray}
\end{theorem}

An upper bound that seems to work in practice is the expression obtained by replacing $J_m/m$ with $r/(1-f_0)$ 
in equation \ref{eqn:lwrbddeqn3} in the proof of Theorem \ref{thm:LowerBdd}.




\section{Simulation Study}
We conducted four simulation studies. The first three of these had fixed $m$ and ranges of the other parameters chosen
based upon relevance to subject matter areas. The first, with $m=200$, was meant to model biomarker studies. In a second
simulation study, we varied $m$ in order to study characteristics of the FDF distribution as $m$ grows. In the third,  
in which $m=\numprint{54675}$, was meant to model micro-array studies for while the fourth, for which $m=\numprint{1000000}$, 
was meant to model genome wide association (GWA) studies.

In all four cases, the test statistic distributions, $F_{0,n}$ and $F_{\nu,n}$, were
chosen to be t-distributions of $2n - 2$ degrees of freedom. The common non-centrality parameter was fixed at 
$\nu=\sqrt{n/2}\theta$. This corresponds to a two group comparison as is often done. For each of these simulation
studies, we chose subject matter relevant ranges for the four parameters, the expected proportion of non-null tests,
$r$, the location parameter, $\theta$, and the false discovery rate, $f$. Except when set explicitly as in the fourth 
case, a range sample sizes, $n$, in increments of 5, was chosen to result in powers between 60\% and 95\% at each setting
of the other parameters. We conducted a total of four simulation studies. The first, with $m=200$ simultaneous tests, was
meant to model biomarker studies. The second, with varying sizes of $m$ ranging from \numprint{1000} to \numprint{20000}
was done in order to assess the width of the FDF distribution and the adequacy of the CLT approximation to it.
The third, with $m=$\numprint{54675} was meant to model human oligo-nucleotide micro-array experiments, and the
forth with $m=$\numprint{1000000} was meant to model GWA studies. We present the first two of these in the main 
text and the remainder in the supplementary material. 

\ifsuppl
\setcounter{Stbls}{0}
\refstepcounter{Stbls}\label{tbl:avgpwr_tbl_Array}
\refstepcounter{Stbls}\label{tbl:Lpwr_tbl_Array}
\refstepcounter{Stbls}\label{tbl:avgpwr_tbl_GWAS}
\refstepcounter{Stbls}\label{tbl:Lpwr_tbl_GWAS}
\refstepcounter{Stbls}\label{tbl:avgpwr_tbl_Bmkr}
\refstepcounter{Stbls}\label{tbl:Lpwr_tbl_Bmkr}
\refstepcounter{Stbls}\label{tbl:tbl_FDFincrN}
\fi 

For each of simulation studies focused on the distribution of the TPF, we computed, at each combination of these four
parameters, the IST average power, $\pi_{_\mathrm{pi}}$, from line \ref{eqn:E_SoM_conv} of Theorem \ref{thm:SoM_convas},
the oracle power, $\pi_o$, mentioned in remark \ref{rmk:oracle} above, and the lower bound, $\pi^{L}_{\mathrm{av},m}$, from
Theorem \ref{thm:LowerBdd}. We computed the approximate $\lambda_{75}$- and $\lambda_{90}$- powers using expression
\ref{eqn:lpwrCLT} based upon the CLT for the TPF (theorem \ref{thm:CLT}) using the expression for the asymptotic
variance, $\sigma^2$ (expression \ref{eqn:s2} given in the proof). Also, at each combination of parameters, we conducted
\numprint{1000} simulation replicates. At each simulation replicate, we began by generating $m$ i.i.d. Bernoulli
$\{0,1\}$ variables, with success probability, $r$, to assign each of the $m$ test statistics to the null (0) or
non-null (1) populations, recording the number, $M_m$, of non-null distributed test statistics. This was followed next
by drawing $m$ test statistics from $F_{0,n}$ or $F_{\nu,n}$, being the central and non-central (respectively)
t-distribution of $2 n -2$ degrees of freedom, corresponding to the particular value of $\xi_i$. Next, the B-H FDR
procedure was applied and the number of positive calls, $J_m$, and number of true positives, $S_m$ were recorded. At the
conclusion of the \numprint{1000} simulation replicates, we recorded the simulated average power as the mean over
simulation replicate of the TPF, $S_m/M_m$. In addition, the simulated $\lambda_{75}$- and $\lambda_{90}$- powers were
derived as the fraction of simulation replicates of the TPF that exceeded $0.75$ and $0.90$, respectively. Finally we
computed the sample size required for $\lambda_{90}$ power.

In another simulation study, focused on the distribution of the FDF for increasing $m$. At each combination of the 
parameters considered, we computed the reduced FDR, $f\tck$, required to bound the FDF with probability $(1-r)f\tck$
as the unique numerical solution to expression \ref{eqn:fprime}. We also computed the sample sizes $n_{_{0,0}}$ and
$n_{_{0,1}}$ required for specified average power under $\mathrm{BHFDR}(f)$ and under $\mathrm{BHFDR}(f\tck)$, 
respectively. Sample sizes $n_{_{1,0}}$ and $n_{_{1,1}}$ at specified $\lambda$-power under the corresponding procedure 
were also derived. A simulation, conducted in a fashion identical to that described above, under $\mathrm{BHFDR}(f\tck)$,
was done at each combination of parameters, this time including the additional two parameters $m$ and specified
power. From simulation replicates of the FDF, $T_m/J_m$, we computed the probability in excess of $f_0$.

All calculations were done in R, version 3.5.0 (\cite{CiteR:2016}) using a R package, \textbf{pwrFDR}, 
written by the author \cite{IzmirlianG:2018}, available for download on \textbf{cran}. Simulation was conducted on 
the NIH Biowulf cluster (\cite{Biowulf:2017}), using the swarm facility,  whereby each of 50 nodes was tasked with 
carrying out 20 simulation replicates resulting in \numprint{1000} simulation replicates for each configuration of 
parameters.

\subsection{Biomarker Studies}
For the first simulation study we considered experiments typical of biomarker studies with $m=\numprint{200}$
simultaneous tests. We attempted to cover a broad spectrum of parameters spanning the domain of typical biomarker study
designs. The false discovery rate, $f$, was ranged over the values $1\%$, and from
5\% to 30\% in increments of 5\%. The expected number of tests with non-zero means, $\mathbb{E}[M_m] =
m r$, was varied over the values 5, 10 and 20, and from 10 to 100 in increments of 10,
representing values of $r$ ranging from 0.025 to 0.5. The effect size,
$\theta$, was allowed to vary from 0.6 to 1.5 in increments of 0.1. At each configuration, a range of sample sizes were
chosen to result in powers between 50\% and 98\% as mentioned above. This resulted in
2,648 configurations of the parameters, $f, \mathbb{E}[M_m],\theta$, and $n$
(full set of parameter combinations). The job took roughly 7
minutes on the NIH Biowulf cluster.

Table \ref{tbl:avgpwr_tbl_Bmkr} tabulates the IST average power, the oracle power and the simulated mean of the TPF at
28 different parameter settings excerpted from the full set of
2,648 parameter combinations. Over the full set of parameter settings, the both
the IST $\pi_{_\mathrm{pi}}$, and oracle $\pi_o$ powers are very close to the simulated average power. The
difference between the IST power, $\pi_{_\mathrm{pi}}$, and the simulated power was less than
0.15, 0.95 and 2.00 at 50\%, 90\% and 99\% of the parameter settings, respectively. As remarked
earlier, the oracle power is actually the average power at the oracle threshold. Since it borders on feasible, we
allowed $r$ to take values as large as 0.5 for which the oracle
threshold has a substantial gain in power.  The oracle power differed from the IST power by
1.8\%, 8.5\% and 16\% at 50\%, 90\% and 99\% of the parameter settings, respectively, suggesting
that the oracle threshold is worth considering.  Recall that our IST power, $\pi_{_\mathrm{pi}}$ can be set to the oracle
threshold by setting the FDR to $f/(1-r)$. However, careful consideration must be taken if using the oracle threshold to
design a study, since when its time to actually threshold the data one needs a plug-in estimate of $r$ and as discussed
extensively in the literature, this can be problematic. The lower bound comes within roughly 10\% of the simulated
power, with differences with the simulated power less than 36\%, 45\%, 50\%, 56\% and 71\% at 20\%, 40\%,
50\%, 60\% and 80\% of the parameter settings, respectively.

Table \ref{tbl:Lpwr_tbl_Bmkr} displays, at threshold 0.75 and at threshold 0.90, the $\lambda$ power as derived from
the CLT \ref{thm:CLT} and estimated from simulation replicates (hatted version), respectively, excerpted from the full
set of 2,648 parameter combinatons as before. In the last column is the ratio 
of the sample size required for $\lambda_{90}$-power to the original sample size. First, we note that when restricted to
powers strictly between 50\% and 100\%, occurring at 1,488
parameter combinations, the CLT approximate- and simulated- $\lambda_{75}$-power were within the following relative error
of one another (median over parameter conditions (lower quartile, upper quartile)): 
2\% (0.7\%, 4.5\%), with 23.1\% over 5\%.  
Corresponding results for the simulated and CLT approximate $\lambda_{90}$-power for powers strictly between 50\% and
100\% occurring at 1275 of the parameter values, were within the following relativer
error of one another 3.3\% (1.3\%, 9.2\%), with 
37.6\% over 5\%. The greater discrepancy between CLT approximate
$\lambda$-powers and simulated values is due to the lack of accuracy of the CLT asymptotic approximation at such
small sample sizes, $n$.  Note that for sample sizes in excess of $n=20$ the degree of accuracy starts to improved
dramatically, especially at higher powers. Also noteworthy is corroboration in ordering of the
average power and $\lambda_{k}$-power based upon the size of $k$ relative to $100 \lambda_{eq}$. All values of
$\lambda_{eq}$ are less than 90\%, but some are between 75\% and 90\%, and the ordering of average power and $\lambda$
powers is in accordance with expression \ref{eqn:LPwr_AvgPwr}. 

Furthermore the discrepancy between the average power and
the $\lambda$-power is reflective difference between $\lambda$ and $\lambda_{eq}$. This trend is echoed in the magnitude
of the sample size ratio, with magnitude increasing in the discrepancy between $\lambda_{eq}$ and $0.90$. Note that in this
case, as the number of simultaneous tests, $m$, is relatively ``small'', the distribution of the TPF, $S_m/M_m$, 
is more dispersed and therefore, growth in the $\lambda$-powers is more gradual with increasing sample size.

\subsection{The false discovery fraction, intermediate number of simultaneous tests}
The second simulation study was focused on the use of the CLT for the FDF to find a bound for the FDF with large
probability. We varied the number of simultaneous tests, $m$, over 1000, 2500, 5000, 7500, 10000 and 20000. The effect size, $\theta$,
was varied over 2/3, 5/6 and 1. The proportion of statistics drawn from the non-null distributed population, $r$, 
ranged over 0.025, 0.05 and 0.075 and the FDR, $f$, ranged over the values 0.1, 0.15 and 0.2. At each set of values of 
these parameters, we used expression \ref{eqn:fprime} based upon the CLT for the FDF to find a reduced FDR
at which the BH-FDR procedure would result in an FDF of no more than $f_0$ with large probability.
We calculated the sample sizes required for specified average power under the original and reduced FDR.
Sample sizes required for specified $\lambda_{90}$-power under the original and reduced FDR were also calculated.
Finally, the probability that the FDF exceeded $f_0$ under the reduced FDR was estimated from simulation replicates.

Table \ref{tbl:tbl_FDFincrN} tabulates the reduced FDR, $f\tck$, required to bound the FDF by $f_0$ with probability
$1-(1-r)f\tck$, and the sample sizes $n_{_{0,0}}$ and $n_{_{0,1}}$, required for specified average power under 
$\mathrm{BHFDR}(f)$ and under $\mathrm{BHFDR}(f\tck)$, respectively. Also shown are the sample sizes, $n_{_{1,0}}$ and 
$n_{_{1,1}}$, required for specified $\lambda_{90}$-power under $\mathrm{BHFDR}(f)$ and under $\mathrm{BHFDR}(f\tck)$, 
respectively, as well as the simulated tail probability, $\hat\pi_{_{T/J}}(f_0)$, in excess of $f_0$.
The general trend for increasing $m$ as the distribution of $T_m/J_m$ collapses to a point mass at $f_0$ are
a value of $f\tck$ closer to $f$, and sample sizes under $\mathrm{BHFDR}(f\tck)$ that are less inflated relative to
corresponding sample sizes under $\mathrm{BHFDR}(f)$. The simulated right tail probability, $\hat\pi_{_{T/J}}$ 
should in theory have only simulation error about its theoretical value, $(1-r)f\tck$. However, as $f\tck$ is 
derived from an approximation based upon a CLT, we expect the accuracy of the approximation to improve with larger $m$.
Not surprisingly, the results are consistent with these observations. Over the full set of 324 parameter
settings we obtained the following results (median, (lower quartile, upper quartile)). The ratio of the reduced FDR, 
$f\tck$, to the original FDR, $f$: 0.79 (0.69, 0.86) when $m\leq 10,000$, and 
0.91 (0.88, 0.93) when $m>10,000$, showing that the reduced FDR gets closer in value to the original 
FDR with increasing $m$. The ratio of sample size required for average power at $\mathrm{BHFDR}(f\tck)$ to that at 
$\mathrm{BHFDR}(f)$: 1.06 (1.04, 1.1) when $m\leq 10,000$ and 1.03 (1.02, 1.04) 
when $m>10,000$, showing that the inflation factor reduces with increasing $m$. This is also the case when the 
sample sizes are derived for given $\lambda_{90}$-power: 1.05 (1.03, 1.08) when $m\leq 10,000$ and 
1.02 (1.0025, 1.03) when $m>10,000$, respectively. Finally, the ratio of the simulated tail probability, 
$\hat\pi_{_{T/J}}(f_0)$ to the CLT approximated value: 1.02 (0.95, 1.11) when $m\leq 10,000$ and
0.995 (0.94, 1.0675) when $m>10,000$ respectively, highlighting that the CLT approximation
gets better with increasing $m$.

In order to judge the relative impact changes in the parameters had, especially those unique to this setting of
multiple testing, we computed numerical partial derivatives of the power function with respect to the proportion of
test statistics distributed as the non-null distribution $r$, the effect size $\theta$, and the FDR $f$.  The
partials were then scaled to the range of the relevant parameter (max minus min) so that unit changes were
comparable and corresponded to the ranges of the parameters considered. Numerical partial derivatives were computed
at all 10,020 configurations of the parameters. These results were
summarized separately for each of the three parameters by calculating quartiles of the respective numerical partial
derivative at each given level of the respective parameter, over all configurations of all other parameters
resulting in powers of 50\%, 60\%, 70\%, 80\% and 90\%. The results corresponding to median values at 70\% power are
displayed in Figure \ref{fig:MA-derivs}.

\section{Discussion}
We proved LLNs for the PCF, FDF and TPF as well as CLTs for $\sqrt{m}$ scaled versions of them. Our LLN result for the 
TPF allowed characterization of the large $m$ limit and this in turn allowed a proper interpretation of the power 
discussed in \cite{JungSH:2005} and in \cite{LiuP:2007}, being nearly identical to the average power. Our CLT result
for the TPF allowed us to introduce the $\lambda$-power, similar in nature to the $k$-power discussed by previous 
authors. The $\lambda$-power allows tighter control over the TPF in the design of multiple testing experiments by
bounding the distribution of the TPF by an acceptable threshold, rather than just its mean, as is the case with the
average power. Our CLT result for the FDF provides a technique whereby an investigator can determine a reduced FDR
at which the usual BH-FDR procedure will result in a FDF no greater than a stipulated value with arbitrary large
probability. This latter technique is useful both at the design phase as well as the analysis phase because the
asymptotic variance depends only upon the limiting PCF, $\gamma$, and proportion belonging to the null distributed
population, $1-r$ and one can use $J_m/m$ as an estimate of $\gamma$ and consider $1-r\approx 1$ when faced with
a data analysis. Key to the proofs of the LLN results was, first, the LLN for the PCF, $J_m/m$, which was proved directly
via a simple argument. Prior results by \cite{GenoveseC:2004} obtained convergence only in probability. Once
established we applied a result of \cite{TaylorRL:1985} for a.s. convergence of triangular arrays of 
finite exchangeable sequences The proofs of the CLT results was made possible by building on the work of 
\cite{GenoveseC:2004} which considered $p$-values thresholded at a deterministic $t$, treating them as
stochastic processes. We applied a result of \cite{SilvestrovD:2004} for weak convergence of stopped stochastic processes.

In a very large and thorough simulation study, we investigated three major domains of the space of operating
characteristics typically encountered in the design of multiple testing experiments: two larger $m$ domains,
$m=\numprint{54675}$ typical to human RNA expression micro-array studies, and $m=\numprint{1000000}$ typical to GWA
studies and a smaller $m$ domain, $m=200$ which is typical of biomarker studies. In each case, we compared the average
power derived from the TPF LLN limit to simulated values and observed at all ranges of sample sizes that the agreement
was quite good.  We also used the CLT result for the TPF to compute approximate $\lambda$ powers and compared these with
the simulation distribution. Agreement in this case was overall very good, but there was some breakdown in the level of
accuracy in the asymptotic approximation at simultaneous tests $m<100$. The last simulation study was focused upon the
procedure for bounding the FDF with large probability and its behavior as the number of simultaneous tests, $m$, grows
from hundreds to tens of thousands. We noted that overall, the method is feasible, even when the asymptotic
approximation begins to fail, as it always offers tighter control of the FDF than the BH-FDR procedure alone.

We investigated departures from the assumption of independent hypothesis tests by conducting a simulation in which
tests were correlated within blocks according to a compound symmetry structure under a multivariate normal, having
marginal variances equal to 1. For the purposes of this investigation, we fixed the block size to 100,
effect size 1.25, FDR at 15\%, proportion of non-null distributed tests, $r$, at
5\% for 2000 simultaneous tests. We varied sample sizes from 14 to 16 and block
correlation from 0 to 80\% in increments of 10\%. The average power, $\lambda_{75}$ power, empirical FDR
and probability that the FDF exceeds 18\% are tabulated in table \ref{tbl:tbl_corr_tests} over
the ranges of sample sizes and block correlations considered. As we can see, comparing the independent tests lines
for each of the two sample sizes, 14 and 16, with corresponding values for correlated test statistics, a very
important point can be made. From the standpoint of the mean, there is virtually no difference. This is to say that
the empirical FDR and average power are virtually unaffected when there are correlated blocks of tests. Notable
differences do occur in the distributions of the TPF and FDF as the $\lambda_{75}$-power for independent test
statistics is 30\% in a sample of 14, and 87\% in a sample of 16, respectively, while the values when there are
correlated blocks of tests are substantially greater for a sample of 14 and substantially less in a sample of 16,
respectively. Discrepancies between the independent tests versus correlated blocks of tests in the same direction
are also observed in the probability that the FDF exceeds 18\%. The reason for this is that correlated blocks of
test statistics result in a reduced effective number of tests. Apparently, the observed effective number of test
statistics is large enough that the empirical means are still very good estimates of their almost sure limiting
values, but not great enough for stability in the distribution of empirical means at the ranges of parameters under
consideration.  The conclusion to be drawn is not that the BH-FDR procedure is to be avoided because it is not
completely imune to departures from the independent test statistics assumption. By analogy, is any limit theorem
meaningless because it doesn't apply to a sample size of 3? Quite not. The first conclusion to be drawn is that the
empirical means appear to be unaffected in the ranges of parameters considered here. If one is truly comfortable
controlling the false discovery rate and powering studies using the average power, then one can ignore the
appropriateness of the independence assumption. Problems start to occur when one uses the tails of the distribution
of the FDF and TPF, as we are making the case for use here. However, rather then give up on the use of the BH-FDR
procedure altogether, the phenomenon should be viewed from the lens of the effective number of simultaneous
tests. So, ironically, the problem is solved if one can simply increase the number of simultaneous tests.

Drawing away from the specific discussion of correlated tests and widening the focus to the conclusions to be drawn
from the paper as a whole, the point to be made is that the quantities arising in the BH-FDR procedure, the expected
FDF which is controlled, and the expected TPF which forms the basis of a power calculation, should be seen for what
they are, location parameters. Because first and second order asymptotics in the FDF and TPF occur as the number of
simultaneous tests tends to infinity, then within the scope of reasonable ranges of parameters, e.g. effect size no
more than 1 or so, and sample sizes within the ranges seen for equipment that is either very expensive per replicate
or just starting to get a bit cheaper, say a few tens of replicates, then the following generalizations can be
made. For more than 20,000 simultaneous tests, the means and the distributions effectively coincide so that
controlling the FDR and using the average power to derive sample sizes is well supported. This is great news 
for GWAS and RNA-seq studies for example. However, for less than one or two thousand simultaneous tests, one must 
use second order asymptotics to control the type I-like error and calculate sample sizes using the CLT's for the 
FDF and TPF in the manner outlined here. For on the order of a hundred or so simultaneous tests, asymptotic approximation 
using the CLT's may not be appropriate. In this case, simulation is advised. This cautionary note is of particular
importance in many biomarker studies. 
\fi 

\ifsuppl 
\section{Appendix: Further simulation studies}
\subsection{RNA Expression Micro-array Studies}
The third simulation study we considered experiments typical of human RNA expression micro-array studies using the
Affymetrix Hgu133plus2 oligonucleotide mRNA gene chip. In this case, there are $m=\numprint{54675}$ simultaneous
tests. We attempted to cover a broad spectrum of parameters spanning the domain typical of micro-array study
designs. The false discovery rate, $f$, was ranged over the values 1\%, and from
5\% to 30\% in increments of 5\%.  The expected number of tests with non-zero means, $\mathbb{E}[M_m]
= m r$, was varied from 100 to 2500 in increments of 100 representing values of $r$ ranging from
0.0018 to 0.046. The effect size, $\theta$, was allowed to vary from
0.6 to 1.5 in increments of 0.1. At each configuration, a range of sample sizes were chosen to result in powers between
60\% and 95\% as mentioned above. This resulted in 10,020 configurations of the
parameters, $f, \mathbb{E}[M_m],\theta$, and $n$ (full set of parameter combinations).  The job took roughly
12 hours on the NIH Biowulf cluster.

Table \ref{tbl:avgpwr_tbl_Array} tabulates the IST average power, the oracle power and the simulated mean of the TPF
at 28 different parameter settings excerpted from the full set of
10,020 parameter combinations. Over the full set of parameter settings, the
both the IST, $\pi_{_\mathrm{pi}}$, and oracle, $\pi_o$, powers are very close to the simulated average power. The
difference between the IST power, $\pi_{_\mathrm{pi}}$, and the simulated power was less than
0.022\%, 0.077\% and 0.19\% at 50\%, 90\% and 99\% of the parameter settings, respectively. The oracle
power differed from the IST power by less than 0.15\%, 0.5\% and 0.92\% at 50\%, 90\% and 99\% of
the parameter settings, respectively. As remarked earlier, the oracle power is actually the average power at the
oracle threshold, but for such small values of $r \leq 0.046$ there
is not much gain in power to be had. The lower bound comes within roughly 10\% of the simulated power, with differences
with the simulated power less than 2.1\%, 4.5\%, 6.2\%, 8.5\% and 15\% at 20\%, 40\%, 50\%, 60\% and 80\% of the
parameter settings, respectively.

Table \ref{tbl:Lpwr_tbl_Array} displays, at threshold 0.75 and at threshold 0.90, the $\lambda$ power as derived from
the CLT \ref{thm:CLT} and estimated from simulation replicates (hatted version), respectively, excerpted from the full
set of 10,020 parameter combinatons as before. In the last column is the ratio 
of the sample size required for $\lambda_{90}$-power to the original sample size. First, we note that when restricted to
powers strictly between 50\% and 100\%, occurring at 1,066
parameter combinations, the CLT approximate- and simulated- $\lambda_{75}$-power were within the following relative error
of one another (median over parameter conditions (lower quartile, upper quartile)): 
0.4\% (0.2\%, 1.2\%), with 2.4\% over 5\%.  
Corresponding results for the simulated and CLT approximate $\lambda_{90}$-power for powers strictly between 50\% and
100\% occurring at 1476 of the parameter values, were within the following relativer
error of one another 0.5\% (0.2\%, 1.3\%), with 
1.8\% over 5\%. Also noteworthy is corroboration in ordering of the
average power and $\lambda_{k}$-power based upon the size of $k$ relative to $100 \lambda_{eq}$. All values of
$\lambda_{eq}$ are less than 90\%, but some are between 75\% and 90\%, and the ordering of average power and $\lambda$
powers is in accordance with expression \ref{eqn:LPwr_AvgPwr}. Furthermore the discrepancy between the average power and
the $\lambda$-power is reflective difference between $\lambda$ and $\lambda_{eq}$. This trend is echoed in the magnitude
of the sample size ratio, with magnitude increasing in the discrepancy between $\lambda_{eq}$ and $0.90$. The relatively
rapid rise in sample size, $n$, of all $\lambda$-powers is an indication of the degree to which the distribution of 
the TPF, $S_m/M_m$, is spiked. 

\subsection{GWA Studies}
The last simulation study we considered experiments typical of GWA studies with $m=\numprint{1000000}$
simultaneous tests. We attempted to cover a broad spectrum of parameters spanning the domain typical of GWA study
designs. The false discovery rate, $f$, was ranged over the values 0.5\%, 1\%, 5\% and 10\%. The
expected number of tests with non-zero means, $\mathbb{E}[M_m] = m r$, was varied from 400 to 1000 in increments of 200 
representing values of $r$ ranging from 4e-04 to 0.001. The effect
size, $\theta$, was allowed to vary from 0.08 to 0.68 in increments of 0.2. At each configuration, a range of sample sizes
were chosen to result in powers between 50\% and 98\% as mentioned above. This resulted in
512 configurations of the parameters, $f, \mathbb{E}[M_m],\theta$, and $n$
(full set of parameter combinations). The job took roughly 5 hours on
the NIH Biowulf cluster.

Table \ref{tbl:avgpwr_tbl_GWAS} tabulates the IST average power, the oracle power and the simulated mean of the TPF at
32 different parameter settings excerpted from the full set of
512 parameter combinations. Over the full set of parameter settings, the both
the IST, $\pi_{_\mathrm{pi}}$, and oracle, $\pi_o$, powers are very close to the simulated average power. The
difference between the IST power, $\pi_{_\mathrm{pi}}$, and the simulated power was less than
0.031, 0.087 and 0.147 at 50\%, 90\% and 99\% of the parameter settings, respectively. The oracle power
differed from the IST power by 0.0038\%, 0.0081\% and 0.011\% at 50\%, 90\% and 99\% of the parameter
settings, respectively. As remarked earlier, the oracle power is actually the average power at the oracle threshold,
but for such small values of $r \leq 0.001$ the gain in power is now
less than 1\%. The lower bound comes within roughly 10\% of the simulated power, with differences with the simulated
power less than 0.83\%, 3.8\%, 6.9\%, 9.6\% and 15\% at 20\%, 40\%, 50\%, 60\% and 80\% of the parameter settings,
respectively.

Table \ref{tbl:Lpwr_tbl_GWAS} displays, at threshold 0.75 and at threshold 0.90, the $\lambda$ power as derived from
the CLT \ref{thm:CLT} and estimated from simulation replicates (hatted version), respectively, excerpted from the full
set of 512 parameter combinatons as before. In the last column is the ratio 
of the sample size required for $\lambda_{90}$-power to the original sample size. First, we note that when restricted to
powers strictly between 50\% and 100\%, occurring at 68
parameter combinations, the CLT approximate- and simulated- $\lambda_{75}$-power were within the following relative error
of one another (median over parameter conditions (lower quartile, upper quartile)): 
0.5\% (0.2\%, 1.3\%), with 1.5\% over 5\%.  
Corresponding results for the simulated and CLT approximate $\lambda_{90}$-power for powers strictly between 50\% and
100\% occurring at 87 of the parameter values, were within the following relativer
error of one another 0.6\% (0.2\%, 1.2\%), with 
1.1\% over 5\%. Also noteworthy is corroboration in ordering of the
average power and $\lambda_{k}$-power based upon the size of $k$ relative to $100 \lambda_{eq}$. All values of
$\lambda_{eq}$ are less than 90\%, but some are between 75\% and 90\%, and the ordering of average power and $\lambda$
powers is in accordance with expression \ref{eqn:LPwr_AvgPwr}. Furthermore the discrepancy between the average power and
the $\lambda$-power is reflective difference between $\lambda$ and $\lambda_{eq}$. This trend is echoed in the magnitude
of the sample size ratio, with magnitude increasing in the discrepancy between $\lambda_{eq}$ and $0.90$. 
Notice that over values considered the ranges of $\sqrt{n}\theta$ are comparable among the the micro-array, GWAS and
biomarker simulation studies. Therefore, the ``all'' or ``nothing'' rapid rise in the $\lambda$-powers with increasing 
sample size here must be solely due to the distribution of the TPF, $S_m/M_m$, being even more dramatically spiked, since 
the number of simultaneous tests, $m$, is in this case, considerably larger.

\section{Appendix: Proofs}
\begin{proof}[Proof of Theorem \ref{thm:Jom_convas}]
\noindent The author wishes to thank Professor Thomas G. Kurtz \citep{KurtzT:2016} for assistance with this proof.
Recall the nominal p-values, $P_i = \bar F_{0,n}^{-1}(X_{i,n})$, their common CDF, $G$, listed in expression 
\ref{eqn:Gdefnd} in the text and their order statistics $P^m_{(i)}$. Let $G_m$ be the empirical C.D.F. of 
$\{P_1, P_2,\ldots,P_m\}$. 
\begin{equation}
  G_m(u) = m^{-1} \sum_{i=1}^m I(P_i \leq u)  \label{eqn:empcdfGm}
\end{equation}
By Kolmogorov's theorem, $G_m(u)\convas G(u)$ at all continuity points, $u$, of $G$. By assumption 
the family $\{F_{\nu,n}^{-1}:\nu \geq 0\}$ is absolutely continuous and has the monotone likelihood
ratio property. It follows that each of the ratios $f_{\nu_{\ell}, n}/f_{0,n}$ is monotone and hence
the mixture of likelihood ratios, $f_{A,n}/f_{0,n} = \sum_{\ell} s_{\ell} f_{\nu_{\ell},n}/f_{0,n}$ is monotone.
It follows that G is concave and therefore $G(uf) = u$ has one non-zero solution which we will call $\gamma$. 
Let $\mathcal{N}\subset \Omega$ be the set of measure zero such that $G_m(\gamma f) \rightarrow G(\gamma f)$
for all $\omega \in \Omega \setminus \mathcal{N}$, and consider $\omega$ fixed in this set of measure 1 
for the remainder of this proof. Substituting $m^{-1} J_m \,f$ for $u$ in expression \ref{eqn:empcdfGm} shows that 
\begin{equation}
  m^{-1} J_m = G_m(m^{-1} J_m f) \label{eqn:Jom_eq_GmofJomf}
\end{equation}
Let $H_m(u) = G_m(u f) - u$ and $H(u) = G(u f) - u$. While $H^{-1}(0) = \{0, \gamma \}$ contains only 0 and
a unique non-zero solution, $G_m$ is a step function and therefore, $H_m^{-1}(0) = \{0, u_1, u_2,\ldots,u_k\}$
can contain multiple non-zero solutions.  None-the-less, for each $m$, $H_m^{-1}(0)$ is a finite set. By definition,
$J_m/m$ is an element of the set $H_m^{-1}(0)$. It follows that 
\begin{eqnarray*}
  m^{-1} J_m  &\leq& \sup H_m^{-1}(0)\nonumber\\
             &=& \max H_m^{-1}(0).\nonumber  
\end{eqnarray*}
where the second line follows because the set is finite. Thus, taking limsup with respect to $m$ on both sides 
above gives:
\begin{eqnarray*}
\limsup_m m^{-1} J_m &\leq& \limsup_m \max H_m^{-1}(0) \\
                    &=& H^{-1}(0) = \gamma\,.
\end{eqnarray*}
where the last equality follows because we can interchange the order of the limsup and maximum and because the 
limit exists. In the other direction, next note that because $m^{-1} J_m$ is a solution to $u = G_m(u f)$, it 
also follows that 
$m^{-1} J_m \geq u$ for every $u$ such that $u < G_m(u f)$. Thus, 
\begin{equation}
m^{-1} J_m \geq \sup \{u : u < G_m(u f) \} = \sup H_m^{-1}((0,\infty))\,. \label{eqn:Jom_geq_Gm}
\end{equation}
Because of the convexity of the limiting function, $G$, it follows, for $m$ large enough, that 
$\sup H_m^{-1}((0,\infty)) = \max H_m^{-1}(0)$. Therefore, upon taking taking liminf with respect to $m$ on both 
sides we have:
\begin{eqnarray*}
\liminf_m m^{-1} J_m &\geq& \liminf_m \sup H_m^{-1}((0,\infty))\\
                    &=& \liminf_m \max H_m^{-1}(0)\\
                    &=& H^{-1}(0) = \gamma\\
\end{eqnarray*}
where the last equality follows because we can interchange the order of liminf and the maximum and because the limit exists.
This completes the proof.
\end{proof}

\begin{proof}[Proof of Theorem \ref{thm:SoM_convas}]
First, we note that
\begin{equation}
m^{-1} S_m = m^{-1} \sum_{i=1}^m \xi_i \, I\left( P_i \leq m^{-1} J_m f \right)\,,
\end{equation}
is the average of row $m$ in a triangular array of finite exchangeable sequences. We will apply Theorem 1 of 
\cite{TaylorRL:1985}. Let $W_{m,i}=\xi_i \,I\left(P_i\leq m^{-1} J_m f\right), \mu_m = \E[W_{m,1}]$ and 
$Z_{m,i} = W_{m,i} - \mu_m$. We must show that (i) the increments on the $m^{th}$ row, $W_{m,i}$, each converge almost 
surely to respective elements of a sequence $W_{\infty, i}$; (ii) the increments $W_{m,i}$ have variances tending to a 
limit and (iii) for each $m, i$ and $j$, the covariance of increments $W_{m,i}$ and $W_{m,j}$ tend to zero. 
\begin{remark}
Our condition (i), element-wise almost sure convergence, which on surface appears weaker than the corresponding first 
condition in the cited reference, almost sure monotone decreasing distances to the limit, is sufficient in the 
context of the other assumptions. See the remark following the proof of Theorem 3 in that reference.
\end{remark}

Verification of condition (i) is trivial, as it follows by Theorem \ref{thm:Jom_convas} that 
$W_{m,i}\rightarrow W_{\infty,i} = \xi_i\, I\left( P_i \leq \gamma f \right)$ 
almost surely as $m\rightarrow \infty $ for each $i$. Let $\mu = \E[W_{\infty,1}]$. Condition (ii) follows easily 
since $Z_{m,i}$ is bounded, so that by the LDCT, for each $i$, 
$\E\left[Z_{m,i}^2 \right] \rightarrow \E\left[Z_{\infty,i}^2 \right]$ as $m\rightarrow\infty$. Note that the same 
argument verifies that $\mu_m \rightarrow \mu$. Next, to verify that condition (iii) is satisfied, note first, for 
$1 \leq i_1 < i_2 \leq m$, that $(W_{m,i_1} - \mu_m)(W_{m,i_2} - \mu_m)$ is bounded above by $4$, and converges almost 
surely to $(W_{\infty,i_1} - \mu)(W_{\infty,i_2} - \mu)$, by Theorem \ref{thm:Jom_convas}. Thus, condition (iii) follows by 
the LDCT. We may now apply Theorem 1 of \cite{TaylorRL:1985} to conclude that $m^{-1} \sum_{i=1}^m Z_{m,i} \rightarrow 0$ 
almost surely as $m\rightarrow \infty$. Therefore,
\begin{eqnarray*}
  \lim_{m\rightarrow\infty} m^{-1} S_m &=&
  \lim_{m\rightarrow\infty} \mu_m + m^{-1} \sum_{i=1}^m Z_{m,i}\\
  \\
&=& \mu\,, \;\mathrm{with~probability~one,}
\end{eqnarray*}
and the last written expectation is equal to $r\,\P \left\{ P_i \leq \gamma f \mid \xi_i=1\right\}$. Because 
$m^{-1} M_m \rightarrow r$ almost surely as $m\rightarrow\infty$, it follows that 
$M_m^{-1}\,S_m\rightarrow\P\left\{ P_i \leq \gamma f \mid \xi_i=1\right\} = \pi_{_\mathrm{pi}}$ almost surely as 
$m\rightarrow\infty$. Because $m^{-1} S_m$ is bounded by 1, the average power, its expectation, also converges to 
$\pi_{_\mathrm{pi}}$ as $m\rightarrow\infty$ by the LDCT.
\end{proof}

\begin{proof}[Proof of Corollary \ref{cor:FDF_convas}]
The first and second statements follow immediately from Theorems \ref{thm:Jom_convas} and \ref{thm:SoM_convas}:
\begin{equation}
m^{-1} T_m = m^{-1} (J_m - S_m) \convas \gamma - r\,\pi_{_\mathrm{pi}} = (1-r)\,f\,\gamma\,,
\end{equation}
and
\begin{equation}
J_m^{-1} T_m = 1 - J_m^{-1}S_m) \convas 1  - r\pi_{_\mathrm{pi}}/\gamma = (1-r)\,f\,.
\end{equation}
where the last equality in each of the expressions above follow since $\gamma = G(\gamma \,f)$. The third statement 
follows by the LDCT.
\end{proof}

\begin{proof}[Proof of Corollary \ref{cor:pi1Alt}]
  In the definition of the IST power function, $\pi_{_\mathrm{pi}}$, appearing in the statement of Theorem 
  \ref{thm:SoM_convas}, $\pi_{_\mathrm{pi}} = \bar F_{\nu, n}(\bar F_{0,n}^{-1}(\gamma\,f))$ we substitute 
  $\gamma = (1-f_0)^{-1} \,r \pi_{_\mathrm{pi}}$ from expression \ref{eqn:gamma_expr} obtaining the result.
\end{proof}

\begin{proof}[Proof of Theorem \ref{thm:CLT}]
\noindent The proof of both statements is made possible by considering each as a stopped stochastic process. We 
first revisit the empirical CDF's defined in the proofs of Theorems \ref{thm:Jom_convas} and \ref{thm:SoM_convas}. 
In the case of $J_m/m$, we have, 
\begin{eqnarray}
  G_{_{m}}(t) &=& m^{-1}\sum_{i=1}^m I\left( P_i \leq t \right)\,,\label{eqn:Gm_t}\\ 
  &&\nonumber\\
  G(t) &=& (1- r) \, t\, + \,r \, \bar F_{\nu,n}(\bar F_{0,n}^{-1}(t)\,. \label{eqn:G_t}
\end{eqnarray}
\noindent First, by the standard theory of empirical distributions, see for example \cite{ShorackWellner:1986}
\begin{equation}
W_m(t) = \sqrt{m} \left(G_m(t) - G(t)\right) \convind W(t)\,,\label{eqn:Wm_t_convind}
\end{equation}
a Gaussian process with covariance function 
\begin{equation}
\rho(s,t) = G(s\wedge t) \, - \, G(s)\,G(t)\,. \label{eqn:covW}
\end{equation}
Having shown that the paths of the centered and scaled stochastic process $W_m$ converge in distribution, we can 
obtain the CLT for the centered and scaled version of the positive fraction, $J_m/m$, claimed in 
expression \ref{eqn:CLT_Jortm}, by appealing to a result concerning weak limits of stopped stochastic processes. 
Towards this end, define the family of filtrations, 
\begin{equation}
\mathcal{F}_t = \sigma\left(\{ \xi_i\,I(P_i \leq t),\,(1-\xi_i)\,I(P_i \leq t), i\geq 1\}\right)\,. \nonumber
\end{equation}
Note that $W$ and $W_m$ are adapted to $\mathcal{F}_t$ for all $m\geq 1$, and that $\tau_m = m^{-1} J_m f$ is a 
stopping-time with respect to this filtration since, clearly, $\{\tau_m \leq t\} \in \mathcal{F}_t$. We will apply 
Theorem 4.2.1 of \cite{SilvestrovD:2004} to conclude that $W_m(\tau_m)$ converges in distribution to $W(\gamma\,f)$.  
To do so, we must verify the following three conditions.
\begin{itemize}
\item[i.~] $(W_m, \tau_m) \convind (W, \gamma\,f)$
\item[ii.~] $\P\{ \lim_{t\rightarrow 0} W(\gamma\,f + t) = W(\gamma\,f)\} = 1$
\item[iii.~] For all $\delta >0, \lim_{c\rightarrow 0} \limsup_{m\rightarrow\infty} \P\{ \Delta(W_m, c, 1) > \delta \} = 0\,,$
\end{itemize}
where $\Delta(x, c, 1)$ is the Skorohod modulus of compactness, 
\begin{equation}
\Delta(x, c, 1) = \sup_{t, t', t'' \in [0,1]} \sup_{ t-c < t' < t < t'' < t+c} |x(t) - x(t')| + |x(t'') - x(t)|
\end{equation}
Having already established that the paths of $W_m$ converge in distribution to those of $W$ above \ref{eqn:Wm_t_convind}, 
as well as the almost sure convergence of $\tau_m = J_m f/m$ to the constant $\tau = \gamma f$ in theorem \ref{thm:Jom_convas},
then part (i) is satisfied by Slutsky's theorem. Item (ii) is true because the limiting process, $W$, a Gaussian process, 
is almost surely continuous at every $t \in [0,1]$.  The almost sure continuity of the limiting process, $W$, also 
guarantees that the third condition, (iii), holds as well. Thus  
\begin{eqnarray}
W_m(m^{-1} \,J_m\,f) &=& \sqrt{m} \left(G_m(m^{-1} \,J_m\,f) - G(m^{-1} \,J_m\,f)\right)\nonumber\\
                      &\convind& W(\gamma\,f)\,,\label{eqn:Wm_taum_convind}
\end{eqnarray}
where the limiting random variable is normally distributed, of mean zero, and variance 
\begin{equation}
\rho(\gamma\,f, \, \gamma\,f) = G(\gamma\,f) \, - \, G^2(\gamma\,f) = \gamma\,(1-\gamma)\label{eqn:varW}
\end{equation}
\noindent The statement \ref{eqn:Wm_taum_convind} is nearly statement \ref{eqn:CLT_Jortm}, except that in 
\ref{eqn:CLT_Jortm}, centering is with respect to the deterministic limit, $\gamma = G(\gamma\,f)$. Thus, starting 
with $\sqrt{m}(J_m/m - \gamma))$ we add and subtract, obtaining a ``delta-method'' term. We can now write
\begin{eqnarray}
X_m &\equiv& \sqrt{m} \left(m^{-1} J_m \, - \,\gamma \right) \nonumber\\
      &=& \sqrt{m} \left(G_m(m^{-1} \,J_m\,f) - G(\gamma \,f)\right) \nonumber\\
      &=& \sqrt{m} \left(G_m(m^{-1} \,J_m\,f) - G(m^{-1} \,J_m\,f)\right) \nonumber\\
      &&  \;\;+\,\sqrt{m} \left(G(m^{-1} \,J_m\,f) - G(\gamma\,f)\right)\nonumber\\
      &=& W_m(m^{-1} \,J_m\,f) + f\,\dot{G}(\gamma\,f)\, X_m + \epsilon_m\label{eqn:Xm}
\end{eqnarray}
where $\epsilon_m = o_p(1)$. The conclusion of this portion of the proof requires that $f\, \dot{G}(\gamma\,f) <1 $.
By the monotone likelihood ratio property, it follows that $G$ is concave as is the function $f\, G$.  Because 
$f\,G(u) = u$ when $u=0$ and when $u=\gamma\, f$, then there is exactly one $u_1 \in (0, \gamma\,f)$ for which
$f\,\dot{G}(u_1) = 1$. By the concavity of $f\, G$, $f\,\dot{G}(u) > 1$ for $0 < u < u_1$ and $f\,\dot{G}(u) < 1$
for $u_1 < u < \gamma\,f$. Thus $f\,\dot{G}(\gamma\, f) < 1$. With this bound in hand, the steps above leading to
\ref{eqn:Xm} can be iterated ad-infinitum, yielding
\begin{eqnarray*}
X_m &=& \left(W_m(m^{-1} \,J_m\,f) + \epsilon_m \right) \, \sum_{k=0}^{\infty} f^k\,\dot{G}(\gamma\,f)^k\\
      &=& \frac{W_m(m^{-1} \,J_m\,f) + o_p(1)}{1 - f\,\dot{G}(\gamma\,f)}\\
      &\convind& X \equiv \frac{W(\gamma\,f)}{1 - f\,\dot{G}(\gamma\,f)}\,,
\end{eqnarray*}
which establishes claim \ref{eqn:CLT_Jortm} above and identifies the form of the limiting mean zero normal random 
variable. Its variance, $\tau^2$, is given by 
\begin{equation}
\tau^2 = \frac{\gamma (1 - \gamma)}{\left(1-\dot{G}(\gamma\, f)\, f\right)^2}\,.\label{eqn:varX}
\end{equation}

\noindent Next, we turn our attention towards verification of the CLT for the centered and scaled TPF, $S_m/M_m$, 
which is claim \ref{eqn:CLT_rtmSoM} above. We first revisit the empirical 
sub CDF's corresponding to the joint outcome of the indicator $\xi_i$ and the indicator $I(P_i \leq t)$ and their 
almost sure deterministic limits.
\begin{eqnarray}
  G_{_{m,0}}(t) &=& m^{-1}\sum_{i=1}^m (1- \xi_i ) \, I\left( P_i \leq t \right)\,,\label{eqn:Gm0_t}\\
  &&\nonumber\\
  G_{_{m,1}}(t) &=& m^{-1}\sum_{i=1}^m \xi_i \, I\left( P_i \leq t \right)\,,\label{eqn:Gm1_t}\\
  &&\nonumber\\
  G_{_{0}}(t) &=& (1- r) \, t\,,\label{eqn:G0_t}\\
  &&\nonumber\\
  G_{_{1}}(t) &=& r \, \bar F_{\nu,n}(\bar F_{0,n}^{-1}(t)\,.\label{eqn:G1_t}
\end{eqnarray}
\noindent This time we look at the bivariate process with components scaled and centered versions of $G_{_{m,0}}(t)$ 
and $G_{_{m,1}}(t)$. Again, from the standard results concerning empirical CDF's, 
\citep{ShorackWellner:1986, GenoveseC:2004} the following bivariate process converges in distribution.
\begin{equation}
\left[\begin{array}{c}
W_{m,0}(t)\\
W_{m,1}(t)
\end{array}\right] =
\sqrt{m}\left[\begin{array}{c}
G_{m,0}(t) - G_0(t)\\
G_{m,1}(t) - G_1(t)
\end{array}\right] 
\convind \left[\begin{array}{c} 
W_0(t)\\
W_1(t)
\end{array}\right]\,,\nonumber
\end{equation}
\noindent where the limit is a bivariate Gaussian process with covariance kernel 
\begin{equation}
R(s,t) = \left[\begin{array}{cc}
         G_0(s\wedge t) - G_0(s) \,G_0(t)  & -G_0(s)\,G_1(t)                  \\
        -G_0(t)\,G_1(s)                     & G_1(s \wedge t) - G_1(s) G_1(t)
         \end{array}\right]\,.\nonumber
\end{equation}
We remark in passing, something which should be already clear, that $W_m(t) = W_{m,0}(t) + W_{m,1}(t)$ and 
$W(t) = W_0(t) + W_1(t)$. If follows from some algebra and the fact that $\gamma = G(\gamma\,f)$ that this 
``new characterization'' of $W$ is consistent with the characterization of the process given 
above. This allows us to compute covariances between $W(t)$ and either $W_0(t)$ or $W_1(t)$
according to the covariance kernel, $R(s,t)$, as needed below. Note that $W_{0}, W_{1}$ and for all $m$, 
$W_{m,0}, W_{m, 1}$ are all adapted to the filtration, $\mathcal{F}_t$ and that $\tau_m = m^{-1}\, J_m \,f$ is a
stopping time with respect to the it, so that once again we apply the result of \cite{SilvestrovD:2004} 
to obtain convergence of the stopped bivariate process. As remarked above, the conditions 
are satisfied since convergence is already established and the limit is almost surely continuous. 
\begin{equation}
\left[\begin{array}{c}
W_{m,0}(J_m\,f/m)\\
W_{m,1}(J_m\,f/m)
\end{array}\right] =
\sqrt{m}\left[\begin{array}{c}
G_{m,0}(J_m\,f/m) - G_0(J_m\,f/m)\\
G_{m,1}(J_m\,f/m) - G_1(J_m\,f/m)
\end{array}\right] 
\convind \left[\begin{array}{c} 
W_0(\gamma\,f)\\
W_1(\gamma\,f)
\end{array}\right]\,,\label{eqn:bvrt}
\end{equation}
Focusing for the moment on the second component above in \ref{eqn:bvrt} and adding and subtracting as before, 
\begin{eqnarray*}
X_{m,1} &\equiv& \sqrt{m}\left(m^{-1} S_m - r \,\pi_{_\mathrm{pi}}\right)\\
        &=& \sqrt{m}\left(G_{m,1}(J_m\,f/m) - G_1(\gamma\,f)\right)\\
        &=& \sqrt{m}\left(G_{m,1}(J_m\,f/m) - G_1(J_m\,f/m)\right)\\
        &&\;\;+\,\sqrt{m}\left(G_1(J_m\,f/m) - G_1(\gamma\,f)\right)
\end{eqnarray*}
Thus, 
\begin{eqnarray*}
X_{m,1} &\convind& X_1 =W_1(\gamma\,f) +f\,\dot{G}_1(\gamma\, f)\, X\\
  &=& W_1(\gamma\,f)+\frac{\dot{G}_1(\gamma\,f)\,f}{1-f\,\dot{G}(\gamma\,f)}\,\left(W_0(\gamma\,f)+W_1(\gamma\,f)\right)
\end{eqnarray*}
which is a mean zero normal random variable having variance equal to 
\begin{equation}
\var[X_1] = 
   v_1+\dot{G}_1^2(\gamma\,f)\,f^2\tau^2+2 f\,\frac{\dot{G}_1(\gamma\,f)\,(v_1+c_{0,1})}{1-f\,\dot{G}(\gamma\,f)}
\end{equation}
where $v_1 = r \,\pi_{_\mathrm{pi}} - r^2 \, \pi_{_\mathrm{pi}}^2$ and $c_{_{0,1}} = -r \,(1-r)\,\gamma \,f \,\pi_{_\mathrm{pi}}$. To 
complete the proof of statement \ref{eqn:CLT_rtmSoM} we need only apply the delta method once more, for a ratio 
estimate. Before proceeding, we note that 
\begin{eqnarray*}
\sqrt{m}(M_m/m -r) &=& \sqrt{m}\left(G_{m,1}(1) - G_1(1)\right)\\
                   &\convind& W_1(1) 
\end{eqnarray*}
Thus, 
\begin{eqnarray*}
Z_{m,1} &\equiv& \sqrt{m}\left(\frac{S_m}{M_m} - \pi_{_\mathrm{pi}}\right)\\
         &=& \sqrt{m}\left(\frac{S_m/m}{M_m/m} - \pi_{_\mathrm{pi}}\right)\\
  &=&\frac{1}{r}\,\sqrt{m}\left(\frac{S_m}{m}-r\pi_{_\mathrm{pi}}\right)-\frac{r\pi_{_\mathrm{pi}}}{r^2}\,
                                                         \sqrt{m}\left(\frac{M_m}{m}-r\right) + \epsilon_m\\
  &\convind& \frac{1}{r}\,\left\{W_1(\gamma\,f) + \frac{f\,\dot{G}_1(\gamma\, f)}{1- f\,\dot{G}(\gamma\, f)} \, 
                 \left(W_0(\gamma\,f) + W_1(\gamma\,f)\right)\right\} - \frac{r\pi_{_\mathrm{pi}}}{r^2}\,W_1(1) \\
         &\equiv& Z_1 = N(0, \sigma^2)\,,
\end{eqnarray*}
where $\epsilon_m$ above is a new term that is $o_p(1)$ and and this completes the proof of statement \ref{eqn:CLT_rtmSoM} 
above and identifies the form of the limiting mean zero normal random variable, $Z_1$. Its variance is given by 
\begin{equation}
\sigma^2 = r^{-2} \, \left(\var[X_1] -2\, \pi_{_\mathrm{pi}} \,\cov[X_1, W_1(1)] +\pi_{_\mathrm{pi}}^2 \,\var[W_1(1)] \right)\,,\label{eqn:s2}
\end{equation}
where $\var[X_1]$ was given above, $\var[W_1(1)] = r\,(1-r)$, and 
\begin{equation}
\cov[X_1, W_1(1)]=r\,(1-r)\,\left\{\pi_{_\mathrm{pi}}+f\,\dot{G}_1(\gamma\,f)
                                  \frac{\gamma\,f +\pi_{_\mathrm{pi}}}{1-f\,\dot{G}(\gamma\,f)}\right\}
\end{equation}

The proof of the CLT for the centered and scaled version of the false discovery fraction follows fairly easily from the parts proved above. 
First, re-writing the centered and scaled difference in terms of the TPF, gives the first line, for which we again invoke 
the delta method, which yields the second line.
\begin{eqnarray*}
X_{m,0} &\equiv& \sqrt{m}\left(\frac{T_m}{J_m} - (1-r) \, f \right) = -\sqrt{m}\left(\frac{S_m}{J_m} - \frac{r \pi_{_\mathrm{pi}}}{\gamma} \right)\\
       &=& -\gamma^{-1} \left\{\sqrt{m}\left(m^{-1}S_m - r\pi_{_\mathrm{pi}}\right) - \gamma^{-1} r \pi_{_\mathrm{pi}}\left(m^{-1}J_m - \gamma\right) + \epsilon_m \right\}\\
       &\convind& -\gamma^{-1} \left\{ X_1 + \gamma^{-1} r \pi_{_\mathrm{pi}} X \right\}\\
       &=& -\gamma^{-1} \left\{\frac{f \dot{G}_1(\gamma f) - \gamma^{-1} r \pi_{_\mathrm{pi}}}{1 - f \dot{G}(\gamma f)} W_0 +
           \left(1 + \frac{f \dot{G}_1(\gamma f) - \gamma^{-1} r \pi_{_\mathrm{pi}}}{1 - f \dot{G}(\gamma f)}\right) W_1 \right\}\\
       &=& -\gamma^{-1} \left\{\frac{f \dot{G}(\gamma f) - (1-r)\,f - \gamma^{-1} r \pi_{_\mathrm{pi}}}{1 - f \dot{G}(\gamma f)} W_0 + 
           \left(1 + \frac{f \dot{G}(\gamma f) - (1-r)\,f- \gamma^{-1} r \pi_{_\mathrm{pi}}}{1 - f \dot{G}(\gamma f)}\right) W_1 \right\}\\
       &=& -\gamma^{-1} \left\{\frac{f \dot{G}(\gamma f) - 1}{1 - f \dot{G}(\gamma f)} W_0 + 
           \left(1 + \frac{f \dot{G}(\gamma f) - 1}{1 - f \dot{G}(\gamma f)}\right) W_1 \right\}\\
       &=& \gamma^{-1} W_0
\end{eqnarray*}
Convergence of all quantities in the second line was established above. The remaining lines are algebraic, and make use of the fact that
$\dot{G}_1(t) = \dot{G}(t) - (1-r)$ (line 5) and $G(\gamma \,f) = \gamma$ (line 6). As before, $\epsilon_m$ is a new term that is $o_p(1)$. 
The limiting random variable is of mean zero and normally distributed, having variance equal to 
\begin{equation}
\alpha^2 = \frac{(1-r)\,f\,\left( 1 - (1-r)\,f\,\gamma\right)}{\gamma}\label{eqn:a2}
\end{equation}
\end{proof}


\begin{proof}[Proof of Theorem \ref{thm:LowerBdd}]

Before we begin, we present an alternate expression for the event that the number of true positives is $s$ or greater.
\begin{equation}
    \{S_m \geq s \} = \left\{ P_{1,(s)} \leq \frac{J_m f}{m}\right\}\nonumber
\end{equation}
This is clearly the case, since there can be $s$ or more true positives if and only if the $s^{th}$ order statistic
in the non-null distributed population is less than the threshold $J_m f/m$. Now, towards obtaining a lower bound, 
we begin with the fact that the expected value of discrete non-negative variable can be derived as the sum of its 
cCDF. This can be used to write an expression of the FST average power by first conditioning on $M_m$:
\begin{eqnarray} 
  \pi_{_{\mathrm{av},m}} &=& \E[ S_m / M_m ] \nonumber \\
  &=& \sum_{\ell=1}^{m} \ell^{-1} \,\E [ S_m \mid M_m={\ell}]\,\P\{ M_m=\ell\}\nonumber\\
  &=& \sum_{\ell=1}^{m} \ell^{-1} \,\sum_{s=1}^{m} \P\{ S_m \geq s \mid M_m=\ell \}\,\P\{ M_m=\ell\}\nonumber\\
  &=& \sum_{\ell=1}^{m} \ell^{-1} \,\sum_{s=1}^{m} \P\{P_{1,(s)} \leq f\,J_m/m\}\,\P\{ M_m=\ell\}
  \label{eqn:lwrbddeqn3}\\
  &\geq& \sum_{\ell=1}^{\infty} m^{-1} \,\sum_{s=1}^{m} \P\{P_{1,(s)} \leq f s/m\}\,\P\{ M_m=\ell\}\nonumber\,,
\end{eqnarray}
where the first equality is just the law of total probability, conditioning on values of $M_N$, the second equality is
just the fact that an expectation of a non-negative random variable is the sum over values of $s$ of its cCDF,
the third equality is an application of the alternate expression stated above, and the lower bound in the last line is 
deduced by observing that $P_{1,(s)} \leq f\,J_N/N$ if and only if $J_N \geq s$. The last written line is equal to 
$\pi^{L}_{\mathrm{av},m}$ as presented in Theorem \ref{thm:LowerBdd} because the CDF of the $s^{th}$ order statistic, $P_{1,(s)}$, 
takes the form shown involving the beta distribution.
\end{proof}
\fi 
\vfil\eject


\begin{figure}[b]
\begin{center}
\includegraphics{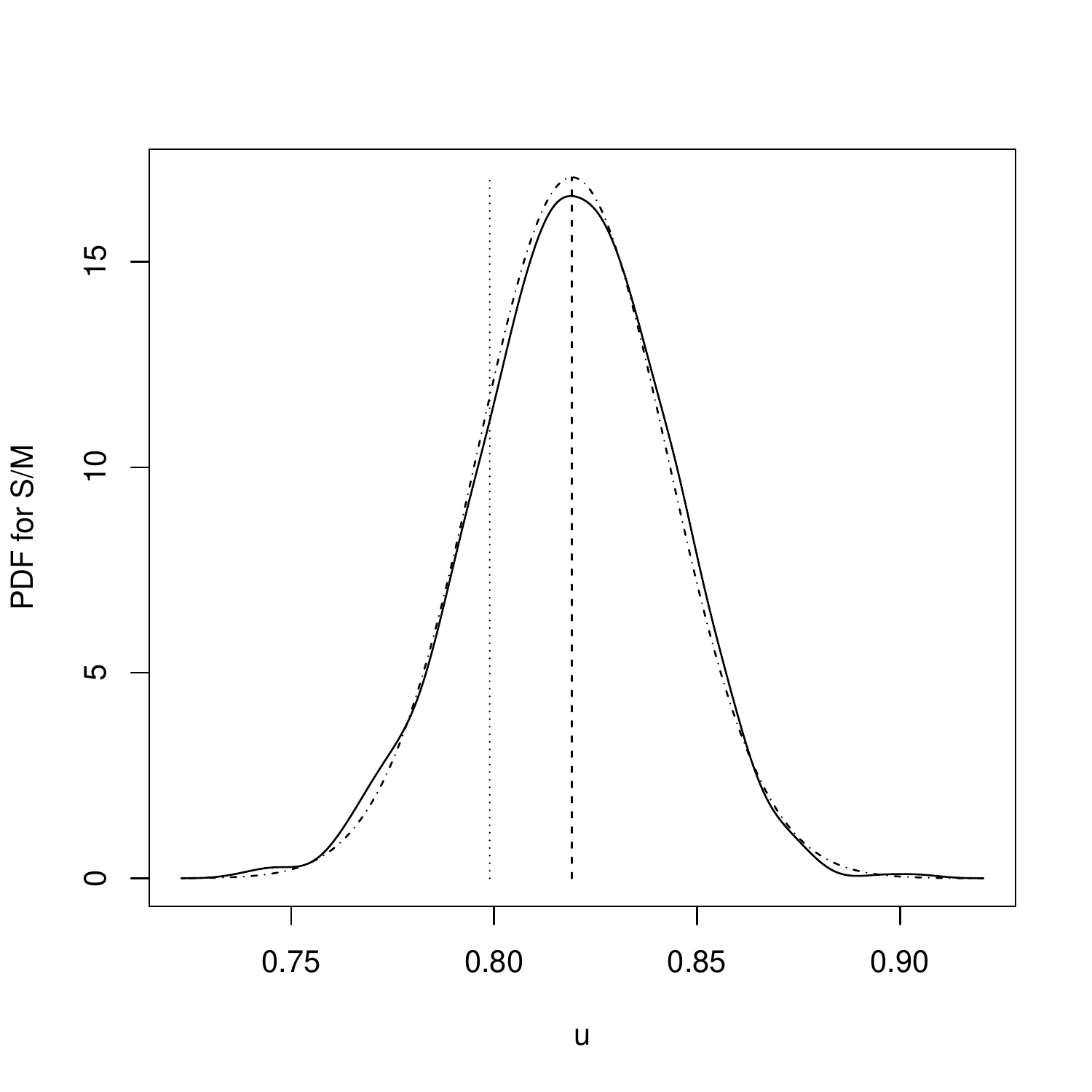}
\end{center}
\caption{Density plot of true positive fraction, $S_m/M_m$, when m=\numprint{10000}, showing
  average power (heavy dashed line) which nearly coincides with the mode and median. The $\lambda$-power is the tail
  probability to the right of a given threshhold, $\lambda$. Also shown (light dotted line) are
  $\lambda_{_{S/M}}(\pi_{_\mathrm{pi}})$, the threshhold at which the $\lambda$-power is equal to the average power and 
  the approximate normal asymptotic distribution, (dot-dash line).}
\label{fig:SoM-pdf}
\end{figure}

\begin{figure}[b]
\begin{center}
\includegraphics{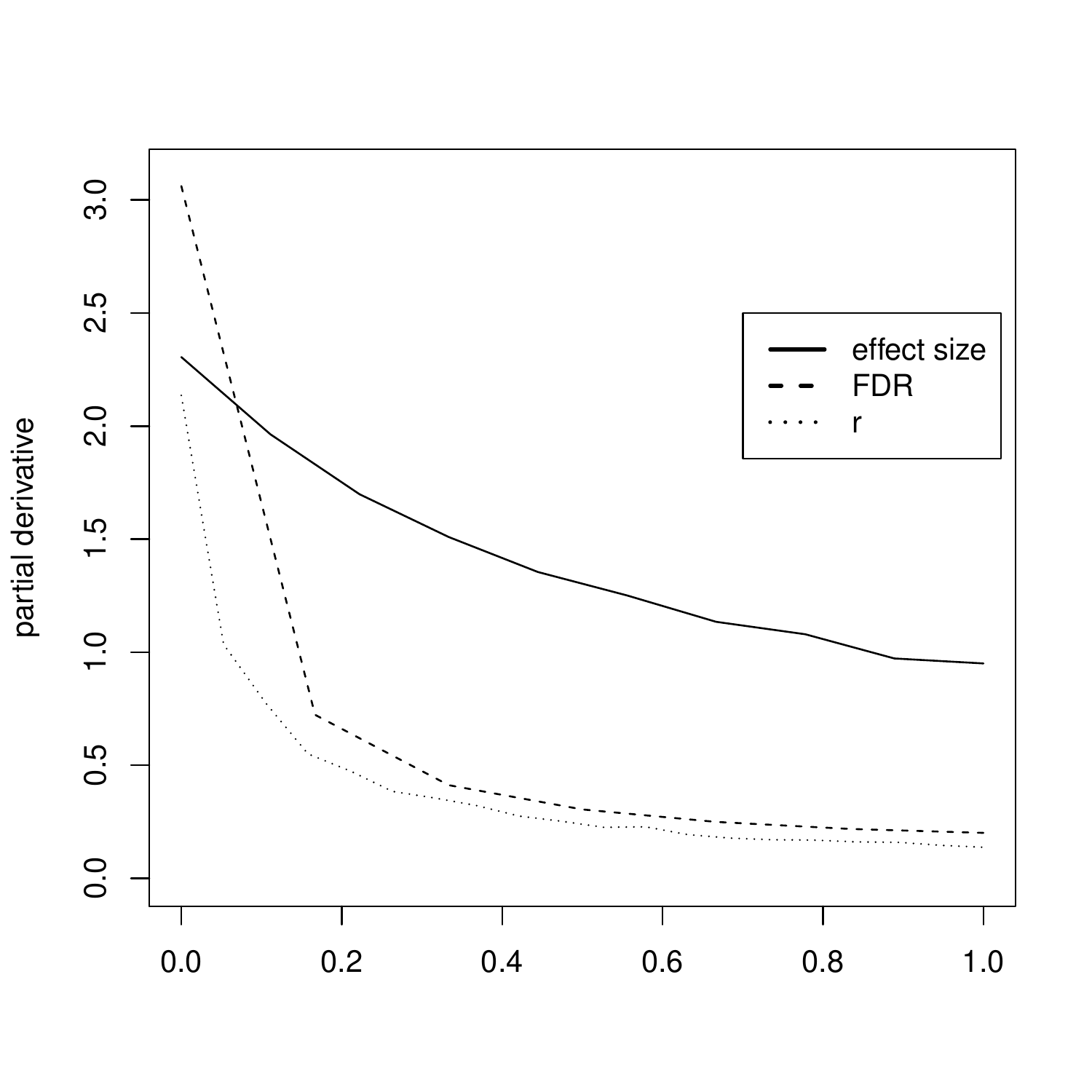}
\end{center}
\caption{Numerical partial derivatives of power function with respect to $\theta, f$ and $r$, with ranges considered 
         scaled to unity.}
\label{fig:MA-derivs}
\end{figure}

\ifmnscpt
\begin{table}[b]
\centering
\begingroup\tiny
\begin{tabular}{lrrrrrrr}
  \toprule
 Eff Sz&$\E[M_m]$&n&FDR&$\pi^{L}_{\mathrm{av},m}$&$\pi_{_\mathrm{pi}}$&$\pi_o$&$\hat\pi_{_{\mathrm{av},m}}$\\
  \cmidrule(r){1-1}\cmidrule(lr){2-2}\cmidrule(lr){3-3}\cmidrule(lr){4-4}\cmidrule(lr){5-5}\cmidrule(lr){6-6}\cmidrule(lr){7-7}\cmidrule(l){8-8}
0.60 & 5 & 70 & 0.15 & 0.224 & 0.704 & 0.707 & 0.692 \\ 
  0.60 & 5 & 80 & 0.15 & 0.235 & 0.795 & 0.798 & 0.790 \\ 
  0.60 & 5 & 90 & 0.15 & 0.241 & 0.860 & 0.863 & 0.850 \\ 
  0.60 & 5 & 100 & 0.15 & 0.246 & 0.906 & 0.908 & 0.896 \\ 
  0.60 & 20 & 50 & 0.15 & 0.052 & 0.664 & 0.685 & 0.668 \\ 
  0.60 & 20 & 60 & 0.15 & 0.052 & 0.781 & 0.796 & 0.774 \\ 
  0.60 & 20 & 70 & 0.15 & 0.053 & 0.859 & 0.870 & 0.857 \\ 
  0.60 & 20 & 80 & 0.15 & 0.053 & 0.911 & 0.919 & 0.905 \\ 
  0.60 & 60 & 40 & 0.15 & 0.388 & 0.706 & 0.777 & 0.708 \\ 
  0.60 & 60 & 50 & 0.15 & 0.388 & 0.823 & 0.872 & 0.823 \\ 
  0.60 & 60 & 60 & 0.15 & 0.388 & 0.895 & 0.927 & 0.894 \\ 
  0.60 & 100 & 30 & 0.15 & 0.546 & 0.632 & 0.796 & 0.634 \\ 
  0.60 & 100 & 40 & 0.15 & 0.563 & 0.788 & 0.895 & 0.788 \\ 
  0.60 & 100 & 50 & 0.15 & 0.563 & 0.879 & 0.946 & 0.875 \\ 
  0.80 & 5 & 40 & 0.15 & 0.222 & 0.697 & 0.701 & 0.678 \\ 
  0.80 & 5 & 50 & 0.15 & 0.239 & 0.844 & 0.847 & 0.842 \\ 
  0.80 & 5 & 60 & 0.15 & 0.247 & 0.924 & 0.925 & 0.909 \\ 
  0.80 & 20 & 30 & 0.15 & 0.052 & 0.692 & 0.712 & 0.685 \\ 
  0.80 & 20 & 40 & 0.15 & 0.052 & 0.859 & 0.870 & 0.857 \\ 
  0.80 & 60 & 20 & 0.15 & 0.385 & 0.616 & 0.703 & 0.617 \\ 
  0.80 & 60 & 30 & 0.15 & 0.388 & 0.845 & 0.889 & 0.843 \\ 
  0.80 & 100 & 20 & 0.15 & 0.561 & 0.717 & 0.855 & 0.715 \\ 
  0.80 & 100 & 30 & 0.15 & 0.563 & 0.896 & 0.955 & 0.897 \\ 
  1.00 & 5 & 30 & 0.15 & 0.233 & 0.790 & 0.794 & 0.774 \\ 
  1.00 & 20 & 20 & 0.15 & 0.052 & 0.699 & 0.720 & 0.704 \\ 
  1.00 & 20 & 30 & 0.15 & 0.053 & 0.913 & 0.921 & 0.908 \\ 
  1.00 & 60 & 20 & 0.15 & 0.388 & 0.853 & 0.897 & 0.853 \\ 
  1.00 & 100 & 20 & 0.15 & 0.563 & 0.904 & 0.960 & 0.903 \\ 
   \bottomrule
\end{tabular}
\endgroup
\caption{Excerpted results from a simulation study modelling biomarker studies, $m=200$. IST average power, oracle power, lower 
bound and simulated average power for a selection of effect sizes, values of $\E[M_m]$, FDR, and $n$.} 
\label{tbl:avgpwr_tbl_Bmkr}
\end{table}
\begin{table}[b]
\centering
\begingroup\tiny
\begin{tabular}{lrrrrrrrrrr}
  \toprule
 Eff Sz&$\E[M_m]$&n&FDR&$\pi_{_\mathrm{pi}}$&$\lambda_{75}$-pwr&$\hat\lambda_{75}$-pwr&$\lambda_{90}$-pwr&$\hat\lambda_{90}$-pwr&$\lambda_{eq}$&SS Ratio\\
  \cmidrule(r){1-1}\cmidrule(lr){2-2}\cmidrule(lr){3-3}\cmidrule(lr){4-4}\cmidrule(lr){5-5}\cmidrule(lr){6-6}\cmidrule(lr){7-7}\cmidrule(lr){8-8}\cmidrule(lr){9-9}\cmidrule(lr){10-10}\cmidrule(l){11-11}
0.60 & 5 & 70 & 0.15 & 0.704 & 0.426 & 0.524 & 0.216 & 0.249 & 0.570 & 1.500 \\ 
  0.60 & 5 & 80 & 0.15 & 0.795 & 0.584 & 0.681 & 0.308 & 0.396 & 0.622 & 1.500 \\ 
  0.60 & 5 & 90 & 0.15 & 0.860 & 0.738 & 0.797 & 0.409 & 0.538 & 0.673 & 1.467 \\ 
  0.60 & 5 & 100 & 0.15 & 0.906 & 0.867 & 0.875 & 0.518 & 0.657 & 0.721 & 1.380 \\ 
  0.60 & 20 & 50 & 0.15 & 0.664 & 0.265 & 0.307 & 0.042 & 0.028 & 0.607 & 1.500 \\ 
  0.60 & 20 & 60 & 0.15 & 0.781 & 0.610 & 0.630 & 0.139 & 0.157 & 0.695 & 1.500 \\ 
  0.60 & 20 & 70 & 0.15 & 0.859 & 0.894 & 0.877 & 0.320 & 0.378 & 0.765 & 1.343 \\ 
  0.60 & 20 & 80 & 0.15 & 0.911 & 0.991 & 0.961 & 0.563 & 0.599 & 0.819 & 1.225 \\ 
  0.60 & 60 & 40 & 0.15 & 0.706 & 0.280 & 0.315 & 0.005 & 0.002 & 0.666 & 1.500 \\ 
  0.60 & 60 & 50 & 0.15 & 0.823 & 0.903 & 0.891 & 0.088 & 0.099 & 0.771 & 1.360 \\ 
  0.60 & 60 & 60 & 0.15 & 0.895 & 1.000 & 0.995 & 0.453 & 0.492 & 0.842 & 1.183 \\ 
  0.60 & 100 & 30 & 0.15 & 0.632 & 0.037 & 0.031 & 0.000 & 0.000 & 0.609 & 1.500 \\ 
  0.60 & 100 & 40 & 0.15 & 0.788 & 0.786 & 0.789 & 0.011 & 0.006 & 0.750 & 1.450 \\ 
  0.60 & 100 & 50 & 0.15 & 0.879 & 1.000 & 0.999 & 0.278 & 0.270 & 0.838 & 1.200 \\ 
  0.80 & 5 & 40 & 0.15 & 0.697 & 0.417 & 0.496 & 0.212 & 0.252 & 0.566 & 1.500 \\ 
  0.80 & 5 & 50 & 0.15 & 0.844 & 0.696 & 0.784 & 0.380 & 0.502 & 0.659 & 1.500 \\ 
  0.80 & 5 & 60 & 0.15 & 0.924 & 0.915 & 0.895 & 0.574 & 0.704 & 0.742 & 1.333 \\ 
  0.80 & 20 & 30 & 0.15 & 0.692 & 0.330 & 0.352 & 0.057 & 0.036 & 0.626 & 1.500 \\ 
  0.80 & 20 & 40 & 0.15 & 0.859 & 0.892 & 0.893 & 0.319 & 0.357 & 0.764 & 1.350 \\ 
  0.80 & 60 & 20 & 0.15 & 0.616 & 0.062 & 0.060 & 0.001 & 0.000 & 0.590 & 1.500 \\ 
  0.80 & 60 & 30 & 0.15 & 0.845 & 0.963 & 0.952 & 0.148 & 0.147 & 0.791 & 1.333 \\ 
  0.80 & 100 & 20 & 0.15 & 0.717 & 0.286 & 0.284 & 0.001 & 0.003 & 0.684 & 1.500 \\ 
  0.80 & 100 & 30 & 0.15 & 0.896 & 1.000 & 1.000 & 0.453 & 0.506 & 0.855 & 1.167 \\ 
  1.00 & 5 & 30 & 0.15 & 0.790 & 0.574 & 0.680 & 0.304 & 0.392 & 0.617 & 1.500 \\ 
  1.00 & 20 & 20 & 0.15 & 0.699 & 0.350 & 0.431 & 0.064 & 0.045 & 0.630 & 1.500 \\ 
  1.00 & 20 & 30 & 0.15 & 0.913 & 0.992 & 0.966 & 0.579 & 0.614 & 0.821 & 1.200 \\ 
  1.00 & 60 & 20 & 0.15 & 0.853 & 0.978 & 0.959 & 0.183 & 0.225 & 0.799 & 1.300 \\ 
  1.00 & 100 & 20 & 0.15 & 0.904 & 1.000 & 0.999 & 0.549 & 0.586 & 0.863 & 1.150 \\ 
   \bottomrule
\end{tabular}
\endgroup
\caption{Excerpted results from a simulation study modelling biomarker studies, $m=200$. Shown are the $\lambda$-power at$\lambda=75\%$ and at $90\%$ from CLT and from simulations, $\lambda_{_{S/M}}(\pi_{_\mathrm{pi}})$ and samplesize ratio. The IST average power is also shown for comparison.} 
\label{tbl:Lpwr_tbl_Bmkr}
\end{table}
\begin{table}[b]
\centering
\begingroup\tiny
\begin{tabular}{lrrrrrrrr}
  \toprule
 $m$&Eff Sz&Power&$f\tck$&$n_{_{0,0}}$&$n_{_{0,1}}$&$n_{_{1,0}}$&$n_{_{1,1}}$&$\hat\pi_{_{T/J}}(f_0)$\\
  \cmidrule(r){1-1}\cmidrule(lr){2-2}\cmidrule(lr){3-3}\cmidrule(lr){4-4}\cmidrule(lr){5-5}\cmidrule(lr){6-6}\cmidrule(lr){7-7}\cmidrule(lr){8-8}\cmidrule(l){9-9}
1000 & 0.6667 & 0.6 & 0.069 & 51 & 65 & 75 & 89 & 0.0760 \\ 
  1000 & 0.6667 & 0.8 & 0.071 & 66 & 78 & 91 & 102 & 0.0900 \\ 
  1000 & 0.8333 & 0.6 & 0.069 & 33 & 43 & 50 & 59 & 0.0790 \\ 
  1000 & 0.8333 & 0.8 & 0.071 & 43 & 51 & 59 & 67 & 0.0880 \\ 
  1000 & 1.0000 & 0.6 & 0.069 & 24 & 31 & 36 & 42 & 0.0910 \\ 
  1000 & 1.0000 & 0.8 & 0.071 & 31 & 36 & 42 & 47 & 0.0770 \\ 
  2500 & 0.6667 & 0.6 & 0.097 & 51 & 57 & 75 & 83 & 0.1020 \\ 
  2500 & 0.6667 & 0.8 & 0.098 & 66 & 72 & 87 & 94 & 0.0980 \\ 
  2500 & 0.8333 & 0.6 & 0.097 & 33 & 38 & 50 & 54 & 0.0800 \\ 
  2500 & 0.8333 & 0.8 & 0.098 & 43 & 47 & 57 & 61 & 0.1100 \\ 
  2500 & 1.0000 & 0.6 & 0.097 & 24 & 27 & 36 & 39 & 0.1150 \\ 
  2500 & 1.0000 & 0.8 & 0.098 & 31 & 34 & 40 & 43 & 0.1040 \\ 
  5000 & 0.6667 & 0.6 & 0.112 & 51 & 55 & 75 & 80 & 0.1290 \\ 
  5000 & 0.6667 & 0.8 & 0.113 & 66 & 70 & 85 & 90 & 0.0970 \\ 
  5000 & 0.8333 & 0.6 & 0.113 & 33 & 36 & 50 & 53 & 0.1080 \\ 
  5000 & 0.8333 & 0.8 & 0.113 & 43 & 46 & 56 & 58 & 0.1260 \\ 
  5000 & 1.0000 & 0.6 & 0.113 & 24 & 26 & 36 & 38 & 0.1080 \\ 
  5000 & 1.0000 & 0.8 & 0.113 & 31 & 33 & 39 & 41 & 0.1050 \\ 
  7500 & 0.6667 & 0.6 & 0.120 & 51 & 54 & 75 & 80 & 0.1120 \\ 
  7500 & 0.6667 & 0.8 & 0.120 & 66 & 69 & 85 & 88 & 0.1100 \\ 
  7500 & 0.8333 & 0.6 & 0.120 & 33 & 36 & 50 & 53 & 0.1020 \\ 
  7500 & 0.8333 & 0.8 & 0.120 & 43 & 45 & 55 & 57 & 0.1210 \\ 
  7500 & 1.0000 & 0.6 & 0.120 & 24 & 26 & 36 & 38 & 0.1090 \\ 
  7500 & 1.0000 & 0.8 & 0.120 & 31 & 32 & 39 & 41 & 0.1160 \\ 
  10000 & 0.6667 & 0.6 & 0.124 & 51 & 53 & 75 & 78 & 0.1310 \\ 
  10000 & 0.6667 & 0.8 & 0.124 & 66 & 69 & 84 & 87 & 0.1060 \\ 
  10000 & 0.8333 & 0.6 & 0.124 & 33 & 35 & 50 & 51 & 0.1420 \\ 
  10000 & 0.8333 & 0.8 & 0.124 & 43 & 45 & 55 & 57 & 0.1150 \\ 
  10000 & 1.0000 & 0.6 & 0.124 & 24 & 25 & 36 & 38 & 0.1250 \\ 
  10000 & 1.0000 & 0.8 & 0.124 & 31 & 32 & 39 & 40 & 0.1040 \\ 
  20000 & 0.6667 & 0.6 & 0.131 & 51 & 52 & 75 & 78 & 0.1240 \\ 
  20000 & 0.6667 & 0.8 & 0.132 & 66 & 68 & 83 & 85 & 0.1300 \\ 
  20000 & 0.8333 & 0.6 & 0.131 & 33 & 35 & 50 & 51 & 0.1360 \\ 
  20000 & 0.8333 & 0.8 & 0.132 & 43 & 45 & 54 & 55 & 0.1160 \\ 
  20000 & 1.0000 & 0.6 & 0.131 & 24 & 25 & 36 & 36 & 0.1240 \\ 
  20000 & 1.0000 & 0.8 & 0.132 & 31 & 32 & 38 & 39 & 0.1190 \\ 
   \bottomrule
\end{tabular}
\endgroup
\caption{Excerpted results from the simulation study on the use of the CLT for the FDF to bound the FDF with large probability with    
FDR and $r$ fixed at 15\% and 2.5\%, respectively. Displayed are the parameter settigns, $m$, $\theta$, and power,       
followed by the value of the reduced FDR, $f\tck$ required to bound the FDF with probability $1-f_0$, the sample sizes       
$n_{_{0,0}}, n_{_{0,1}}, n_{_{1,0}},$ and $n_{_{1,1}}$, required for specified average power at $\mathrm{BHFDR}(f)$          
specified average power at $\mathrm{BHFDR}(f\tck)$, specified $\lambda_{90}$-power at $\mathrm{BHFDR}(f)$, and specified  
$\lambda_{90}$-power at $\mathrm{BHFDR}(f\tck)$. The last column is the simulated value, $\hat\pi_{_{T/J}}(f_0)$, of the 
tail probability of the FDF under $\mathrm{BHFDR}(f\tck)$.} 
\label{tbl:tbl_FDFincrN}
\end{table}
\begin{table}[b]
\centering
\begingroup\small
\begin{tabular}{lrrrrr}
  \toprule
 $n$&$\rho$&$\pi_{_\mathrm{pi}}$&$\lambda_{75}$-pwr&$\widehat{\mathrm{FDR}}$&$\pi_{_{T/J}}(0.18)$\\
  \cmidrule(r){1-1}\cmidrule(lr){2-2}\cmidrule(lr){3-3}\cmidrule(lr){4-4}\cmidrule(lr){5-5}\cmidrule(l){6-6}
14 & 0.00 & 0.7172 & 0.3000 & 0.1421 & 0.1910 \\ 
  14 & 0.10 & 0.7080 & 0.5020 & 0.1402 & 0.2080 \\ 
  14 & 0.20 & 0.7173 & 0.5260 & 0.1432 & 0.2320 \\ 
  14 & 0.30 & 0.6965 & 0.4780 & 0.1412 & 0.2310 \\ 
  14 & 0.40 & 0.7045 & 0.4930 & 0.1421 & 0.2180 \\ 
  14 & 0.50 & 0.6993 & 0.4860 & 0.1408 & 0.2180 \\ 
  14 & 0.60 & 0.7034 & 0.4830 & 0.1413 & 0.2280 \\ 
  14 & 0.70 & 0.6954 & 0.4910 & 0.1447 & 0.2500 \\ 
  14 & 0.80 & 0.7107 & 0.4900 & 0.1378 & 0.2220 \\ 
  16 & 0.00 & 0.7989 & 0.8680 & 0.1419 & 0.1670 \\ 
  16 & 0.10 & 0.7924 & 0.6950 & 0.1420 & 0.2370 \\ 
  16 & 0.20 & 0.7870 & 0.6880 & 0.1432 & 0.2270 \\ 
  16 & 0.30 & 0.7882 & 0.6940 & 0.1426 & 0.2350 \\ 
  16 & 0.40 & 0.7926 & 0.7020 & 0.1420 & 0.2270 \\ 
  16 & 0.50 & 0.7871 & 0.6940 & 0.1427 & 0.2420 \\ 
  16 & 0.60 & 0.7966 & 0.7050 & 0.1439 & 0.2350 \\ 
  16 & 0.70 & 0.7967 & 0.7060 & 0.1423 & 0.2250 \\ 
  16 & 0.80 & 0.7884 & 0.6780 & 0.1422 & 0.2150 \\ 
   \bottomrule
\end{tabular}
\endgroup
\caption{Average power, $\lambda_{75}$ power, empirical FDR, and $Pr( FDF > 0.18)$ at fixed effect size=1.25, FDR=15\%, r=5\% and 2000simultaneous tests, for sample sizes 14 and 16 and correlation varies over 0 to 80\% in increments of 10\%, with block size 100.} 
\label{tbl:tbl_corr_tests}
\end{table}\fi 

\ifsuppl 
\begin{table}[b]
\centering
\begingroup\tiny
\begin{tabular}{lrrrrrrr}
  \toprule
 Eff Sz&$\E[M_m]$&n&FDR&$\pi^{L}_{\mathrm{av},m}$&$\pi_{_\mathrm{pi}}$&$\pi_o$&$\hat\pi_{_{\mathrm{av},m}}$\\
  \cmidrule(r){1-1}\cmidrule(lr){2-2}\cmidrule(lr){3-3}\cmidrule(lr){4-4}\cmidrule(lr){5-5}\cmidrule(lr){6-6}\cmidrule(lr){7-7}\cmidrule(l){8-8}
0.60 & 100 & 100 & 0.15 & 0.444 & 0.683 & 0.683 & 0.681 \\ 
  0.60 & 100 & 110 & 0.15 & 0.444 & 0.763 & 0.763 & 0.762 \\ 
  0.60 & 100 & 120 & 0.15 & 0.444 & 0.826 & 0.826 & 0.825 \\ 
  0.60 & 100 & 130 & 0.15 & 0.444 & 0.874 & 0.874 & 0.874 \\ 
  0.60 & 100 & 140 & 0.15 & 0.444 & 0.910 & 0.910 & 0.910 \\ 
  0.60 & 1000 & 70 & 0.15 & 0.640 & 0.662 & 0.665 & 0.662 \\ 
  0.60 & 1000 & 80 & 0.15 & 0.743 & 0.761 & 0.764 & 0.761 \\ 
  0.60 & 1000 & 90 & 0.15 & 0.805 & 0.835 & 0.836 & 0.834 \\ 
  0.60 & 1000 & 100 & 0.15 & 0.815 & 0.887 & 0.889 & 0.887 \\ 
  0.60 & 1000 & 110 & 0.15 & 0.816 & 0.924 & 0.925 & 0.924 \\ 
  0.60 & 2000 & 60 & 0.15 & 0.616 & 0.640 & 0.647 & 0.640 \\ 
  0.60 & 2000 & 70 & 0.15 & 0.732 & 0.752 & 0.757 & 0.752 \\ 
  0.60 & 2000 & 80 & 0.15 & 0.817 & 0.832 & 0.836 & 0.832 \\ 
  0.60 & 2000 & 90 & 0.15 & 0.864 & 0.888 & 0.891 & 0.888 \\ 
  0.60 & 2000 & 100 & 0.15 & 0.870 & 0.927 & 0.929 & 0.927 \\ 
  0.80 & 100 & 60 & 0.15 & 0.444 & 0.717 & 0.717 & 0.718 \\ 
  0.80 & 100 & 70 & 0.15 & 0.444 & 0.835 & 0.836 & 0.837 \\ 
  0.80 & 100 & 80 & 0.15 & 0.444 & 0.909 & 0.909 & 0.908 \\ 
  0.80 & 1000 & 40 & 0.15 & 0.630 & 0.653 & 0.656 & 0.654 \\ 
  0.80 & 1000 & 50 & 0.15 & 0.794 & 0.816 & 0.818 & 0.815 \\ 
  0.80 & 1000 & 60 & 0.15 & 0.816 & 0.907 & 0.908 & 0.907 \\ 
  0.80 & 2000 & 40 & 0.15 & 0.727 & 0.747 & 0.753 & 0.747 \\ 
  0.80 & 2000 & 50 & 0.15 & 0.857 & 0.875 & 0.879 & 0.876 \\ 
  1.00 & 100 & 40 & 0.15 & 0.444 & 0.725 & 0.726 & 0.726 \\ 
  1.00 & 100 & 50 & 0.15 & 0.444 & 0.886 & 0.886 & 0.885 \\ 
  1.00 & 1000 & 30 & 0.15 & 0.734 & 0.754 & 0.756 & 0.753 \\ 
  1.00 & 1000 & 40 & 0.15 & 0.816 & 0.915 & 0.916 & 0.914 \\ 
  1.00 & 2000 & 30 & 0.15 & 0.814 & 0.830 & 0.835 & 0.830 \\ 
   \bottomrule
\end{tabular}
\endgroup
\caption{Excerpted results from a simulation study modelling micro-array studies, $m=\numprint{54675}$. IST average power, oracle power, lower bound and simulated average power for a selection of effect sizes, values of $\E[M_m]$, FDR, and $n$.} 
\label{tbl:avgpwr_tbl_Array}
\end{table}
\begin{table}[b]
\centering
\begingroup\tiny
\begin{tabular}{lrrrrrrrrrr}
  \toprule
 Eff Sz&$\E[M_m]$&n&FDR&$\pi_{_\mathrm{pi}}$&$\lambda_{75}$-pwr&$\hat\lambda_{75}$-pwr&$\lambda_{90}$-pwr&$\hat\lambda_{90}$-pwr&$\lambda_{eq}$&SS Ratio\\
  \cmidrule(r){1-1}\cmidrule(lr){2-2}\cmidrule(lr){3-3}\cmidrule(lr){4-4}\cmidrule(lr){5-5}\cmidrule(lr){6-6}\cmidrule(lr){7-7}\cmidrule(lr){8-8}\cmidrule(lr){9-9}\cmidrule(lr){10-10}\cmidrule(l){11-11}
0.60 & 100 & 100 & 0.15 & 0.683 & 0.112 & 0.098 & 0.000 & 0.000 & 0.656 & 1.420 \\ 
  0.60 & 100 & 110 & 0.15 & 0.763 & 0.602 & 0.611 & 0.002 & 0.000 & 0.728 & 1.309 \\ 
  0.60 & 100 & 120 & 0.15 & 0.826 & 0.963 & 0.953 & 0.039 & 0.037 & 0.786 & 1.217 \\ 
  0.60 & 100 & 130 & 0.15 & 0.874 & 1.000 & 0.998 & 0.233 & 0.248 & 0.832 & 1.138 \\ 
  0.60 & 100 & 140 & 0.15 & 0.910 & 1.000 & 1.000 & 0.627 & 0.661 & 0.869 & 1.071 \\ 
  0.60 & 1000 & 70 & 0.15 & 0.662 & 0.000 & 0.000 & 0.000 & 0.000 & 0.654 & 1.500 \\ 
  0.60 & 1000 & 80 & 0.15 & 0.761 & 0.758 & 0.754 & 0.000 & 0.000 & 0.750 & 1.312 \\ 
  0.60 & 1000 & 90 & 0.15 & 0.835 & 1.000 & 1.000 & 0.000 & 0.000 & 0.822 & 1.178 \\ 
  0.60 & 1000 & 100 & 0.15 & 0.887 & 1.000 & 1.000 & 0.121 & 0.110 & 0.874 & 1.070 \\ 
  0.60 & 1000 & 110 & 0.15 & 0.924 & 1.000 & 1.000 & 0.997 & 0.993 & 0.911 & 0.973 \\ 
  0.60 & 2000 & 60 & 0.15 & 0.640 & 0.000 & 0.000 & 0.000 & 0.000 & 0.635 & 1.500 \\ 
  0.60 & 2000 & 70 & 0.15 & 0.752 & 0.557 & 0.549 & 0.000 & 0.000 & 0.744 & 1.343 \\ 
  0.60 & 2000 & 80 & 0.15 & 0.832 & 1.000 & 1.000 & 0.000 & 0.000 & 0.823 & 1.188 \\ 
  0.60 & 2000 & 90 & 0.15 & 0.888 & 1.000 & 1.000 & 0.063 & 0.054 & 0.879 & 1.056 \\ 
  0.60 & 2000 & 100 & 0.15 & 0.927 & 1.000 & 1.000 & 1.000 & 1.000 & 0.918 & 0.960 \\ 
  0.80 & 100 & 60 & 0.15 & 0.717 & 0.266 & 0.311 & 0.000 & 0.000 & 0.687 & 1.367 \\ 
  0.80 & 100 & 70 & 0.15 & 0.835 & 0.981 & 0.984 & 0.058 & 0.055 & 0.795 & 1.200 \\ 
  0.80 & 100 & 80 & 0.15 & 0.909 & 1.000 & 1.000 & 0.612 & 0.625 & 0.868 & 1.075 \\ 
  0.80 & 1000 & 40 & 0.15 & 0.653 & 0.000 & 0.000 & 0.000 & 0.000 & 0.646 & 1.500 \\ 
  0.80 & 1000 & 50 & 0.15 & 0.816 & 1.000 & 1.000 & 0.000 & 0.000 & 0.803 & 1.220 \\ 
  0.80 & 1000 & 60 & 0.15 & 0.907 & 1.000 & 1.000 & 0.762 & 0.780 & 0.894 & 1.017 \\ 
  0.80 & 2000 & 40 & 0.15 & 0.747 & 0.404 & 0.393 & 0.000 & 0.000 & 0.739 & 1.350 \\ 
  0.80 & 2000 & 50 & 0.15 & 0.875 & 1.000 & 1.000 & 0.001 & 0.001 & 0.866 & 1.100 \\ 
  1.00 & 100 & 40 & 0.15 & 0.725 & 0.320 & 0.346 & 0.000 & 0.000 & 0.694 & 1.350 \\ 
  1.00 & 100 & 50 & 0.15 & 0.886 & 1.000 & 0.999 & 0.338 & 0.352 & 0.844 & 1.120 \\ 
  1.00 & 1000 & 30 & 0.15 & 0.754 & 0.592 & 0.593 & 0.000 & 0.000 & 0.743 & 1.333 \\ 
  1.00 & 1000 & 40 & 0.15 & 0.915 & 1.000 & 1.000 & 0.940 & 0.942 & 0.902 & 1.175 \\ 
  1.00 & 2000 & 30 & 0.15 & 0.830 & 1.000 & 1.000 & 0.000 & 0.000 & 0.821 & 1.200 \\ 
   \bottomrule
\end{tabular}
\endgroup
\caption{Excerpted results from a simulation study modelling micro-array studies, $m=\numprint{54675}$, for various values of effect sizes, values of $\E[M_m]$, FDR, and $n$. Shown are the $\lambda$-power at $\lambda=75\%$ and at $90\%$ from CLT and from simulations, $\lambda_{_{S/M}}(\pi_{_\mathrm{pi}})$ and sample size ratio. The IST average power is also shown for comparison.} 
\label{tbl:Lpwr_tbl_Array}
\end{table}
\begin{table}[b]
\centering
\begingroup\tiny
\begin{tabular}{lrrrrrrr}
  \toprule
 Eff Sz&$\E[M_m]$&n&FDR&$\pi^{L}_{\mathrm{av},m}$&$\pi_{_\mathrm{pi}}$&$\pi_o$&$\hat\pi_{_{\mathrm{av},m}}$\\
  \cmidrule(r){1-1}\cmidrule(lr){2-2}\cmidrule(lr){3-3}\cmidrule(lr){4-4}\cmidrule(lr){5-5}\cmidrule(lr){6-6}\cmidrule(lr){7-7}\cmidrule(l){8-8}
0.08 & 400 & 7800 & 0.01 & 0.610 & 0.612 & 0.612 & 0.612 \\ 
  0.08 & 400 & 8400 & 0.01 & 0.679 & 0.690 & 0.691 & 0.691 \\ 
  0.08 & 400 & 9000 & 0.01 & 0.705 & 0.757 & 0.758 & 0.757 \\ 
  0.08 & 400 & 9600 & 0.01 & 0.709 & 0.813 & 0.813 & 0.813 \\ 
  0.08 & 400 & 10200 & 0.01 & 0.709 & 0.858 & 0.858 & 0.857 \\ 
  0.08 & 400 & 10800 & 0.01 & 0.709 & 0.893 & 0.893 & 0.893 \\ 
  0.08 & 400 & 11400 & 0.01 & 0.709 & 0.921 & 0.921 & 0.921 \\ 
  0.28 & 400 & 650 & 0.01 & 0.624 & 0.626 & 0.626 & 0.626 \\ 
  0.28 & 400 & 700 & 0.01 & 0.687 & 0.704 & 0.704 & 0.705 \\ 
  0.28 & 400 & 750 & 0.01 & 0.707 & 0.770 & 0.770 & 0.771 \\ 
  0.28 & 400 & 800 & 0.01 & 0.709 & 0.824 & 0.824 & 0.824 \\ 
  0.28 & 400 & 850 & 0.01 & 0.709 & 0.868 & 0.868 & 0.868 \\ 
  0.28 & 400 & 900 & 0.01 & 0.709 & 0.902 & 0.902 & 0.902 \\ 
  0.28 & 400 & 950 & 0.01 & 0.709 & 0.928 & 0.928 & 0.929 \\ 
  0.68 & 400 & 120 & 0.01 & 0.668 & 0.676 & 0.676 & 0.677 \\ 
  0.68 & 400 & 160 & 0.01 & 0.709 & 0.912 & 0.912 & 0.911 \\ 
  0.08 & 1000 & 7500 & 0.01 & 0.651 & 0.652 & 0.652 & 0.651 \\ 
  0.08 & 1000 & 8000 & 0.01 & 0.715 & 0.716 & 0.716 & 0.716 \\ 
  0.08 & 1000 & 8500 & 0.01 & 0.769 & 0.771 & 0.771 & 0.772 \\ 
  0.08 & 1000 & 9000 & 0.01 & 0.803 & 0.818 & 0.818 & 0.818 \\ 
  0.08 & 1000 & 9500 & 0.01 & 0.813 & 0.856 & 0.856 & 0.856 \\ 
  0.08 & 1000 & 10000 & 0.01 & 0.814 & 0.888 & 0.888 & 0.888 \\ 
  0.08 & 1000 & 10500 & 0.01 & 0.814 & 0.913 & 0.913 & 0.913 \\ 
  0.28 & 1000 & 600 & 0.01 & 0.622 & 0.623 & 0.623 & 0.621 \\ 
  0.28 & 1000 & 640 & 0.01 & 0.689 & 0.689 & 0.690 & 0.689 \\ 
  0.28 & 1000 & 680 & 0.01 & 0.746 & 0.747 & 0.747 & 0.748 \\ 
  0.28 & 1000 & 720 & 0.01 & 0.791 & 0.797 & 0.797 & 0.797 \\ 
  0.28 & 1000 & 760 & 0.01 & 0.810 & 0.838 & 0.838 & 0.838 \\ 
  0.28 & 1000 & 800 & 0.01 & 0.813 & 0.872 & 0.873 & 0.873 \\ 
  0.28 & 1000 & 840 & 0.01 & 0.814 & 0.900 & 0.900 & 0.900 \\ 
  0.28 & 1000 & 880 & 0.01 & 0.814 & 0.923 & 0.923 & 0.923 \\ 
  0.68 & 1000 & 120 & 0.01 & 0.751 & 0.752 & 0.752 & 0.753 \\ 
   \bottomrule
\end{tabular}
\endgroup
\caption{Excerpted results from a simulation study modelling GWA studies, $m=\numprint{1000000}$. IST average power, oracle power, lower bound and simulated average power for a selection of effect sizes, values of $\E[M_m]$, FDR, and $n$.} 
\label{tbl:avgpwr_tbl_GWAS}
\end{table}
\begin{table}[ht]
\centering
\begingroup\tiny
\begin{tabular}{lrrrrrrrrrr}
  \toprule
 Eff Sz&$\E[M_m]$&n&FDR&$\pi_{_\mathrm{pi}}$&$\lambda_{75}$-pwr&$\hat\lambda_{75}$-pwr&$\lambda_{90}$-pwr&$\hat\lambda_{90}$-pwr&$\lambda_{eq}$&SS Ratio\\
  \cmidrule(r){1-1}\cmidrule(lr){2-2}\cmidrule(lr){3-3}\cmidrule(lr){4-4}\cmidrule(lr){5-5}\cmidrule(lr){6-6}\cmidrule(lr){7-7}\cmidrule(lr){8-8}\cmidrule(lr){9-9}\cmidrule(lr){10-10}\cmidrule(l){11-11}
0.08 & 400 & 7800 & 0.01 & 0.612 & 0.000 & 0.000 & 0.000 & 0.000 & 0.604 & 1.413 \\ 
  0.08 & 400 & 8400 & 0.01 & 0.690 & 0.011 & 0.012 & 0.000 & 0.000 & 0.677 & 1.320 \\ 
  0.08 & 400 & 9000 & 0.01 & 0.757 & 0.624 & 0.627 & 0.000 & 0.000 & 0.741 & 1.239 \\ 
  0.08 & 400 & 9600 & 0.01 & 0.813 & 0.999 & 0.999 & 0.000 & 0.000 & 0.794 & 1.168 \\ 
  0.08 & 400 & 10200 & 0.01 & 0.858 & 1.000 & 1.000 & 0.012 & 0.011 & 0.838 & 1.105 \\ 
  0.08 & 400 & 10800 & 0.01 & 0.893 & 1.000 & 1.000 & 0.341 & 0.347 & 0.873 & 1.048 \\ 
  0.08 & 400 & 11400 & 0.01 & 0.921 & 1.000 & 1.000 & 0.933 & 0.927 & 0.901 & 0.998 \\ 
  0.28 & 400 & 650 & 0.01 & 0.626 & 0.000 & 0.000 & 0.000 & 0.000 & 0.617 & 1.394 \\ 
  0.28 & 400 & 700 & 0.01 & 0.704 & 0.037 & 0.040 & 0.000 & 0.000 & 0.690 & 1.303 \\ 
  0.28 & 400 & 750 & 0.01 & 0.770 & 0.807 & 0.827 & 0.000 & 0.000 & 0.753 & 1.223 \\ 
  0.28 & 400 & 800 & 0.01 & 0.824 & 1.000 & 1.000 & 0.000 & 0.000 & 0.805 & 1.153 \\ 
  0.28 & 400 & 850 & 0.01 & 0.868 & 1.000 & 1.000 & 0.036 & 0.031 & 0.848 & 1.089 \\ 
  0.28 & 400 & 900 & 0.01 & 0.902 & 1.000 & 1.000 & 0.543 & 0.548 & 0.882 & 1.034 \\ 
  0.28 & 400 & 950 & 0.01 & 0.928 & 1.000 & 1.000 & 0.981 & 0.972 & 0.908 & 0.984 \\ 
  0.68 & 400 & 120 & 0.01 & 0.676 & 0.003 & 0.001 & 0.000 & 0.000 & 0.663 & 1.325 \\ 
  0.68 & 400 & 160 & 0.01 & 0.912 & 1.000 & 1.000 & 0.784 & 0.785 & 0.892 & 1.019 \\ 
  0.08 & 1000 & 7500 & 0.01 & 0.652 & 0.000 & 0.000 & 0.000 & 0.000 & 0.645 & 1.374 \\ 
  0.08 & 1000 & 8000 & 0.01 & 0.716 & 0.017 & 0.008 & 0.000 & 0.000 & 0.707 & 1.292 \\ 
  0.08 & 1000 & 8500 & 0.01 & 0.771 & 0.926 & 0.938 & 0.000 & 0.000 & 0.760 & 1.220 \\ 
  0.08 & 1000 & 9000 & 0.01 & 0.818 & 1.000 & 1.000 & 0.000 & 0.000 & 0.806 & 1.156 \\ 
  0.08 & 1000 & 9500 & 0.01 & 0.856 & 1.000 & 1.000 & 0.000 & 0.000 & 0.844 & 1.098 \\ 
  0.08 & 1000 & 10000 & 0.01 & 0.888 & 1.000 & 1.000 & 0.123 & 0.130 & 0.875 & 1.046 \\ 
  0.08 & 1000 & 10500 & 0.01 & 0.913 & 1.000 & 1.000 & 0.921 & 0.913 & 0.901 & 1.151 \\ 
  0.28 & 1000 & 600 & 0.01 & 0.623 & 0.000 & 0.000 & 0.000 & 0.000 & 0.617 & 1.408 \\ 
  0.28 & 1000 & 640 & 0.01 & 0.689 & 0.000 & 0.000 & 0.000 & 0.000 & 0.681 & 1.325 \\ 
  0.28 & 1000 & 680 & 0.01 & 0.747 & 0.432 & 0.474 & 0.000 & 0.000 & 0.737 & 1.250 \\ 
  0.28 & 1000 & 720 & 0.01 & 0.797 & 1.000 & 1.000 & 0.000 & 0.000 & 0.785 & 1.185 \\ 
  0.28 & 1000 & 760 & 0.01 & 0.838 & 1.000 & 1.000 & 0.000 & 0.000 & 0.826 & 1.125 \\ 
  0.28 & 1000 & 800 & 0.01 & 0.872 & 1.000 & 1.000 & 0.007 & 0.006 & 0.860 & 1.072 \\ 
  0.28 & 1000 & 840 & 0.01 & 0.900 & 1.000 & 1.000 & 0.516 & 0.506 & 0.888 & 1.151 \\ 
  0.28 & 1000 & 880 & 0.01 & 0.923 & 1.000 & 1.000 & 0.995 & 0.993 & 0.910 & 0.980 \\ 
  0.68 & 1000 & 120 & 0.01 & 0.752 & 0.546 & 0.586 & 0.000 & 0.000 & 0.741 & 1.242 \\ 
   \bottomrule
\end{tabular}
\endgroup
\caption{Excerpted results from a simulation study modelling GWA studies, $m=\numprint{1000000}$. Shown are the $\lambda$-powerat $\lambda=75\%$ and at $90\%$ from CLT and from simulations, $\lambda_{_{S/M}}(\pi_{_\mathrm{pi}})$ and samplesize ratio. The IST average power is also shown for comparison.} 
\label{tbl:Lpwr_tbl_GWAS}
\end{table}\fi 

\newif\ifmnscpt  
\newif\ifsuppl
\newif\ifnboth
\newif\ifblind
\newif\ifunblind

\mnscptfalse
\suppltrue
\nbothtrue
\blindfalse
\unblindtrue

\ifmnscpt
\ifnboth
\externaldocument{2018-08-13-pwrFDR-JASA-suppl-blind}
\fi 
\fi 

\ifsuppl
\ifnboth
\externaldocument{2018-08-13-pwrFDR-JASA-mnscpt-blind}
\fi 
\fi 


\ifsuppl 
\ifnboth
\title{Supplementary Material Accompanying ``Average- and $\lambda$- powers under BH-FDR}
\author{~~~}
\maketitle
\fi 
\fi 

\ifmnscpt 
\ifunblind
{
  \title{Average Power and $\lambda$-power in Multiple Testing Scenarios when the Benjamini-Hochberg 
    False Discovery Rate Procedure is Used\footnote{\arxiv{1801.03989}}}
  \author{Grant Izmirlian\thanks{This article is a U.S. Government work and is in the public domain in the U.S.A.}\hspace{.2cm}\\
    Biometry Research Group, Div. of Cancer Prev., NCI\\
    9609 Medical Center Dr; Bethesda, MD 20892-9789\\
    \href{mailto:izmirlig@mail.nih.gov}{izmirlig@mail.nih.gov}}
  \maketitle
} 
\fi 
\ifblind
{
 \title{Average Power and $\lambda$-power in Multiple Testing Scenarios when the Benjamini-Hochberg 
    False Discovery Rate Procedure is Used}
  \author{~~~}
  \maketitle
} 
\fi 

\bigskip
\begin{abstract}
We discuss several approaches to defining power in studies designed around the Benjamini-Hochberg (BH) false discovery
rate (FDR) procedure. We focus primarily on the \textit{average power} and the $\lambda$-\textit{power}, which are the
expected true positive fraction and the probability that the true positive fraction exceeds $\lambda$, respectively.
We prove results concerning strong consistency and asymptotic normality for the positive call fraction (PCF), the true
positive fraction (TPF) and false discovery fraction (FDF). Convergence of their corresponding expected values, including a
convergence result for the average power, follow as a corollaries. After reviewing what is known about convergence in
distribution of the errors of the plugin procedure, \cite{GenoveseC:2004}, we prove central limit theorems for fully
empirical versions of the PCF, TPF, and FDF, using a result for stopped stochastic processes. The central limit theorem
(CLT) for the TPF is used to obtain an approximate expression for the $\lambda$-power, while the CLT for the FDF is used
to introduce an approximate procedure for determining a suitably small nominal FDR that results in a speicified bound on
the FDF with stipulated high probability. The paper also contains the results of a large simulation study covering a
fairly substantial portion of the space of possible inputs encountered in application of the results in the design of a
biomarker study, a micro-array experiment and a GWAS study.
\end{abstract}

\thispagestyle{empty}
\noindent%
{\it Keywords:} MSC-2008 Primary: 62L12; MSC-2008 Secondary: 62N022; Multiple testing; false discovery rate; average power; k power; LLN; CLT\\
\ifunblind
\arxiv{1801.03989}
\fi 
\vfill

\newpage
\spacingset{1.45} 

\pagestyle{plain}
\setcounter{page}{1}

\fi 


\ifmnscpt 
\section{Introduction}
%
%
%
%
The explosion of available high-throughput technological pipelines in the biological and medical sciences over the past
20 years has opened up many new avenues of research that were previously unthinkable. The need to understand the role of
power in the era of ``omics'' studies cannot be overstated.  As a case in point, consider the promise and pitfalls of
RNA expression micro-arrays. One of the take away themes is that new technologies such as this one enjoy initial
exuberance and early victories \citep{AlizadehA:2000}, followed by calls for caution from epidemiologists and
statisticians \citep{IonnidisJPA:2005, BaggerlyKA:2009}. Some of the most constructive things gleaned from this journey,
in hindsight, have been a thorough re-evaluation of what should constitute the ``bar for science''. More and more
researchers are starting to realize that some of the blame for lack of reproducibility is owed to lack of power
\citep{IonnidisJPA:2005}. Central to these renewed calls for scientific vetting has been the concept of multiple
testing. Very early in the history of ``omics'' researchers realized that correction for multiple testing should be done
and that the most commonly used method, Bonferroni correction, was by far too conservative for tens of thousands of
simultaneous tests. Somewhat prophetically, half a decade before, \cite{BenjaminiY:1995} and colleagues introduced a new
testing paradigm, that, in contrast to the Bonferroni procedure which controls the probability that one or more type I
errors are committed, instead controls the proportion of false discoveries among the tests called significant.  By now,
use of the Benjamini-Hochberg (BH) false discovery rate (FDR) procedure for making statistical significance calls in
multiple testing scenarios is widespread. 

\section{Definitions and Notation}
Before describing the model of the data distribution, we need the following definition.
\begin{definition}
  A family of pdf's, $\{f_{\nu}:\nu \geq 0\}$, has the \textit{monotone likelihood ratio} property if and only if  
  \begin{equation}
    \nu' > \nu \mathrm{~implies~that~} f_{\nu'}/f_{\nu} \mathrm{~is~monotone~increasing}. 
  \end{equation}
  A family of CDF's has the monotone likelihood ratio property if its members are absolutely continuous and the 
  corresponding family of pdf's has the property. 
\end{definition}
Now, consider $m$ simultaneous tests of hypotheses $i=1,2,\ldots,m$, each a test whether a location parameter is 0, ($H_{0,i}$)
or non-zero ($H_{A,i}$). We start by supposing that an expected proportion, $0 < r < 1$, of the test statistics are
distributed about non-zero location parameters. The test statistic distributions are modeled according to a mixture model,
first introduced by \cite{StoreyJD:2002} and others \cite{BaldiP:2001, IbrahimJG:2002}.  First, let $\{\xi_i\}_{i=1}^{m}$
be an i.i.d. Bernouli $\{0,1\}$ sequence, with success probability, $r$. Denote the binomially distributed sum, 
$M_m = \sum_{i=1}^m \xi_i$, which is the number of test statistics belonging to the non-zero location parameter population. 
For a sample of size $n$ replicate outcomes resulting in $m$ simultaneous tests, let $X_{i,n}$ be the $i^{th}$ test statistic. 
We assume that conditional upon $\xi_i$, that its CDF is of the form: 
\begin{equation}
F_{X_{i,n}|\xi_i} = (1-\xi_i) F_{0, n} + \xi_i F_{A,n}.
\end{equation}
In the above, $F_{0,n}$ is the common distribution of all null distributed tests, and $F_{A,n}$ is the common distribution
of all non-null distributed tests. We assume that $F_{0,n}$ is the ``minimal'' element of a class of CDF's, 
$\mathbb{F}_n = \{F_{\nu, n} : \nu \geq 0\}$, satisfying the monotone likelihood ratio principle and that 
$F_{A,n} = \sum_{\ell=1}^h s_{\ell} F_{\nu_{\ell}, n}$ is a finite mixture of elements $F_{\nu_{\ell}, n}\in \mathbb{F}_n$, 
including the possibility that the mixing proportions are degenerate at a single distribution. Note that since we will 
only consider only two-sided tests, our scope is solely focused on non-negative test statistics, $X_{i,n}$, since they 
represent the the absolute value of some intermediate quantity. We consider $X_{i,n}$ to be non-negative in the remainder 
of the paper. Let $\bar F$ denote the complementary CDF (cCDF), so that $\bar F_{0,n}(x) = \P\{ X_{i,n} > x \mid \xi_i=0\}$
and $\bar F_{A,n} = \P\{ X_{i,n} > x \mid \xi_i = 1\}$. We will for nearly the entire paper make the assumption of
independent hypothesis tests. While the Benjamini-Hochberg false discovery rate procedure is not imune to departures
from the assumption of independent tests, the effect of departures from independence (i) do not affect the expected
false discovery fraction or expected true positive fraction (defined precisely below) and (ii) discrepencies which do 
result from such departures are seen in the distribution of the FDF and TPF and are caused by a reduced effective number 
of simultaneous tests. For these reasons, results obtained under the assumption of independent hypothesis tests are still 
of great utility. We return to this discussion in the final section.

\subsection{Mixed distribution of the P-values}
For $i=1,2,\ldots,m$, let $P_i = \bar F_{0,n}(X_{i,n})$ denote the two-sided nominal p-values corresponding to the test
statistics, $X_{i,n}$, and let $P^m_{(i)}$ denote their order statistics. Notice that the nominal p-values, 
$P_i, i=1,2,\ldots,m$ are i.i.d. having mixed CDF $G(u) = \P\{ P_i \leq u \}$, 
\begin{equation}
  G(u) = (1-r) u + r \bar F_{A,n}(\bar F_{0,n}^{-1}(u))\,.\label{eqn:Gdefnd}
\end{equation}
As we shall see in the proofs of Theorems \ref{thm:Jom_convas}, \ref{thm:SoM_convas}, and \ref{thm:CLT}, below,
the requirement that the family $\{F_{\nu, n}: \nu \geq 0 \}$ satisfies the monotone likelihood ratio
principle guarantees that $G$ is concave. Next in the original unsorted list of nominal p-values, 
$\{P_i: i=1,2,\ldots,m\}$, let $\{P_{1,i}:i=1,2,\ldots, M_m\}$ be the subset of nominal p-values corresponding to test 
statistics from the non-central population, in the order that they occur in the original unsorted list, e.g., if 
$N_{m,i} = \min\{j:i=\sum_{\ell=1}^j \xi_{\ell}\}$ counts the number of non-centrally located statistics among the first $i$ in the
original unsorted list, then $P_{1, i} = P_{_{N_{m,i}}}$.  

\subsection{Numbers of significant calls, true postives and false positives}
The Benjamini and Hochberg (B-H) false discovery rate (FDR) procedure \citep{BenjaminiY:1995} provides a simultaneous test
of all $m$ null hypotheses, that controls for multiplicity in a less conservative way than Bonferroni adjustment by
changing the paradigm. Instead of controlling the probability that one or more null hypotheses is erroneously rejected, it
controls the expected proportion of null hypotheses rejected that were true, or equivalently, the posterior probability
that a test statistic has null location parameter given it was called significant. The algorithm is implemented by
specifying a tolerable false discovery rate, $f$, and then finding the largest row number, $i$, for which the
corresponding order statistic, $P^m_{(i)}$, is less than $i f/m$. The total number of test statistics in the rejection
region, $J_m$, is given by the following expression:
\begin{definition}
  \label{def:J}
\begin{equation}
J_m = \max\left\{ i : P^m_{(i)} \leq \frac{i f}{m} \right\}\label{eqn:Jdef}\,.\nonumber
\end{equation}
\end{definition}
We will refer to $J_m$ as the number of \textit{positive calls} or \textit{discoveries} which is consistent with the 
terminology of Benjamini and Hochberg, and we call the ratio  $J_m/m$ the \textit{positive call fraction}. Notice that 
expression \ref{eqn:Jdef} has the following alternate form:
\begin{equation}
J_m = \sum_{i=1}^m I\left\{ P_i \leq m^{-1}  J_m \,f\right\}\label{eqn:JmsumExch}
\end{equation}

The number of positive calls partitions into \emph{true positve calls} and \emph{false positive calls}
\begin{definition}
  \label{def:S}
Let $S_m$ denote the \emph{number of true positive calls}:
\begin{equation}
  S_m = \sum_{i=1}^m \xi_i \, I\left( P_i \leq m^{-1} J_m f \right)\,.\label{eqn:SmsumExch}
\end{equation}
\end{definition}
\begin{definition}
  \label{def:T}
Let $T_m$ denote the \emph{number of false positive calls}:
  \begin{equation}
  T_m = \sum_{i=1}^m (1-\xi_i) \, I\left( P_i \leq m^{-1} J_m f \right)\,.\label{eqn:TmsumExch}
\end{equation}
\end{definition}
There are several possible choices of normalizers for $S_m$ and $T_m$, depending upon the popultion value being 
estimated.  Because power in the single hypothesis test scenario is the probability of rejection conditional
upon the alternative distribution, it is natural to normalize by the number of non-null distributed statistics, $M_m$:
\begin{definition}
We define the true positive fraction as the ratio $S_m/M_m$.
\end{definition}
Results concerning the false discovery rate will follow as corollaries to our other results. For this reason, we
normalize the number of false positive calls, $T_m$, by the number of positive callse, $J_m$:
\begin{definition}
We define the false discovery fraction as the ratio $T_m/J_m$.
\end{definition}
In general we will use the term \textit{fraction} for a ratio that is a random quantity and \textit{rate} for its
expectation.

Table \ref{tbl:SensSpec} shows rows partitioning the test statistics into those that are non-null distributed, 
and those that are null distributed, numbering $M_m$ and $m - M_m$, respectively, and columns 
partitioning the results of hypothesis testing into the positives and negatives calls, numbering $J_m$, and $m-J_m$,
respectively. 

\subsection{False discovery rate}
Let $f_0 = (1-r) f$. \cite{BenjaminiY:1995} showed in their original paper that their procedure 
controls the expected false discovery fraction, which they called the false discovery rate:
\begin{equation}
\E\left[\frac{T_m}{J_m}\right] = f_0 \leq f \label{eqn:BHFDR}
\end{equation}
In keeping with pervasive terminology, the phrase ``false discovery rate'' is applied to both the expected false discovery
fraction, $f_0=\E\left[J_m^{-1} T_m \right]$, and in addition, the nominal value, $f$ which is used to set the threshold
on the p-value scale. We will use the symbol $\mathrm{BHFDR}(f)$ to denote the BH-FDR procedure at nominal false discovery
rate, $\mathrm{FDR}=f$.  In this paper, whenever a random variable occurs in the denominator, we tacitly define the
indeterminate 0/0 to be 0, which has the effect that all such ratios are defined jointly with the event that the
denominator is non-zero.
\vspace{0.5truein}
\begin{table}[b]
\centering
\begin{tabular}{lrrr}
  \toprule
  &rej $H_{0,i}$&acc $H_{0,i}$&row Total\\
  \cmidrule(r){1-1}\cmidrule(lr){2-2}\cmidrule(lr){3-3}\cmidrule(l){4-4}
$H_{0,i}$ is FALSE & $S_m$ & $M_m - S_m$ & $M_m$ \\ 
  $H_{0,i}$ is TRUE & $T_m$ & $(m-M_m)-T_m$ & $m - M_m$ \\ 
   \cmidrule(r){1-1}\cmidrule(lr){2-2}\cmidrule(lr){3-3}\cmidrule(l){4-4}
col Total & $J_m$ & $m-J_m$ & $m$ \\ 
   \bottomrule
\end{tabular}
\caption{Counts of true positives, false positives, false negatives and true negatives.} 
\label{tbl:SensSpec}
\end{table}
\section{The distribution of $S_m$ and notions of power in multiple testing scenarios}
In the single hypothesis test situation, $m=1$, we consider probabilities of rejection given $H_0$ is true or false.
Under the setup introduced here, when $m=1$, the BH-FDR, $f$ becomes the type I error probability, and the power as it is
usually defined for a single hypothesis test, is the conditional expectation of $S_1$ given that $\xi_1=1$. In the case of
multiple tests, $S_m$ is distributed over values from zero to as high as $m$ so that naturally there are a multitude of
avenues for conceptualizing the power. Consider first, that had we been using the Bonferroni procedure for multiple tests
adjustment to thresh-hold the test statistics arriving at $J_m$ positives and $S_m$ true positives, the distribution of
$S_m$ would have been binomial with common success probability equal to the per-test power. The fact that the distribution
of $S_m$ is not binomial when the BH-FDR criterion is used is what makes discussion of power more difficult. However, the
common thread is that any discussion of power in the multiple testing scenario must be based upon some summary of the
distribution of $S_m$, e.g. a right tail or a moment.

\subsection{Various definitions of power in multiple testing scenarios}
One of the first approaches was to use the probability that $S_m$ is non-zero: $\P\left\{S_m > 0\right\}$. 
\cite{LeeMLT:2002} used a Poisson approximation to derive a closed-form expression for the probability to observe
one or more true positives.  This kind of power, the family-wise power, is arguably not a meaningful target of
optimization for experiments built around a large number of simultaneous tests, especially when there are typically
complex underlying hypotheses relying on positive calls for a sizable portion of those tests for which the
alternative is true. For example, consider that in a micro-array experiment in which there will be downstream
pathways analysis, we would start by assuming that there are around 3\% or more of the $m$ tests for which the 
alternate hypothesis is true, and hope to make significant calls at an FDR of 15\% for at least 80\% of these non-null 
distributed statistics, so as to have a thresholded list of roughly $1600$ genes to send to an analysis of pathways.

\subsection{The Average Power and $\lambda$-power}\hfil\break
\subsection*{Average Power}
In the BH-FDR procedure for multiple testing, the role of the type I error played in the single testing scenario is 
assumed by the FDR which is an expected proportion. Therefore, it is natural in the multiple testing scenario to 
consider a power that is also defined as an expected proportion. One interpretation of power in the setting of
multiple testing that falls along this line of reasoning is the ``average power''.
\begin{definition}
The average power is the expected true positive fraction, i.e. the expected proportion of all non-null distributed 
statistics that are declared significant by the BH FDR procedure.
\begin{equation}
    \pi_{_{\mathrm{av},m}} = \E\left[\frac{S_m}{M_m}\right] \label{eqn:avgpwr}
\end{equation}
\end{definition}
Notice that here the dependence upon $m$ is made explicit, so that the average power depends upon the number of
simultaneous tests in addition to quantities named above. \cite{GlueckD:2008} provided an explicit formula for the
average power in a finite number, $m$, of simultaneous test, but its complexity grows as the factorial of the number of
simultaneous tests, and this is clearly intractable in the realm of micro-array studies and GWAS where there are tens of
thousands or even a million simultaneous tests in question.\hfil\break

\subsection{The plug-in estimate of the average power}
Thinking heuristically for the moment, if $m$ is very large as will be the case in many ``omics'' scenarios, then
the positive call fraction $J_m/m$ could be considered very close to a limiting value, $\gamma$, if such a limiting
value existed either in probability or almost surely. Continuing along heuristic lines then, we could replace the
positive call fraction, $J_m/m$, with the limiting constant, $\gamma$, in the right hand side of expression 
\ref{eqn:JmsumExch}, as well as in expressions \ref{eqn:SmsumExch} and \ref{eqn:TmsumExch}, which define $S_m$ and 
$T_m$. If such a treatment were legitimate, the resulting analysis would be extremely easy as $J_m, S_m$ and $T_m$ 
would be sums of i.i.d's and the usual L.L.N. holds, with limits given by the expected value of a single increment 
in the corresponding sum. This gives rise to the plug-in estimate of the average power,
\begin{equation}
  \pi_{\mathrm{pi}} = \P\{ P_i < \gamma f | \xi_i = 1\}  = \bar F_{A,n}(\bar F_{0,n}^{-1}(\gamma f))
\end{equation}
Several authors have discussed this plug-in estimate of average power, for example \cite{GenoveseC:2004,
  StoreyJD:2002}.  Independently, \cite{JungSH:2005} and \cite{LiuP:2007} discuss sample size and power in the
setting of multiple testing based upon the BH FDR procedure. Without actually ever calling it the average power,
they derive an expression very close to the above plugin estimate. Their derivation starts with the posterior
probability that a statistic was drawn from the null-distributed population, given that it was called
significant. Bayes theorem is used to express this in terms of the prior, $1-r$ and $r$, and conditional, $\bar
F_{0,n}$ and $\bar F_{A,n}$. They mistakenly equate this to the nominal false discovery rate, $f$, when in actuality
it is the observed false discovery rate, $f_0 = (1-r) f$. Not withstanding, their methodology is valid because the
resulting power is, as we shall see below, the average power at the ``oracle'' threshold (\cite{GenoveseC:2004}) on
the p-value scale , $\gamma f/(1-r)$. This is the largest cut-off that is still valid at the nominal false discovery
rate, $f$. Around the same time, \cite{SunWG:2007}, discussed a generalization of of the FDR procedure based upon
the local FDR, and showed via decision theoretic techniques, that the false non-discovery rate, a quantity related
to the average power, has better performance characteristics than the FDR under many circumstances.  The paper also
provided a very good survey of the then currently available results. While the ramifications of that work are great,
the results provided here have merit in that the results and methodology supplied in the form of second order
asymptotics for the TDF and the FDF are entirely new and have important ramifications in of themselves.

\subsection*{The $\lambda$-Power}
Use of the average power in designing studies or deriving operating characteristics of them makes sense only when the
width of the distribution of $S_m/M_m$ is very narrow. To have more definitive control over the true positive fraction,
some authors have introduced the ``K-power''. This was originally introduced in a model where the number of non-null
distributed tests was fixed and was defined as the probability that the number of true positives exceeded a given
integral threshold, $k$.  Under the current setup in which the number of non-null distributed tests is a binomial random
variable this no longer makes sense. We introduce instead the $\lambda$-power, which is the probability that the true
positive fraction, $S_m/M_m$ exceeds a given threshold, $\lambda \in (0, 1)$:
\begin{definition}
We define the $\lambda$-power:
\begin{equation}
\pi_{_{S/M}}(\lambda) = \P\left\{\frac{S_m}{M_m} \geq \lambda\right\}.\label{eqn:Lpower}
\end{equation}
We will also use the term ``$\lambda_{k}$-power'' to denote $\pi_{_{S/M}}(k/100)$, the $\lambda$-power at threshold $k/100$.
The associated quantile function is denoted:
\begin{equation}
\lambda_{_{S/M}}(\pi) = \pi_{_{S/M}}^{-1}(\pi).\label{eqn:leq}
\end{equation}
\end{definition}
As mentioned above, the $\lambda$-power becomes especially meaningful in experiments for which there are a small to intermediate 
number of simultaneous tests and for which the distribution of the TPF, $S_m/M_m$, becomes non-negligibly dispersed.  We
prove a CLT for the true positive fraction which we use to approximate the $\lambda$-power. The accuracy of this approximation 
will be investigated in a simulation study.

\begin{remark}
Because the distribution of the TPF is nearly symetric for even relatively small values of $m>50$, the mean and median
nearly coincide. Thus 
\begin{equation}
\pi_{_{S/M}}(\lambda) \approx 1/2 \mathrm{~when~} \lambda=\pi_{_\mathrm{pi}} \label{eqn:LpwrEqHlf}
\end{equation}
Because the $\lambda$-power takes the values $1$ when $\lambda=0$ and $0$ when $\lambda=1$ and is continuous by assumption,
there exists a quantile, $\lambda_{eq} = \pi_{_{S/M}}^{-1}(\pi_{_\mathrm{pi}})$, at which the $\lambda$ power equals the average power:
\begin{equation}
\pi_{_{S/M}}(\lambda_{eq}) = \pi_{_\mathrm{pi}}\,. \label{eqn:Lambda_eq}
\end{equation}
Because the $\lambda$-power is a cCDF it is a non-increasing function of $\lambda$,
\begin{equation}
\pi_{_{S/M}}(\lambda) < \pi_{_\mathrm{pi}} \mathrm{~for~} \lambda > \lambda_{eq} \mathrm{~and~} \pi_{_{S/M}}(\lambda) > \pi_{_\mathrm{pi}} \mathrm{~for~} \lambda < \lambda_{eq} 
\label{eqn:LPwr_AvgPwr}
\end{equation}
\end{remark}

\subsection*{Bounding the FDF}
We conclude the section on various notions of power with a brief diversion. The fact that the TPF may be non-negligibly
dispersed at small to intermediate values of $m$ leads to concern that the FDF distribution is similarly dispersed at
these small to intermediate values of $m$. This concern is addressed in one of the CLT results and in the simulation
study. We introduce some notation for its cCDF and quantile function.
\begin{definition}
At BH-FDR $f$, denote the FDF tail probability: 
\begin{equation}
\pi_{_{T/J}}(\lambda) = \P\left\{\frac{T_m}{J_m} \geq \lambda\right\}\,.\label{eqn:FDFcCDF}
\end{equation}
Denote its quantile function:
\begin{equation}
\lambda_{_{T/J}}(p) = \pi_{_{T/J}}^{-1}(p)\,.\label{eqn:lToJp}
\end{equation}
\end{definition}
At small and intermediate values of $m$, the value of $\lambda_{_{T/J}}$ required to bound the FDF by $f_0$ with probability
bounded by $f_0$ can be as much as 100\% larger than the FDR. As remarked above, this will be discussed further in the 
context of our CLT results and simulation studies below.

The remainder of the paper proceeds according to the following plan. Section 4 is a presentation of the main theoretical
results, and this is done two subsections. In subsection 4.1, almost sure limits of the positive call fraction, true
positive fraction and false discovery fraction, as the number of simultaneous tests tends to infinity, are shown to
exist and are fully characterized. Convergence of the corresponding expectations, the true positive rate or average
power, and false positive rate, follow as a corollaries. Subsection 4.2 contains central limit theorems (CLT's) for the
positive call fraction, true positive fraction and false discovery fraction.  We also provide a lower bound for the
average power at a finite number, $m$, of simultaneous tests. We show how these CLT results can be used to approximate
the $\lambda$-power allowing tighter control over the TPF in power and sample size calculations, as well as how the
approximate distirbution of the FDF can be used to tighten down control over the FDF at both the design and analysis
stage. Section 5 is devoted to a simulation study, in which we study the regions of the parameter space that are typical
to small biomarker studies, micro-array studies and GWAS studies. We also focus on characteristics of the distribution
of the FDF as the number of simultaneous tests grows. Weak consistency of $J_m/m, S_m/M_m$ and $T_m/J_m$ to $\gamma$, 
$\pi_{\mathrm{pi}}$ and $(1-r)f$ was proved in \cite{StoreyJD:2002} and \cite{GenoveseC:2002}. The paper \cite{GenoveseC:2004} is 
a study of consistency and convergence in distribution of the paths of the plug-in estimator, considered a stochastic
process in the p-value plugin criterion, $t\in(0,1)$. The strong consistency results and weak convergence results for
the observed positive call fraction, true positive fraction and false discovery fraction presented here are new. Note that
almost sure convergence of the positive call fraction is necessary for almost sure convergence of the true positive and 
false discovery fractions.c
\section{Theoretical Results}
\subsection{Law of Large Numbers}\hfil\break
\subsection*{LLN for Positive Call Fraction, $J_m/m$}
\begin{theorem}
  \label{thm:Jom_convas}
If the family $\{F_{\nu,n}:\nu \geq 0\}$ is absolutely continuous and has the monotone likelihood
ratio property, then 
\begin{equation}
\lim_{m\rightarrow\infty} m^{-1}\,J_m = \sup \{ u: u = G(u f) \} \equiv \gamma \mathrm{~almost~surely,}
\end{equation}
\end{theorem}
Proofs of this and all other results are contained in the accompanying supplemental material. 

\begin{remark}
  When the family $\{F_{\nu,n}: \nu \geq 0\}$ has the monotone likelihood ratio property, $\gamma$ will be the unique 
  non-zero solution of $G(u f) = u$. 
\end{remark}
Once the almost sure convergence of $J_m/m$ is established we can apply a result of \cite{TaylorRL:1985} 
to establish convergence results for the true positive fraction. 
\subsection*{LLN for the True Positive Fraction, $S_m/M_m$}
\begin{theorem}
  \label{thm:SoM_convas}
Under the conditions of theorem \ref{thm:Jom_convas},
\begin{eqnarray}
\lim_{m\rightarrow\infty} m^{-1}\,S_m &=& \P\{P_{i} \leq \gamma f\,, \xi_i=1\} = r\,\bar F_{A,n}(\bar F_{0,n}^{-1}(\gamma f)) 
                                     \mathrm{~a.s.}\,,\label{eqn:Som_convas}\\
&&\nonumber\\
\lim_{m\rightarrow\infty} M_m^{-1}\,S_m &=& \P\{P_{i} \leq \gamma f \mid \xi_i=1\} = \bar F_{A,n}(\bar F_{0,n}^{-1}(\gamma f))
                                      \equiv \pi_{_\mathrm{pi}}\mathrm{~a.s.}\label{eqn:SoM_convas}\\
\mathrm{and}&&\nonumber\\
\lim_{m\rightarrow\infty} \pi_{_{\mathrm{av},m}} &=& \lim_{m\rightarrow\infty}\E\left[M_m^{-1}\,S_m\right] =\pi_{_\mathrm{pi}}\label{eqn:E_SoM_conv}
\end{eqnarray}
\end{theorem}

Corresponding convergence results for the false discovery fraction and its expected value follow as a corollary.

\subsection*{LLN for the False Discovery Fraction, $T_m/J_m$}
\begin{corollary}
  \label{cor:FDF_convas}
Under the conditions of theorem \ref{thm:Jom_convas},
\begin{eqnarray}
\lim_{m\rightarrow\infty} m^{-1}\,T_m &=& \P\{P_{i} \leq \gamma f\,, \xi_i=0\} = (1-r)\,\gamma f 
                                     \mathrm{~a.s.}\,,\label{eqn:Tom_convas}\\
&&\nonumber\\
\lim_{m\rightarrow\infty} J_m^{-1}\,T_m &=& \P\{\xi_i = 0 \mid P_{i} \leq \gamma f \} = (1-r) f 
                                      \mathrm{~a.s.}\label{eqn:FDF_convas}\\
\mathrm{and}&&\nonumber\\
\lim_{m\rightarrow\infty} \E\left[J_m^{-1}\,T_m\right] &=& (1-r) f
\end{eqnarray}
\end{corollary}

\begin{remark}
Because $T_m = J_m - S_m$ then by Theorem \ref{cor:FDF_convas} and its corollary \ref{thm:SoM_convas}, we obtain the
identity $(1-r) f = 1 - r \pi_{_\mathrm{pi}}/\gamma$, which can be rearranged to obtain an expression for the limiting
positive call fraction:
\begin{equation} 
  \gamma = \frac{r \pi_{_\mathrm{pi}}}{1-f_0}.\label{eqn:gamma_expr}
\end{equation}
\end{remark}

\begin{remark}
  \label{rmk:oracle}
If the nominal false discovery rate, $f$, is replaced by the inflated value, $f/(1-r)$, resulting in the 
$\mathrm{BHFDR}(f/(1-r))$ procedure, note that the FDR is still controlled at the nominal value, $f$, since in this 
case, $\mathbb{E}[ J - S / J ] = f$ due to cancellation.  This threshold, $\gamma\,f/(1-r)$, on the p-value scale, has
been called the oracle threshold by some authors, \cite{GenoveseC:2004}, because it is the criterion resulting in the 
largest power which is still valid for a given FDR, $f$. Call this the oracle average power, $\pi_o$. The actual 
difference only begins to get appreciable as $r$ increases in size. In practice, as we will see in our simulation 
study, $r$ must be as large as 50\% or more before this has a dramatic effect on the average power. Keep in mind that 
in practice when analyzing a given dataset, this increased power is only attainable at the stage of estimation if 
a reasonably good estimate of $r$ is possible. The fact that this is very problematic has also been a topic of 
much discussion.
\end{remark}

\noindent Next, if we replace $\gamma$ in the definition of the IST average power \ref{eqn:SoM_convas} by the expression
\ref{eqn:gamma_expr}, we arrive at a new equation which gives an implicit definition for the IST average power.
\begin{corollary}
  \label{cor:pi1Alt}
Under the conditions of theorem \ref{thm:Jom_convas},
  \begin{equation}
\pi_{_\mathrm{pi}} = \bar F_{A,n}\left(\bar F_{0,n}^{-1}\left((1-f_0)^{-1} r \pi_{_\mathrm{pi}} f\right)\right)\,.
\end{equation}
\end{corollary}

\noindent The almost sure convergence results given in Theorems \ref{thm:Jom_convas}, \ref{cor:FDF_convas} and 
\ref{thm:SoM_convas} above each have corresponding central limit results which we state now. The first, a CLT for 
the centered and $\sqrt{m}$-scaled positive call fraction, $J_m/m$, is needed in the proof of the second 
and third results, CLT's for centered and scaled versions of the false discovery fraction and the true positive 
fraction.

\subsection{CLTs for the PCF, FDF, and TPF; Lower Bound for Average Power}
\begin{theorem}
  \label{thm:CLT}
Under the conditions of theorem \ref{thm:Jom_convas},
  \begin{eqnarray}
  \sqrt{m} \Big(m^{-1} J_m \kern-0.75em &-&\kern-0.75em \gamma \Big) \convind N(0, \tau^2)\,.\label{eqn:CLT_Jortm}\\
  &&\nonumber\\
  \sqrt{m} \Big(J_m^{-1} T_m \kern-0.75em &-&\kern-0.75em f_0 \Big) \convind N(0, \alpha^2)\,.\label{eqn:CLT_rtmFDF}\\
  &&\nonumber\\
  \sqrt{m} \Big(M_m^{-1} S_m\kern-0.75em &-&\kern-0.75em \pi_{_\mathrm{pi}}\Big)\convind N(0,\sigma^2)\,.\label{eqn:CLT_rtmSoM}
  \end{eqnarray}
\end{theorem}
The proof, which uses results on convergence of stopped stochastic processes, is constructive in nature producing fully
characterized limiting distributions yielding asymptotic variance formulae. We reiterate the practical
implications of these results.  

\subsection*{Approximating the $\lambda$-power via the CLT for the TPF}
Currently, multiple testing experiments are designed using the average power, $\pi_{_\mathrm{pi}}$, which is the mean of
the distribution of $S_m/M_m$. In cases in which the width of this distribution is non-negligible, e.g. $m<1000$
simultaneous tests or so, we recommend using the $\lambda$-power instead of the average power. As we defined above, in
equation \ref{eqn:Lpower}, the $\lambda$-power is the probability that the TPF exceeds a given $\lambda$. We will see in 
our simulation study that in the ranges of the parameter space investigated, this CLT approximation is quite good and can
be used to approximate the $\lambda$-power:
\begin{equation}
 \pi_{_{S/M}}(\lambda) = \P\{ S_m/M_m \geq \lambda\} \approx \Phi(\sqrt{m}/\sigma (\pi_{_\mathrm{pi}} - \lambda))\,,\label{eqn:lpwrCLT}
\end{equation}
where $\sigma$ above is the square-root of the asymptotic variance, $\sigma^2$, given in formula \ref{eqn:s2} in the proof 
of theorem \ref{thm:CLT}. 

\subsection*{Enhanced control of the FDF via its CLT}
As defined above in expression \ref{eqn:lToJp} and the text leading up to it, $\lambda_{_{T/J}}(p) = \pi_{_{T/J}}^{-1}(p)$ is the
quantile of the FDF distribution at upper tail probability, $p$. The CLT for the FDF can be used to approximate it:
\begin{equation}
\lambda_{_{T/J}}(p) \approx  f_0 + \alpha/\sqrt{m}  \Phi^{-1}(1 - p)\,.
\end{equation}
Here, $f_0 = (1-r) f$ as above, $\alpha$ is the square root of the asymptotic variance, $\alpha^2$ given in formula \ref{eqn:a2}
in the proof of \ref{thm:CLT}, and $\Phi^{-1}$ is the standard normal quantile function. This can be used in several different
ways to bound the FDF with specified probability. Three possibilities are as follows. First, as a kind of loss function on 
lack of control inherent in the use of the $\mathrm{BHFDR}(f)$ procedure, we could determine how large a threshhold is required 
so that the FDF is bounded by $\lambda$ except for a tail probability of $f_0$
\begin{equation}
\lambda_{_{T/J}}(f_0) =  f_0 + \alpha/\sqrt{m}  \Phi^{-1}(1 - f_0)
\end{equation}
A second way would to find the solution $f\tck$ to the following equation.
\begin{equation}
f_0 =  (1-r) f\tck + \alpha/\sqrt{m}  \Phi^{-1}(1 - (1-r) f_0)\,.\label{eqn:fpr}
\end{equation}
This would produce a reduced FDR, $f\tck < f$, at which the $\mathrm{BHFDR}(f\tck)$ procedure would result in a FDF, $T_m/J_m$
of no more than $f_0$ with probability $1-f_0$. The solution is 
\begin{equation}
f\tck = f - \alpha/(\sqrt{m}(1-r))  \Phi^{-1}(1 - (1-r) f_0)
\end{equation}
A third way to do this, and the most conservative of the three, would be to determine the value of a reduced FDR, $f\tck$, at which 
the $\mathrm{BHFDR}(f\tck)$ procedure would result in a FDF no more than $f_0$ with probability $1-(1-r) f\tck$, by solving the 
following equation numerically: 
\begin{equation}
f_0 =  (1-r) f\tck + \alpha/\sqrt{m}  \Phi^{-1}(1 - (1-r) f\tck)\,.\label{eqn:fprime}
\end{equation}

\begin{remark}
The farther apart $f\tck$ is from $f_0$ is an indication of the dispersion of the distribution of $T_m/J_m$. 
\end{remark}
  
\begin{remark}
The procedure summarized in equation \ref{eqn:fprime}, for finding a reduced FDR, $f\tck$, at which
$\mathrm{BHFDR}(f\tck)$ would produce an FDF of no more than $f_0$ with probability $1-(1-r)f\tck$, can also be used at the
analysis phase. Note that expression \ref{eqn:a2} in the proof of \ref{thm:CLT} for the asymptotic variance, $\alpha^2$,
of $T_m/J_m$ depends only upon $f_0=(1-r)f$ and $\gamma$. Thus we can replace $f_0$ with $f$ and estimate $\gamma$
from the data using the plug-in estimate, $J_m/m$. This has important ramifications for the setting of small to intermediate 
number of simultaneous tests, $m\leq 1000$.
\end{remark}

\subsection*{Lower Bound for finite simultaneous tests average power}
As we will see in the simulation study which follows, the IST average power is in fact extremely close to 
the finite simultaneous tests (FST) average power for the broad ranges of the parameters studied. Nevertheless, it is 
still useful to have bounds for the FST average power. 

\begin{theorem}
\label{thm:LowerBdd}
The FST average power, $\pi_{_{\mathrm{av},m}}$, is bounded below by the following quantity, $\pi^{L}_{\mathrm{av},m}$, given below. 
\begin{eqnarray}
\pi_{_{\mathrm{av},m}} &\geq& \sum_{{\ell}=1}^{m} \binom{m}{{\ell}} r^{\ell} (1-r)^{m-{\ell}} \frac{1}{{\ell}} \sum_{s=1}^{{\ell}} 
             \bar B_{{\ell}-s+1,s}\left(\bar F_{\nu,n}\left(1-\bar F_{0,n}^{-1}\left(\frac{s f}{m}\right)\right)\right) \nonumber\\
    &\equiv& \pi^{L}_{\mathrm{av},m} \nonumber
\end{eqnarray}
\end{theorem}

An upper bound that seems to work in practice is the expression obtained by replacing $J_m/m$ with $r/(1-f_0)$ 
in equation \ref{eqn:lwrbddeqn3} in the proof of Theorem \ref{thm:LowerBdd}.




\section{Simulation Study}
We conducted four simulation studies. The first three of these had fixed $m$ and ranges of the other parameters chosen
based upon relevance to subject matter areas. The first, with $m=200$, was meant to model biomarker studies. In a second
simulation study, we varied $m$ in order to study characteristics of the FDF distribution as $m$ grows. In the third,  
in which $m=\numprint{54675}$, was meant to model micro-array studies for while the fourth, for which $m=\numprint{1000000}$, 
was meant to model genome wide association (GWA) studies.

In all four cases, the test statistic distributions, $F_{0,n}$ and $F_{\nu,n}$, were
chosen to be t-distributions of $2n - 2$ degrees of freedom. The common non-centrality parameter was fixed at 
$\nu=\sqrt{n/2}\theta$. This corresponds to a two group comparison as is often done. For each of these simulation
studies, we chose subject matter relevant ranges for the four parameters, the expected proportion of non-null tests,
$r$, the location parameter, $\theta$, and the false discovery rate, $f$. Except when set explicitly as in the fourth 
case, a range sample sizes, $n$, in increments of 5, was chosen to result in powers between 60\% and 95\% at each setting
of the other parameters. We conducted a total of four simulation studies. The first, with $m=200$ simultaneous tests, was
meant to model biomarker studies. The second, with varying sizes of $m$ ranging from \numprint{1000} to \numprint{20000}
was done in order to assess the width of the FDF distribution and the adequacy of the CLT approximation to it.
The third, with $m=$\numprint{54675} was meant to model human oligo-nucleotide micro-array experiments, and the
forth with $m=$\numprint{1000000} was meant to model GWA studies. We present the first two of these in the main 
text and the remainder in the supplementary material. 

\ifsuppl
\setcounter{Stbls}{0}
\refstepcounter{Stbls}\label{tbl:avgpwr_tbl_Array}
\refstepcounter{Stbls}\label{tbl:Lpwr_tbl_Array}
\refstepcounter{Stbls}\label{tbl:avgpwr_tbl_GWAS}
\refstepcounter{Stbls}\label{tbl:Lpwr_tbl_GWAS}
\refstepcounter{Stbls}\label{tbl:avgpwr_tbl_Bmkr}
\refstepcounter{Stbls}\label{tbl:Lpwr_tbl_Bmkr}
\refstepcounter{Stbls}\label{tbl:tbl_FDFincrN}
\fi 

For each of simulation studies focused on the distribution of the TPF, we computed, at each combination of these four
parameters, the IST average power, $\pi_{_\mathrm{pi}}$, from line \ref{eqn:E_SoM_conv} of Theorem \ref{thm:SoM_convas},
the oracle power, $\pi_o$, mentioned in remark \ref{rmk:oracle} above, and the lower bound, $\pi^{L}_{\mathrm{av},m}$, from
Theorem \ref{thm:LowerBdd}. We computed the approximate $\lambda_{75}$- and $\lambda_{90}$- powers using expression
\ref{eqn:lpwrCLT} based upon the CLT for the TPF (theorem \ref{thm:CLT}) using the expression for the asymptotic
variance, $\sigma^2$ (expression \ref{eqn:s2} given in the proof). Also, at each combination of parameters, we conducted
\numprint{1000} simulation replicates. At each simulation replicate, we began by generating $m$ i.i.d. Bernoulli
$\{0,1\}$ variables, with success probability, $r$, to assign each of the $m$ test statistics to the null (0) or
non-null (1) populations, recording the number, $M_m$, of non-null distributed test statistics. This was followed next
by drawing $m$ test statistics from $F_{0,n}$ or $F_{\nu,n}$, being the central and non-central (respectively)
t-distribution of $2 n -2$ degrees of freedom, corresponding to the particular value of $\xi_i$. Next, the B-H FDR
procedure was applied and the number of positive calls, $J_m$, and number of true positives, $S_m$ were recorded. At the
conclusion of the \numprint{1000} simulation replicates, we recorded the simulated average power as the mean over
simulation replicate of the TPF, $S_m/M_m$. In addition, the simulated $\lambda_{75}$- and $\lambda_{90}$- powers were
derived as the fraction of simulation replicates of the TPF that exceeded $0.75$ and $0.90$, respectively. Finally we
computed the sample size required for $\lambda_{90}$ power.

In another simulation study, focused on the distribution of the FDF for increasing $m$. At each combination of the 
parameters considered, we computed the reduced FDR, $f\tck$, required to bound the FDF with probability $(1-r)f\tck$
as the unique numerical solution to expression \ref{eqn:fprime}. We also computed the sample sizes $n_{_{0,0}}$ and
$n_{_{0,1}}$ required for specified average power under $\mathrm{BHFDR}(f)$ and under $\mathrm{BHFDR}(f\tck)$, 
respectively. Sample sizes $n_{_{1,0}}$ and $n_{_{1,1}}$ at specified $\lambda$-power under the corresponding procedure 
were also derived. A simulation, conducted in a fashion identical to that described above, under $\mathrm{BHFDR}(f\tck)$,
was done at each combination of parameters, this time including the additional two parameters $m$ and specified
power. From simulation replicates of the FDF, $T_m/J_m$, we computed the probability in excess of $f_0$.

All calculations were done in R, version 3.5.0 (\cite{CiteR:2016}) using a R package, \textbf{pwrFDR}, 
written by the author \cite{IzmirlianG:2018}, available for download on \textbf{cran}. Simulation was conducted on 
the NIH Biowulf cluster (\cite{Biowulf:2017}), using the swarm facility,  whereby each of 50 nodes was tasked with 
carrying out 20 simulation replicates resulting in \numprint{1000} simulation replicates for each configuration of 
parameters.

\subsection{Biomarker Studies}
For the first simulation study we considered experiments typical of biomarker studies with $m=\numprint{200}$
simultaneous tests. We attempted to cover a broad spectrum of parameters spanning the domain of typical biomarker study
designs. The false discovery rate, $f$, was ranged over the values $1\%$, and from
5\% to 30\% in increments of 5\%. The expected number of tests with non-zero means, $\mathbb{E}[M_m] =
m r$, was varied over the values 5, 10 and 20, and from 10 to 100 in increments of 10,
representing values of $r$ ranging from 0.025 to 0.5. The effect size,
$\theta$, was allowed to vary from 0.6 to 1.5 in increments of 0.1. At each configuration, a range of sample sizes were
chosen to result in powers between 50\% and 98\% as mentioned above. This resulted in
2,648 configurations of the parameters, $f, \mathbb{E}[M_m],\theta$, and $n$
(full set of parameter combinations). The job took roughly 7
minutes on the NIH Biowulf cluster.

Table \ref{tbl:avgpwr_tbl_Bmkr} tabulates the IST average power, the oracle power and the simulated mean of the TPF at
28 different parameter settings excerpted from the full set of
2,648 parameter combinations. Over the full set of parameter settings, the both
the IST $\pi_{_\mathrm{pi}}$, and oracle $\pi_o$ powers are very close to the simulated average power. The
difference between the IST power, $\pi_{_\mathrm{pi}}$, and the simulated power was less than
0.15, 0.95 and 2.00 at 50\%, 90\% and 99\% of the parameter settings, respectively. As remarked
earlier, the oracle power is actually the average power at the oracle threshold. Since it borders on feasible, we
allowed $r$ to take values as large as 0.5 for which the oracle
threshold has a substantial gain in power.  The oracle power differed from the IST power by
1.8\%, 8.5\% and 16\% at 50\%, 90\% and 99\% of the parameter settings, respectively, suggesting
that the oracle threshold is worth considering.  Recall that our IST power, $\pi_{_\mathrm{pi}}$ can be set to the oracle
threshold by setting the FDR to $f/(1-r)$. However, careful consideration must be taken if using the oracle threshold to
design a study, since when its time to actually threshold the data one needs a plug-in estimate of $r$ and as discussed
extensively in the literature, this can be problematic. The lower bound comes within roughly 10\% of the simulated
power, with differences with the simulated power less than 36\%, 45\%, 50\%, 56\% and 71\% at 20\%, 40\%,
50\%, 60\% and 80\% of the parameter settings, respectively.

Table \ref{tbl:Lpwr_tbl_Bmkr} displays, at threshold 0.75 and at threshold 0.90, the $\lambda$ power as derived from
the CLT \ref{thm:CLT} and estimated from simulation replicates (hatted version), respectively, excerpted from the full
set of 2,648 parameter combinatons as before. In the last column is the ratio 
of the sample size required for $\lambda_{90}$-power to the original sample size. First, we note that when restricted to
powers strictly between 50\% and 100\%, occurring at 1,488
parameter combinations, the CLT approximate- and simulated- $\lambda_{75}$-power were within the following relative error
of one another (median over parameter conditions (lower quartile, upper quartile)): 
2\% (0.7\%, 4.5\%), with 23.1\% over 5\%.  
Corresponding results for the simulated and CLT approximate $\lambda_{90}$-power for powers strictly between 50\% and
100\% occurring at 1275 of the parameter values, were within the following relativer
error of one another 3.3\% (1.3\%, 9.2\%), with 
37.6\% over 5\%. The greater discrepancy between CLT approximate
$\lambda$-powers and simulated values is due to the lack of accuracy of the CLT asymptotic approximation at such
small sample sizes, $n$.  Note that for sample sizes in excess of $n=20$ the degree of accuracy starts to improved
dramatically, especially at higher powers. Also noteworthy is corroboration in ordering of the
average power and $\lambda_{k}$-power based upon the size of $k$ relative to $100 \lambda_{eq}$. All values of
$\lambda_{eq}$ are less than 90\%, but some are between 75\% and 90\%, and the ordering of average power and $\lambda$
powers is in accordance with expression \ref{eqn:LPwr_AvgPwr}. 

Furthermore the discrepancy between the average power and
the $\lambda$-power is reflective difference between $\lambda$ and $\lambda_{eq}$. This trend is echoed in the magnitude
of the sample size ratio, with magnitude increasing in the discrepancy between $\lambda_{eq}$ and $0.90$. Note that in this
case, as the number of simultaneous tests, $m$, is relatively ``small'', the distribution of the TPF, $S_m/M_m$, 
is more dispersed and therefore, growth in the $\lambda$-powers is more gradual with increasing sample size.

\subsection{The false discovery fraction, intermediate number of simultaneous tests}
The second simulation study was focused on the use of the CLT for the FDF to find a bound for the FDF with large
probability. We varied the number of simultaneous tests, $m$, over 1000, 2500, 5000, 7500, 10000 and 20000. The effect size, $\theta$,
was varied over 2/3, 5/6 and 1. The proportion of statistics drawn from the non-null distributed population, $r$, 
ranged over 0.025, 0.05 and 0.075 and the FDR, $f$, ranged over the values 0.1, 0.15 and 0.2. At each set of values of 
these parameters, we used expression \ref{eqn:fprime} based upon the CLT for the FDF to find a reduced FDR
at which the BH-FDR procedure would result in an FDF of no more than $f_0$ with large probability.
We calculated the sample sizes required for specified average power under the original and reduced FDR.
Sample sizes required for specified $\lambda_{90}$-power under the original and reduced FDR were also calculated.
Finally, the probability that the FDF exceeded $f_0$ under the reduced FDR was estimated from simulation replicates.

Table \ref{tbl:tbl_FDFincrN} tabulates the reduced FDR, $f\tck$, required to bound the FDF by $f_0$ with probability
$1-(1-r)f\tck$, and the sample sizes $n_{_{0,0}}$ and $n_{_{0,1}}$, required for specified average power under 
$\mathrm{BHFDR}(f)$ and under $\mathrm{BHFDR}(f\tck)$, respectively. Also shown are the sample sizes, $n_{_{1,0}}$ and 
$n_{_{1,1}}$, required for specified $\lambda_{90}$-power under $\mathrm{BHFDR}(f)$ and under $\mathrm{BHFDR}(f\tck)$, 
respectively, as well as the simulated tail probability, $\hat\pi_{_{T/J}}(f_0)$, in excess of $f_0$.
The general trend for increasing $m$ as the distribution of $T_m/J_m$ collapses to a point mass at $f_0$ are
a value of $f\tck$ closer to $f$, and sample sizes under $\mathrm{BHFDR}(f\tck)$ that are less inflated relative to
corresponding sample sizes under $\mathrm{BHFDR}(f)$. The simulated right tail probability, $\hat\pi_{_{T/J}}$ 
should in theory have only simulation error about its theoretical value, $(1-r)f\tck$. However, as $f\tck$ is 
derived from an approximation based upon a CLT, we expect the accuracy of the approximation to improve with larger $m$.
Not surprisingly, the results are consistent with these observations. Over the full set of 324 parameter
settings we obtained the following results (median, (lower quartile, upper quartile)). The ratio of the reduced FDR, 
$f\tck$, to the original FDR, $f$: 0.79 (0.69, 0.86) when $m\leq 10,000$, and 
0.91 (0.88, 0.93) when $m>10,000$, showing that the reduced FDR gets closer in value to the original 
FDR with increasing $m$. The ratio of sample size required for average power at $\mathrm{BHFDR}(f\tck)$ to that at 
$\mathrm{BHFDR}(f)$: 1.06 (1.04, 1.1) when $m\leq 10,000$ and 1.03 (1.02, 1.04) 
when $m>10,000$, showing that the inflation factor reduces with increasing $m$. This is also the case when the 
sample sizes are derived for given $\lambda_{90}$-power: 1.05 (1.03, 1.08) when $m\leq 10,000$ and 
1.02 (1.0025, 1.03) when $m>10,000$, respectively. Finally, the ratio of the simulated tail probability, 
$\hat\pi_{_{T/J}}(f_0)$ to the CLT approximated value: 1.02 (0.95, 1.11) when $m\leq 10,000$ and
0.995 (0.94, 1.0675) when $m>10,000$ respectively, highlighting that the CLT approximation
gets better with increasing $m$.

In order to judge the relative impact changes in the parameters had, especially those unique to this setting of
multiple testing, we computed numerical partial derivatives of the power function with respect to the proportion of
test statistics distributed as the non-null distribution $r$, the effect size $\theta$, and the FDR $f$.  The
partials were then scaled to the range of the relevant parameter (max minus min) so that unit changes were
comparable and corresponded to the ranges of the parameters considered. Numerical partial derivatives were computed
at all 10,020 configurations of the parameters. These results were
summarized separately for each of the three parameters by calculating quartiles of the respective numerical partial
derivative at each given level of the respective parameter, over all configurations of all other parameters
resulting in powers of 50\%, 60\%, 70\%, 80\% and 90\%. The results corresponding to median values at 70\% power are
displayed in Figure \ref{fig:MA-derivs}.

\section{Discussion}
We proved LLNs for the PCF, FDF and TPF as well as CLTs for $\sqrt{m}$ scaled versions of them. Our LLN result for the 
TPF allowed characterization of the large $m$ limit and this in turn allowed a proper interpretation of the power 
discussed in \cite{JungSH:2005} and in \cite{LiuP:2007}, being nearly identical to the average power. Our CLT result
for the TPF allowed us to introduce the $\lambda$-power, similar in nature to the $k$-power discussed by previous 
authors. The $\lambda$-power allows tighter control over the TPF in the design of multiple testing experiments by
bounding the distribution of the TPF by an acceptable threshold, rather than just its mean, as is the case with the
average power. Our CLT result for the FDF provides a technique whereby an investigator can determine a reduced FDR
at which the usual BH-FDR procedure will result in a FDF no greater than a stipulated value with arbitrary large
probability. This latter technique is useful both at the design phase as well as the analysis phase because the
asymptotic variance depends only upon the limiting PCF, $\gamma$, and proportion belonging to the null distributed
population, $1-r$ and one can use $J_m/m$ as an estimate of $\gamma$ and consider $1-r\approx 1$ when faced with
a data analysis. Key to the proofs of the LLN results was, first, the LLN for the PCF, $J_m/m$, which was proved directly
via a simple argument. Prior results by \cite{GenoveseC:2004} obtained convergence only in probability. Once
established we applied a result of \cite{TaylorRL:1985} for a.s. convergence of triangular arrays of 
finite exchangeable sequences The proofs of the CLT results was made possible by building on the work of 
\cite{GenoveseC:2004} which considered $p$-values thresholded at a deterministic $t$, treating them as
stochastic processes. We applied a result of \cite{SilvestrovD:2004} for weak convergence of stopped stochastic processes.

In a very large and thorough simulation study, we investigated three major domains of the space of operating
characteristics typically encountered in the design of multiple testing experiments: two larger $m$ domains,
$m=\numprint{54675}$ typical to human RNA expression micro-array studies, and $m=\numprint{1000000}$ typical to GWA
studies and a smaller $m$ domain, $m=200$ which is typical of biomarker studies. In each case, we compared the average
power derived from the TPF LLN limit to simulated values and observed at all ranges of sample sizes that the agreement
was quite good.  We also used the CLT result for the TPF to compute approximate $\lambda$ powers and compared these with
the simulation distribution. Agreement in this case was overall very good, but there was some breakdown in the level of
accuracy in the asymptotic approximation at simultaneous tests $m<100$. The last simulation study was focused upon the
procedure for bounding the FDF with large probability and its behavior as the number of simultaneous tests, $m$, grows
from hundreds to tens of thousands. We noted that overall, the method is feasible, even when the asymptotic
approximation begins to fail, as it always offers tighter control of the FDF than the BH-FDR procedure alone.

We investigated departures from the assumption of independent hypothesis tests by conducting a simulation in which
tests were correlated within blocks according to a compound symmetry structure under a multivariate normal, having
marginal variances equal to 1. For the purposes of this investigation, we fixed the block size to 100,
effect size 1.25, FDR at 15\%, proportion of non-null distributed tests, $r$, at
5\% for 2000 simultaneous tests. We varied sample sizes from 14 to 16 and block
correlation from 0 to 80\% in increments of 10\%. The average power, $\lambda_{75}$ power, empirical FDR
and probability that the FDF exceeds 18\% are tabulated in table \ref{tbl:tbl_corr_tests} over
the ranges of sample sizes and block correlations considered. As we can see, comparing the independent tests lines
for each of the two sample sizes, 14 and 16, with corresponding values for correlated test statistics, a very
important point can be made. From the standpoint of the mean, there is virtually no difference. This is to say that
the empirical FDR and average power are virtually unaffected when there are correlated blocks of tests. Notable
differences do occur in the distributions of the TPF and FDF as the $\lambda_{75}$-power for independent test
statistics is 30\% in a sample of 14, and 87\% in a sample of 16, respectively, while the values when there are
correlated blocks of tests are substantially greater for a sample of 14 and substantially less in a sample of 16,
respectively. Discrepancies between the independent tests versus correlated blocks of tests in the same direction
are also observed in the probability that the FDF exceeds 18\%. The reason for this is that correlated blocks of
test statistics result in a reduced effective number of tests. Apparently, the observed effective number of test
statistics is large enough that the empirical means are still very good estimates of their almost sure limiting
values, but not great enough for stability in the distribution of empirical means at the ranges of parameters under
consideration.  The conclusion to be drawn is not that the BH-FDR procedure is to be avoided because it is not
completely imune to departures from the independent test statistics assumption. By analogy, is any limit theorem
meaningless because it doesn't apply to a sample size of 3? Quite not. The first conclusion to be drawn is that the
empirical means appear to be unaffected in the ranges of parameters considered here. If one is truly comfortable
controlling the false discovery rate and powering studies using the average power, then one can ignore the
appropriateness of the independence assumption. Problems start to occur when one uses the tails of the distribution
of the FDF and TPF, as we are making the case for use here. However, rather then give up on the use of the BH-FDR
procedure altogether, the phenomenon should be viewed from the lens of the effective number of simultaneous
tests. So, ironically, the problem is solved if one can simply increase the number of simultaneous tests.

Drawing away from the specific discussion of correlated tests and widening the focus to the conclusions to be drawn
from the paper as a whole, the point to be made is that the quantities arising in the BH-FDR procedure, the expected
FDF which is controlled, and the expected TPF which forms the basis of a power calculation, should be seen for what
they are, location parameters. Because first and second order asymptotics in the FDF and TPF occur as the number of
simultaneous tests tends to infinity, then within the scope of reasonable ranges of parameters, e.g. effect size no
more than 1 or so, and sample sizes within the ranges seen for equipment that is either very expensive per replicate
or just starting to get a bit cheaper, say a few tens of replicates, then the following generalizations can be
made. For more than 20,000 simultaneous tests, the means and the distributions effectively coincide so that
controlling the FDR and using the average power to derive sample sizes is well supported. This is great news 
for GWAS and RNA-seq studies for example. However, for less than one or two thousand simultaneous tests, one must 
use second order asymptotics to control the type I-like error and calculate sample sizes using the CLT's for the 
FDF and TPF in the manner outlined here. For on the order of a hundred or so simultaneous tests, asymptotic approximation 
using the CLT's may not be appropriate. In this case, simulation is advised. This cautionary note is of particular
importance in many biomarker studies. 
\fi 

\ifsuppl 
\section{Appendix: Further simulation studies}
\subsection{RNA Expression Micro-array Studies}
The third simulation study we considered experiments typical of human RNA expression micro-array studies using the
Affymetrix Hgu133plus2 oligonucleotide mRNA gene chip. In this case, there are $m=\numprint{54675}$ simultaneous
tests. We attempted to cover a broad spectrum of parameters spanning the domain typical of micro-array study
designs. The false discovery rate, $f$, was ranged over the values 1\%, and from
5\% to 30\% in increments of 5\%.  The expected number of tests with non-zero means, $\mathbb{E}[M_m]
= m r$, was varied from 100 to 2500 in increments of 100 representing values of $r$ ranging from
0.0018 to 0.046. The effect size, $\theta$, was allowed to vary from
0.6 to 1.5 in increments of 0.1. At each configuration, a range of sample sizes were chosen to result in powers between
60\% and 95\% as mentioned above. This resulted in 10,020 configurations of the
parameters, $f, \mathbb{E}[M_m],\theta$, and $n$ (full set of parameter combinations).  The job took roughly
12 hours on the NIH Biowulf cluster.

Table \ref{tbl:avgpwr_tbl_Array} tabulates the IST average power, the oracle power and the simulated mean of the TPF
at 28 different parameter settings excerpted from the full set of
10,020 parameter combinations. Over the full set of parameter settings, the
both the IST, $\pi_{_\mathrm{pi}}$, and oracle, $\pi_o$, powers are very close to the simulated average power. The
difference between the IST power, $\pi_{_\mathrm{pi}}$, and the simulated power was less than
0.022\%, 0.077\% and 0.19\% at 50\%, 90\% and 99\% of the parameter settings, respectively. The oracle
power differed from the IST power by less than 0.15\%, 0.5\% and 0.92\% at 50\%, 90\% and 99\% of
the parameter settings, respectively. As remarked earlier, the oracle power is actually the average power at the
oracle threshold, but for such small values of $r \leq 0.046$ there
is not much gain in power to be had. The lower bound comes within roughly 10\% of the simulated power, with differences
with the simulated power less than 2.1\%, 4.5\%, 6.2\%, 8.5\% and 15\% at 20\%, 40\%, 50\%, 60\% and 80\% of the
parameter settings, respectively.

Table \ref{tbl:Lpwr_tbl_Array} displays, at threshold 0.75 and at threshold 0.90, the $\lambda$ power as derived from
the CLT \ref{thm:CLT} and estimated from simulation replicates (hatted version), respectively, excerpted from the full
set of 10,020 parameter combinatons as before. In the last column is the ratio 
of the sample size required for $\lambda_{90}$-power to the original sample size. First, we note that when restricted to
powers strictly between 50\% and 100\%, occurring at 1,066
parameter combinations, the CLT approximate- and simulated- $\lambda_{75}$-power were within the following relative error
of one another (median over parameter conditions (lower quartile, upper quartile)): 
0.4\% (0.2\%, 1.2\%), with 2.4\% over 5\%.  
Corresponding results for the simulated and CLT approximate $\lambda_{90}$-power for powers strictly between 50\% and
100\% occurring at 1476 of the parameter values, were within the following relativer
error of one another 0.5\% (0.2\%, 1.3\%), with 
1.8\% over 5\%. Also noteworthy is corroboration in ordering of the
average power and $\lambda_{k}$-power based upon the size of $k$ relative to $100 \lambda_{eq}$. All values of
$\lambda_{eq}$ are less than 90\%, but some are between 75\% and 90\%, and the ordering of average power and $\lambda$
powers is in accordance with expression \ref{eqn:LPwr_AvgPwr}. Furthermore the discrepancy between the average power and
the $\lambda$-power is reflective difference between $\lambda$ and $\lambda_{eq}$. This trend is echoed in the magnitude
of the sample size ratio, with magnitude increasing in the discrepancy between $\lambda_{eq}$ and $0.90$. The relatively
rapid rise in sample size, $n$, of all $\lambda$-powers is an indication of the degree to which the distribution of 
the TPF, $S_m/M_m$, is spiked. 

\subsection{GWA Studies}
The last simulation study we considered experiments typical of GWA studies with $m=\numprint{1000000}$
simultaneous tests. We attempted to cover a broad spectrum of parameters spanning the domain typical of GWA study
designs. The false discovery rate, $f$, was ranged over the values 0.5\%, 1\%, 5\% and 10\%. The
expected number of tests with non-zero means, $\mathbb{E}[M_m] = m r$, was varied from 400 to 1000 in increments of 200 
representing values of $r$ ranging from 4e-04 to 0.001. The effect
size, $\theta$, was allowed to vary from 0.08 to 0.68 in increments of 0.2. At each configuration, a range of sample sizes
were chosen to result in powers between 50\% and 98\% as mentioned above. This resulted in
512 configurations of the parameters, $f, \mathbb{E}[M_m],\theta$, and $n$
(full set of parameter combinations). The job took roughly 5 hours on
the NIH Biowulf cluster.

Table \ref{tbl:avgpwr_tbl_GWAS} tabulates the IST average power, the oracle power and the simulated mean of the TPF at
32 different parameter settings excerpted from the full set of
512 parameter combinations. Over the full set of parameter settings, the both
the IST, $\pi_{_\mathrm{pi}}$, and oracle, $\pi_o$, powers are very close to the simulated average power. The
difference between the IST power, $\pi_{_\mathrm{pi}}$, and the simulated power was less than
0.031, 0.087 and 0.147 at 50\%, 90\% and 99\% of the parameter settings, respectively. The oracle power
differed from the IST power by 0.0038\%, 0.0081\% and 0.011\% at 50\%, 90\% and 99\% of the parameter
settings, respectively. As remarked earlier, the oracle power is actually the average power at the oracle threshold,
but for such small values of $r \leq 0.001$ the gain in power is now
less than 1\%. The lower bound comes within roughly 10\% of the simulated power, with differences with the simulated
power less than 0.83\%, 3.8\%, 6.9\%, 9.6\% and 15\% at 20\%, 40\%, 50\%, 60\% and 80\% of the parameter settings,
respectively.

Table \ref{tbl:Lpwr_tbl_GWAS} displays, at threshold 0.75 and at threshold 0.90, the $\lambda$ power as derived from
the CLT \ref{thm:CLT} and estimated from simulation replicates (hatted version), respectively, excerpted from the full
set of 512 parameter combinatons as before. In the last column is the ratio 
of the sample size required for $\lambda_{90}$-power to the original sample size. First, we note that when restricted to
powers strictly between 50\% and 100\%, occurring at 68
parameter combinations, the CLT approximate- and simulated- $\lambda_{75}$-power were within the following relative error
of one another (median over parameter conditions (lower quartile, upper quartile)): 
0.5\% (0.2\%, 1.3\%), with 1.5\% over 5\%.  
Corresponding results for the simulated and CLT approximate $\lambda_{90}$-power for powers strictly between 50\% and
100\% occurring at 87 of the parameter values, were within the following relativer
error of one another 0.6\% (0.2\%, 1.2\%), with 
1.1\% over 5\%. Also noteworthy is corroboration in ordering of the
average power and $\lambda_{k}$-power based upon the size of $k$ relative to $100 \lambda_{eq}$. All values of
$\lambda_{eq}$ are less than 90\%, but some are between 75\% and 90\%, and the ordering of average power and $\lambda$
powers is in accordance with expression \ref{eqn:LPwr_AvgPwr}. Furthermore the discrepancy between the average power and
the $\lambda$-power is reflective difference between $\lambda$ and $\lambda_{eq}$. This trend is echoed in the magnitude
of the sample size ratio, with magnitude increasing in the discrepancy between $\lambda_{eq}$ and $0.90$. 
Notice that over values considered the ranges of $\sqrt{n}\theta$ are comparable among the the micro-array, GWAS and
biomarker simulation studies. Therefore, the ``all'' or ``nothing'' rapid rise in the $\lambda$-powers with increasing 
sample size here must be solely due to the distribution of the TPF, $S_m/M_m$, being even more dramatically spiked, since 
the number of simultaneous tests, $m$, is in this case, considerably larger.

\section{Appendix: Proofs}
\begin{proof}[Proof of Theorem \ref{thm:Jom_convas}]
\noindent The author wishes to thank Professor Thomas G. Kurtz \citep{KurtzT:2016} for assistance with this proof.
Recall the nominal p-values, $P_i = \bar F_{0,n}^{-1}(X_{i,n})$, their common CDF, $G$, listed in expression 
\ref{eqn:Gdefnd} in the text and their order statistics $P^m_{(i)}$. Let $G_m$ be the empirical C.D.F. of 
$\{P_1, P_2,\ldots,P_m\}$. 
\begin{equation}
  G_m(u) = m^{-1} \sum_{i=1}^m I(P_i \leq u)  \label{eqn:empcdfGm}
\end{equation}
By Kolmogorov's theorem, $G_m(u)\convas G(u)$ at all continuity points, $u$, of $G$. By assumption 
the family $\{F_{\nu,n}^{-1}:\nu \geq 0\}$ is absolutely continuous and has the monotone likelihood
ratio property. It follows that each of the ratios $f_{\nu_{\ell}, n}/f_{0,n}$ is monotone and hence
the mixture of likelihood ratios, $f_{A,n}/f_{0,n} = \sum_{\ell} s_{\ell} f_{\nu_{\ell},n}/f_{0,n}$ is monotone.
It follows that G is concave and therefore $G(uf) = u$ has one non-zero solution which we will call $\gamma$. 
Let $\mathcal{N}\subset \Omega$ be the set of measure zero such that $G_m(\gamma f) \rightarrow G(\gamma f)$
for all $\omega \in \Omega \setminus \mathcal{N}$, and consider $\omega$ fixed in this set of measure 1 
for the remainder of this proof. Substituting $m^{-1} J_m \,f$ for $u$ in expression \ref{eqn:empcdfGm} shows that 
\begin{equation}
  m^{-1} J_m = G_m(m^{-1} J_m f) \label{eqn:Jom_eq_GmofJomf}
\end{equation}
Let $H_m(u) = G_m(u f) - u$ and $H(u) = G(u f) - u$. While $H^{-1}(0) = \{0, \gamma \}$ contains only 0 and
a unique non-zero solution, $G_m$ is a step function and therefore, $H_m^{-1}(0) = \{0, u_1, u_2,\ldots,u_k\}$
can contain multiple non-zero solutions.  None-the-less, for each $m$, $H_m^{-1}(0)$ is a finite set. By definition,
$J_m/m$ is an element of the set $H_m^{-1}(0)$. It follows that 
\begin{eqnarray*}
  m^{-1} J_m  &\leq& \sup H_m^{-1}(0)\nonumber\\
             &=& \max H_m^{-1}(0).\nonumber  
\end{eqnarray*}
where the second line follows because the set is finite. Thus, taking limsup with respect to $m$ on both sides 
above gives:
\begin{eqnarray*}
\limsup_m m^{-1} J_m &\leq& \limsup_m \max H_m^{-1}(0) \\
                    &=& H^{-1}(0) = \gamma\,.
\end{eqnarray*}
where the last equality follows because we can interchange the order of the limsup and maximum and because the 
limit exists. In the other direction, next note that because $m^{-1} J_m$ is a solution to $u = G_m(u f)$, it 
also follows that 
$m^{-1} J_m \geq u$ for every $u$ such that $u < G_m(u f)$. Thus, 
\begin{equation}
m^{-1} J_m \geq \sup \{u : u < G_m(u f) \} = \sup H_m^{-1}((0,\infty))\,. \label{eqn:Jom_geq_Gm}
\end{equation}
Because of the convexity of the limiting function, $G$, it follows, for $m$ large enough, that 
$\sup H_m^{-1}((0,\infty)) = \max H_m^{-1}(0)$. Therefore, upon taking taking liminf with respect to $m$ on both 
sides we have:
\begin{eqnarray*}
\liminf_m m^{-1} J_m &\geq& \liminf_m \sup H_m^{-1}((0,\infty))\\
                    &=& \liminf_m \max H_m^{-1}(0)\\
                    &=& H^{-1}(0) = \gamma\\
\end{eqnarray*}
where the last equality follows because we can interchange the order of liminf and the maximum and because the limit exists.
This completes the proof.
\end{proof}

\begin{proof}[Proof of Theorem \ref{thm:SoM_convas}]
First, we note that
\begin{equation}
m^{-1} S_m = m^{-1} \sum_{i=1}^m \xi_i \, I\left( P_i \leq m^{-1} J_m f \right)\,,
\end{equation}
is the average of row $m$ in a triangular array of finite exchangeable sequences. We will apply Theorem 1 of 
\cite{TaylorRL:1985}. Let $W_{m,i}=\xi_i \,I\left(P_i\leq m^{-1} J_m f\right), \mu_m = \E[W_{m,1}]$ and 
$Z_{m,i} = W_{m,i} - \mu_m$. We must show that (i) the increments on the $m^{th}$ row, $W_{m,i}$, each converge almost 
surely to respective elements of a sequence $W_{\infty, i}$; (ii) the increments $W_{m,i}$ have variances tending to a 
limit and (iii) for each $m, i$ and $j$, the covariance of increments $W_{m,i}$ and $W_{m,j}$ tend to zero. 
\begin{remark}
Our condition (i), element-wise almost sure convergence, which on surface appears weaker than the corresponding first 
condition in the cited reference, almost sure monotone decreasing distances to the limit, is sufficient in the 
context of the other assumptions. See the remark following the proof of Theorem 3 in that reference.
\end{remark}

Verification of condition (i) is trivial, as it follows by Theorem \ref{thm:Jom_convas} that 
$W_{m,i}\rightarrow W_{\infty,i} = \xi_i\, I\left( P_i \leq \gamma f \right)$ 
almost surely as $m\rightarrow \infty $ for each $i$. Let $\mu = \E[W_{\infty,1}]$. Condition (ii) follows easily 
since $Z_{m,i}$ is bounded, so that by the LDCT, for each $i$, 
$\E\left[Z_{m,i}^2 \right] \rightarrow \E\left[Z_{\infty,i}^2 \right]$ as $m\rightarrow\infty$. Note that the same 
argument verifies that $\mu_m \rightarrow \mu$. Next, to verify that condition (iii) is satisfied, note first, for 
$1 \leq i_1 < i_2 \leq m$, that $(W_{m,i_1} - \mu_m)(W_{m,i_2} - \mu_m)$ is bounded above by $4$, and converges almost 
surely to $(W_{\infty,i_1} - \mu)(W_{\infty,i_2} - \mu)$, by Theorem \ref{thm:Jom_convas}. Thus, condition (iii) follows by 
the LDCT. We may now apply Theorem 1 of \cite{TaylorRL:1985} to conclude that $m^{-1} \sum_{i=1}^m Z_{m,i} \rightarrow 0$ 
almost surely as $m\rightarrow \infty$. Therefore,
\begin{eqnarray*}
  \lim_{m\rightarrow\infty} m^{-1} S_m &=&
  \lim_{m\rightarrow\infty} \mu_m + m^{-1} \sum_{i=1}^m Z_{m,i}\\
  \\
&=& \mu\,, \;\mathrm{with~probability~one,}
\end{eqnarray*}
and the last written expectation is equal to $r\,\P \left\{ P_i \leq \gamma f \mid \xi_i=1\right\}$. Because 
$m^{-1} M_m \rightarrow r$ almost surely as $m\rightarrow\infty$, it follows that 
$M_m^{-1}\,S_m\rightarrow\P\left\{ P_i \leq \gamma f \mid \xi_i=1\right\} = \pi_{_\mathrm{pi}}$ almost surely as 
$m\rightarrow\infty$. Because $m^{-1} S_m$ is bounded by 1, the average power, its expectation, also converges to 
$\pi_{_\mathrm{pi}}$ as $m\rightarrow\infty$ by the LDCT.
\end{proof}

\begin{proof}[Proof of Corollary \ref{cor:FDF_convas}]
The first and second statements follow immediately from Theorems \ref{thm:Jom_convas} and \ref{thm:SoM_convas}:
\begin{equation}
m^{-1} T_m = m^{-1} (J_m - S_m) \convas \gamma - r\,\pi_{_\mathrm{pi}} = (1-r)\,f\,\gamma\,,
\end{equation}
and
\begin{equation}
J_m^{-1} T_m = 1 - J_m^{-1}S_m) \convas 1  - r\pi_{_\mathrm{pi}}/\gamma = (1-r)\,f\,.
\end{equation}
where the last equality in each of the expressions above follow since $\gamma = G(\gamma \,f)$. The third statement 
follows by the LDCT.
\end{proof}

\begin{proof}[Proof of Corollary \ref{cor:pi1Alt}]
  In the definition of the IST power function, $\pi_{_\mathrm{pi}}$, appearing in the statement of Theorem 
  \ref{thm:SoM_convas}, $\pi_{_\mathrm{pi}} = \bar F_{\nu, n}(\bar F_{0,n}^{-1}(\gamma\,f))$ we substitute 
  $\gamma = (1-f_0)^{-1} \,r \pi_{_\mathrm{pi}}$ from expression \ref{eqn:gamma_expr} obtaining the result.
\end{proof}

\begin{proof}[Proof of Theorem \ref{thm:CLT}]
\noindent The proof of both statements is made possible by considering each as a stopped stochastic process. We 
first revisit the empirical CDF's defined in the proofs of Theorems \ref{thm:Jom_convas} and \ref{thm:SoM_convas}. 
In the case of $J_m/m$, we have, 
\begin{eqnarray}
  G_{_{m}}(t) &=& m^{-1}\sum_{i=1}^m I\left( P_i \leq t \right)\,,\label{eqn:Gm_t}\\ 
  &&\nonumber\\
  G(t) &=& (1- r) \, t\, + \,r \, \bar F_{\nu,n}(\bar F_{0,n}^{-1}(t)\,. \label{eqn:G_t}
\end{eqnarray}
\noindent First, by the standard theory of empirical distributions, see for example \cite{ShorackWellner:1986}
\begin{equation}
W_m(t) = \sqrt{m} \left(G_m(t) - G(t)\right) \convind W(t)\,,\label{eqn:Wm_t_convind}
\end{equation}
a Gaussian process with covariance function 
\begin{equation}
\rho(s,t) = G(s\wedge t) \, - \, G(s)\,G(t)\,. \label{eqn:covW}
\end{equation}
Having shown that the paths of the centered and scaled stochastic process $W_m$ converge in distribution, we can 
obtain the CLT for the centered and scaled version of the positive fraction, $J_m/m$, claimed in 
expression \ref{eqn:CLT_Jortm}, by appealing to a result concerning weak limits of stopped stochastic processes. 
Towards this end, define the family of filtrations, 
\begin{equation}
\mathcal{F}_t = \sigma\left(\{ \xi_i\,I(P_i \leq t),\,(1-\xi_i)\,I(P_i \leq t), i\geq 1\}\right)\,. \nonumber
\end{equation}
Note that $W$ and $W_m$ are adapted to $\mathcal{F}_t$ for all $m\geq 1$, and that $\tau_m = m^{-1} J_m f$ is a 
stopping-time with respect to this filtration since, clearly, $\{\tau_m \leq t\} \in \mathcal{F}_t$. We will apply 
Theorem 4.2.1 of \cite{SilvestrovD:2004} to conclude that $W_m(\tau_m)$ converges in distribution to $W(\gamma\,f)$.  
To do so, we must verify the following three conditions.
\begin{itemize}
\item[i.~] $(W_m, \tau_m) \convind (W, \gamma\,f)$
\item[ii.~] $\P\{ \lim_{t\rightarrow 0} W(\gamma\,f + t) = W(\gamma\,f)\} = 1$
\item[iii.~] For all $\delta >0, \lim_{c\rightarrow 0} \limsup_{m\rightarrow\infty} \P\{ \Delta(W_m, c, 1) > \delta \} = 0\,,$
\end{itemize}
where $\Delta(x, c, 1)$ is the Skorohod modulus of compactness, 
\begin{equation}
\Delta(x, c, 1) = \sup_{t, t', t'' \in [0,1]} \sup_{ t-c < t' < t < t'' < t+c} |x(t) - x(t')| + |x(t'') - x(t)|
\end{equation}
Having already established that the paths of $W_m$ converge in distribution to those of $W$ above \ref{eqn:Wm_t_convind}, 
as well as the almost sure convergence of $\tau_m = J_m f/m$ to the constant $\tau = \gamma f$ in theorem \ref{thm:Jom_convas},
then part (i) is satisfied by Slutsky's theorem. Item (ii) is true because the limiting process, $W$, a Gaussian process, 
is almost surely continuous at every $t \in [0,1]$.  The almost sure continuity of the limiting process, $W$, also 
guarantees that the third condition, (iii), holds as well. Thus  
\begin{eqnarray}
W_m(m^{-1} \,J_m\,f) &=& \sqrt{m} \left(G_m(m^{-1} \,J_m\,f) - G(m^{-1} \,J_m\,f)\right)\nonumber\\
                      &\convind& W(\gamma\,f)\,,\label{eqn:Wm_taum_convind}
\end{eqnarray}
where the limiting random variable is normally distributed, of mean zero, and variance 
\begin{equation}
\rho(\gamma\,f, \, \gamma\,f) = G(\gamma\,f) \, - \, G^2(\gamma\,f) = \gamma\,(1-\gamma)\label{eqn:varW}
\end{equation}
\noindent The statement \ref{eqn:Wm_taum_convind} is nearly statement \ref{eqn:CLT_Jortm}, except that in 
\ref{eqn:CLT_Jortm}, centering is with respect to the deterministic limit, $\gamma = G(\gamma\,f)$. Thus, starting 
with $\sqrt{m}(J_m/m - \gamma))$ we add and subtract, obtaining a ``delta-method'' term. We can now write
\begin{eqnarray}
X_m &\equiv& \sqrt{m} \left(m^{-1} J_m \, - \,\gamma \right) \nonumber\\
      &=& \sqrt{m} \left(G_m(m^{-1} \,J_m\,f) - G(\gamma \,f)\right) \nonumber\\
      &=& \sqrt{m} \left(G_m(m^{-1} \,J_m\,f) - G(m^{-1} \,J_m\,f)\right) \nonumber\\
      &&  \;\;+\,\sqrt{m} \left(G(m^{-1} \,J_m\,f) - G(\gamma\,f)\right)\nonumber\\
      &=& W_m(m^{-1} \,J_m\,f) + f\,\dot{G}(\gamma\,f)\, X_m + \epsilon_m\label{eqn:Xm}
\end{eqnarray}
where $\epsilon_m = o_p(1)$. The conclusion of this portion of the proof requires that $f\, \dot{G}(\gamma\,f) <1 $.
By the monotone likelihood ratio property, it follows that $G$ is concave as is the function $f\, G$.  Because 
$f\,G(u) = u$ when $u=0$ and when $u=\gamma\, f$, then there is exactly one $u_1 \in (0, \gamma\,f)$ for which
$f\,\dot{G}(u_1) = 1$. By the concavity of $f\, G$, $f\,\dot{G}(u) > 1$ for $0 < u < u_1$ and $f\,\dot{G}(u) < 1$
for $u_1 < u < \gamma\,f$. Thus $f\,\dot{G}(\gamma\, f) < 1$. With this bound in hand, the steps above leading to
\ref{eqn:Xm} can be iterated ad-infinitum, yielding
\begin{eqnarray*}
X_m &=& \left(W_m(m^{-1} \,J_m\,f) + \epsilon_m \right) \, \sum_{k=0}^{\infty} f^k\,\dot{G}(\gamma\,f)^k\\
      &=& \frac{W_m(m^{-1} \,J_m\,f) + o_p(1)}{1 - f\,\dot{G}(\gamma\,f)}\\
      &\convind& X \equiv \frac{W(\gamma\,f)}{1 - f\,\dot{G}(\gamma\,f)}\,,
\end{eqnarray*}
which establishes claim \ref{eqn:CLT_Jortm} above and identifies the form of the limiting mean zero normal random 
variable. Its variance, $\tau^2$, is given by 
\begin{equation}
\tau^2 = \frac{\gamma (1 - \gamma)}{\left(1-\dot{G}(\gamma\, f)\, f\right)^2}\,.\label{eqn:varX}
\end{equation}

\noindent Next, we turn our attention towards verification of the CLT for the centered and scaled TPF, $S_m/M_m$, 
which is claim \ref{eqn:CLT_rtmSoM} above. We first revisit the empirical 
sub CDF's corresponding to the joint outcome of the indicator $\xi_i$ and the indicator $I(P_i \leq t)$ and their 
almost sure deterministic limits.
\begin{eqnarray}
  G_{_{m,0}}(t) &=& m^{-1}\sum_{i=1}^m (1- \xi_i ) \, I\left( P_i \leq t \right)\,,\label{eqn:Gm0_t}\\
  &&\nonumber\\
  G_{_{m,1}}(t) &=& m^{-1}\sum_{i=1}^m \xi_i \, I\left( P_i \leq t \right)\,,\label{eqn:Gm1_t}\\
  &&\nonumber\\
  G_{_{0}}(t) &=& (1- r) \, t\,,\label{eqn:G0_t}\\
  &&\nonumber\\
  G_{_{1}}(t) &=& r \, \bar F_{\nu,n}(\bar F_{0,n}^{-1}(t)\,.\label{eqn:G1_t}
\end{eqnarray}
\noindent This time we look at the bivariate process with components scaled and centered versions of $G_{_{m,0}}(t)$ 
and $G_{_{m,1}}(t)$. Again, from the standard results concerning empirical CDF's, 
\citep{ShorackWellner:1986, GenoveseC:2004} the following bivariate process converges in distribution.
\begin{equation}
\left[\begin{array}{c}
W_{m,0}(t)\\
W_{m,1}(t)
\end{array}\right] =
\sqrt{m}\left[\begin{array}{c}
G_{m,0}(t) - G_0(t)\\
G_{m,1}(t) - G_1(t)
\end{array}\right] 
\convind \left[\begin{array}{c} 
W_0(t)\\
W_1(t)
\end{array}\right]\,,\nonumber
\end{equation}
\noindent where the limit is a bivariate Gaussian process with covariance kernel 
\begin{equation}
R(s,t) = \left[\begin{array}{cc}
         G_0(s\wedge t) - G_0(s) \,G_0(t)  & -G_0(s)\,G_1(t)                  \\
        -G_0(t)\,G_1(s)                     & G_1(s \wedge t) - G_1(s) G_1(t)
         \end{array}\right]\,.\nonumber
\end{equation}
We remark in passing, something which should be already clear, that $W_m(t) = W_{m,0}(t) + W_{m,1}(t)$ and 
$W(t) = W_0(t) + W_1(t)$. If follows from some algebra and the fact that $\gamma = G(\gamma\,f)$ that this 
``new characterization'' of $W$ is consistent with the characterization of the process given 
above. This allows us to compute covariances between $W(t)$ and either $W_0(t)$ or $W_1(t)$
according to the covariance kernel, $R(s,t)$, as needed below. Note that $W_{0}, W_{1}$ and for all $m$, 
$W_{m,0}, W_{m, 1}$ are all adapted to the filtration, $\mathcal{F}_t$ and that $\tau_m = m^{-1}\, J_m \,f$ is a
stopping time with respect to the it, so that once again we apply the result of \cite{SilvestrovD:2004} 
to obtain convergence of the stopped bivariate process. As remarked above, the conditions 
are satisfied since convergence is already established and the limit is almost surely continuous. 
\begin{equation}
\left[\begin{array}{c}
W_{m,0}(J_m\,f/m)\\
W_{m,1}(J_m\,f/m)
\end{array}\right] =
\sqrt{m}\left[\begin{array}{c}
G_{m,0}(J_m\,f/m) - G_0(J_m\,f/m)\\
G_{m,1}(J_m\,f/m) - G_1(J_m\,f/m)
\end{array}\right] 
\convind \left[\begin{array}{c} 
W_0(\gamma\,f)\\
W_1(\gamma\,f)
\end{array}\right]\,,\label{eqn:bvrt}
\end{equation}
Focusing for the moment on the second component above in \ref{eqn:bvrt} and adding and subtracting as before, 
\begin{eqnarray*}
X_{m,1} &\equiv& \sqrt{m}\left(m^{-1} S_m - r \,\pi_{_\mathrm{pi}}\right)\\
        &=& \sqrt{m}\left(G_{m,1}(J_m\,f/m) - G_1(\gamma\,f)\right)\\
        &=& \sqrt{m}\left(G_{m,1}(J_m\,f/m) - G_1(J_m\,f/m)\right)\\
        &&\;\;+\,\sqrt{m}\left(G_1(J_m\,f/m) - G_1(\gamma\,f)\right)
\end{eqnarray*}
Thus, 
\begin{eqnarray*}
X_{m,1} &\convind& X_1 =W_1(\gamma\,f) +f\,\dot{G}_1(\gamma\, f)\, X\\
  &=& W_1(\gamma\,f)+\frac{\dot{G}_1(\gamma\,f)\,f}{1-f\,\dot{G}(\gamma\,f)}\,\left(W_0(\gamma\,f)+W_1(\gamma\,f)\right)
\end{eqnarray*}
which is a mean zero normal random variable having variance equal to 
\begin{equation}
\var[X_1] = 
   v_1+\dot{G}_1^2(\gamma\,f)\,f^2\tau^2+2 f\,\frac{\dot{G}_1(\gamma\,f)\,(v_1+c_{0,1})}{1-f\,\dot{G}(\gamma\,f)}
\end{equation}
where $v_1 = r \,\pi_{_\mathrm{pi}} - r^2 \, \pi_{_\mathrm{pi}}^2$ and $c_{_{0,1}} = -r \,(1-r)\,\gamma \,f \,\pi_{_\mathrm{pi}}$. To 
complete the proof of statement \ref{eqn:CLT_rtmSoM} we need only apply the delta method once more, for a ratio 
estimate. Before proceeding, we note that 
\begin{eqnarray*}
\sqrt{m}(M_m/m -r) &=& \sqrt{m}\left(G_{m,1}(1) - G_1(1)\right)\\
                   &\convind& W_1(1) 
\end{eqnarray*}
Thus, 
\begin{eqnarray*}
Z_{m,1} &\equiv& \sqrt{m}\left(\frac{S_m}{M_m} - \pi_{_\mathrm{pi}}\right)\\
         &=& \sqrt{m}\left(\frac{S_m/m}{M_m/m} - \pi_{_\mathrm{pi}}\right)\\
  &=&\frac{1}{r}\,\sqrt{m}\left(\frac{S_m}{m}-r\pi_{_\mathrm{pi}}\right)-\frac{r\pi_{_\mathrm{pi}}}{r^2}\,
                                                         \sqrt{m}\left(\frac{M_m}{m}-r\right) + \epsilon_m\\
  &\convind& \frac{1}{r}\,\left\{W_1(\gamma\,f) + \frac{f\,\dot{G}_1(\gamma\, f)}{1- f\,\dot{G}(\gamma\, f)} \, 
                 \left(W_0(\gamma\,f) + W_1(\gamma\,f)\right)\right\} - \frac{r\pi_{_\mathrm{pi}}}{r^2}\,W_1(1) \\
         &\equiv& Z_1 = N(0, \sigma^2)\,,
\end{eqnarray*}
where $\epsilon_m$ above is a new term that is $o_p(1)$ and and this completes the proof of statement \ref{eqn:CLT_rtmSoM} 
above and identifies the form of the limiting mean zero normal random variable, $Z_1$. Its variance is given by 
\begin{equation}
\sigma^2 = r^{-2} \, \left(\var[X_1] -2\, \pi_{_\mathrm{pi}} \,\cov[X_1, W_1(1)] +\pi_{_\mathrm{pi}}^2 \,\var[W_1(1)] \right)\,,\label{eqn:s2}
\end{equation}
where $\var[X_1]$ was given above, $\var[W_1(1)] = r\,(1-r)$, and 
\begin{equation}
\cov[X_1, W_1(1)]=r\,(1-r)\,\left\{\pi_{_\mathrm{pi}}+f\,\dot{G}_1(\gamma\,f)
                                  \frac{\gamma\,f +\pi_{_\mathrm{pi}}}{1-f\,\dot{G}(\gamma\,f)}\right\}
\end{equation}

The proof of the CLT for the centered and scaled version of the false discovery fraction follows fairly easily from the parts proved above. 
First, re-writing the centered and scaled difference in terms of the TPF, gives the first line, for which we again invoke 
the delta method, which yields the second line.
\begin{eqnarray*}
X_{m,0} &\equiv& \sqrt{m}\left(\frac{T_m}{J_m} - (1-r) \, f \right) = -\sqrt{m}\left(\frac{S_m}{J_m} - \frac{r \pi_{_\mathrm{pi}}}{\gamma} \right)\\
       &=& -\gamma^{-1} \left\{\sqrt{m}\left(m^{-1}S_m - r\pi_{_\mathrm{pi}}\right) - \gamma^{-1} r \pi_{_\mathrm{pi}}\left(m^{-1}J_m - \gamma\right) + \epsilon_m \right\}\\
       &\convind& -\gamma^{-1} \left\{ X_1 + \gamma^{-1} r \pi_{_\mathrm{pi}} X \right\}\\
       &=& -\gamma^{-1} \left\{\frac{f \dot{G}_1(\gamma f) - \gamma^{-1} r \pi_{_\mathrm{pi}}}{1 - f \dot{G}(\gamma f)} W_0 +
           \left(1 + \frac{f \dot{G}_1(\gamma f) - \gamma^{-1} r \pi_{_\mathrm{pi}}}{1 - f \dot{G}(\gamma f)}\right) W_1 \right\}\\
       &=& -\gamma^{-1} \left\{\frac{f \dot{G}(\gamma f) - (1-r)\,f - \gamma^{-1} r \pi_{_\mathrm{pi}}}{1 - f \dot{G}(\gamma f)} W_0 + 
           \left(1 + \frac{f \dot{G}(\gamma f) - (1-r)\,f- \gamma^{-1} r \pi_{_\mathrm{pi}}}{1 - f \dot{G}(\gamma f)}\right) W_1 \right\}\\
       &=& -\gamma^{-1} \left\{\frac{f \dot{G}(\gamma f) - 1}{1 - f \dot{G}(\gamma f)} W_0 + 
           \left(1 + \frac{f \dot{G}(\gamma f) - 1}{1 - f \dot{G}(\gamma f)}\right) W_1 \right\}\\
       &=& \gamma^{-1} W_0
\end{eqnarray*}
Convergence of all quantities in the second line was established above. The remaining lines are algebraic, and make use of the fact that
$\dot{G}_1(t) = \dot{G}(t) - (1-r)$ (line 5) and $G(\gamma \,f) = \gamma$ (line 6). As before, $\epsilon_m$ is a new term that is $o_p(1)$. 
The limiting random variable is of mean zero and normally distributed, having variance equal to 
\begin{equation}
\alpha^2 = \frac{(1-r)\,f\,\left( 1 - (1-r)\,f\,\gamma\right)}{\gamma}\label{eqn:a2}
\end{equation}
\end{proof}


\begin{proof}[Proof of Theorem \ref{thm:LowerBdd}]

Before we begin, we present an alternate expression for the event that the number of true positives is $s$ or greater.
\begin{equation}
    \{S_m \geq s \} = \left\{ P_{1,(s)} \leq \frac{J_m f}{m}\right\}\nonumber
\end{equation}
This is clearly the case, since there can be $s$ or more true positives if and only if the $s^{th}$ order statistic
in the non-null distributed population is less than the threshold $J_m f/m$. Now, towards obtaining a lower bound, 
we begin with the fact that the expected value of discrete non-negative variable can be derived as the sum of its 
cCDF. This can be used to write an expression of the FST average power by first conditioning on $M_m$:
\begin{eqnarray} 
  \pi_{_{\mathrm{av},m}} &=& \E[ S_m / M_m ] \nonumber \\
  &=& \sum_{\ell=1}^{m} \ell^{-1} \,\E [ S_m \mid M_m={\ell}]\,\P\{ M_m=\ell\}\nonumber\\
  &=& \sum_{\ell=1}^{m} \ell^{-1} \,\sum_{s=1}^{m} \P\{ S_m \geq s \mid M_m=\ell \}\,\P\{ M_m=\ell\}\nonumber\\
  &=& \sum_{\ell=1}^{m} \ell^{-1} \,\sum_{s=1}^{m} \P\{P_{1,(s)} \leq f\,J_m/m\}\,\P\{ M_m=\ell\}
  \label{eqn:lwrbddeqn3}\\
  &\geq& \sum_{\ell=1}^{\infty} m^{-1} \,\sum_{s=1}^{m} \P\{P_{1,(s)} \leq f s/m\}\,\P\{ M_m=\ell\}\nonumber\,,
\end{eqnarray}
where the first equality is just the law of total probability, conditioning on values of $M_N$, the second equality is
just the fact that an expectation of a non-negative random variable is the sum over values of $s$ of its cCDF,
the third equality is an application of the alternate expression stated above, and the lower bound in the last line is 
deduced by observing that $P_{1,(s)} \leq f\,J_N/N$ if and only if $J_N \geq s$. The last written line is equal to 
$\pi^{L}_{\mathrm{av},m}$ as presented in Theorem \ref{thm:LowerBdd} because the CDF of the $s^{th}$ order statistic, $P_{1,(s)}$, 
takes the form shown involving the beta distribution.
\end{proof}
\fi 
\vfil\eject


\ifmnscpt
\begin{table}[b]
\centering
\begingroup\tiny
\begin{tabular}{lrrrrrrr}
  \toprule
 Eff Sz&$\E[M_m]$&n&FDR&$\pi^{L}_{\mathrm{av},m}$&$\pi_{_\mathrm{pi}}$&$\pi_o$&$\hat\pi_{_{\mathrm{av},m}}$\\
  \cmidrule(r){1-1}\cmidrule(lr){2-2}\cmidrule(lr){3-3}\cmidrule(lr){4-4}\cmidrule(lr){5-5}\cmidrule(lr){6-6}\cmidrule(lr){7-7}\cmidrule(l){8-8}
0.60 & 5 & 70 & 0.15 & 0.224 & 0.704 & 0.707 & 0.692 \\ 
  0.60 & 5 & 80 & 0.15 & 0.235 & 0.795 & 0.798 & 0.790 \\ 
  0.60 & 5 & 90 & 0.15 & 0.241 & 0.860 & 0.863 & 0.850 \\ 
  0.60 & 5 & 100 & 0.15 & 0.246 & 0.906 & 0.908 & 0.896 \\ 
  0.60 & 20 & 50 & 0.15 & 0.052 & 0.664 & 0.685 & 0.668 \\ 
  0.60 & 20 & 60 & 0.15 & 0.052 & 0.781 & 0.796 & 0.774 \\ 
  0.60 & 20 & 70 & 0.15 & 0.053 & 0.859 & 0.870 & 0.857 \\ 
  0.60 & 20 & 80 & 0.15 & 0.053 & 0.911 & 0.919 & 0.905 \\ 
  0.60 & 60 & 40 & 0.15 & 0.388 & 0.706 & 0.777 & 0.708 \\ 
  0.60 & 60 & 50 & 0.15 & 0.388 & 0.823 & 0.872 & 0.823 \\ 
  0.60 & 60 & 60 & 0.15 & 0.388 & 0.895 & 0.927 & 0.894 \\ 
  0.60 & 100 & 30 & 0.15 & 0.546 & 0.632 & 0.796 & 0.634 \\ 
  0.60 & 100 & 40 & 0.15 & 0.563 & 0.788 & 0.895 & 0.788 \\ 
  0.60 & 100 & 50 & 0.15 & 0.563 & 0.879 & 0.946 & 0.875 \\ 
  0.80 & 5 & 40 & 0.15 & 0.222 & 0.697 & 0.701 & 0.678 \\ 
  0.80 & 5 & 50 & 0.15 & 0.239 & 0.844 & 0.847 & 0.842 \\ 
  0.80 & 5 & 60 & 0.15 & 0.247 & 0.924 & 0.925 & 0.909 \\ 
  0.80 & 20 & 30 & 0.15 & 0.052 & 0.692 & 0.712 & 0.685 \\ 
  0.80 & 20 & 40 & 0.15 & 0.052 & 0.859 & 0.870 & 0.857 \\ 
  0.80 & 60 & 20 & 0.15 & 0.385 & 0.616 & 0.703 & 0.617 \\ 
  0.80 & 60 & 30 & 0.15 & 0.388 & 0.845 & 0.889 & 0.843 \\ 
  0.80 & 100 & 20 & 0.15 & 0.561 & 0.717 & 0.855 & 0.715 \\ 
  0.80 & 100 & 30 & 0.15 & 0.563 & 0.896 & 0.955 & 0.897 \\ 
  1.00 & 5 & 30 & 0.15 & 0.233 & 0.790 & 0.794 & 0.774 \\ 
  1.00 & 20 & 20 & 0.15 & 0.052 & 0.699 & 0.720 & 0.704 \\ 
  1.00 & 20 & 30 & 0.15 & 0.053 & 0.913 & 0.921 & 0.908 \\ 
  1.00 & 60 & 20 & 0.15 & 0.388 & 0.853 & 0.897 & 0.853 \\ 
  1.00 & 100 & 20 & 0.15 & 0.563 & 0.904 & 0.960 & 0.903 \\ 
   \bottomrule
\end{tabular}
\endgroup
\caption{Excerpted results from a simulation study modelling biomarker studies, $m=200$. IST average power, oracle power, lower 
bound and simulated average power for a selection of effect sizes, values of $\E[M_m]$, FDR, and $n$.} 
\label{tbl:avgpwr_tbl_Bmkr}
\end{table}
\begin{table}[b]
\centering
\begingroup\tiny
\begin{tabular}{lrrrrrrrrrr}
  \toprule
 Eff Sz&$\E[M_m]$&n&FDR&$\pi_{_\mathrm{pi}}$&$\lambda_{75}$-pwr&$\hat\lambda_{75}$-pwr&$\lambda_{90}$-pwr&$\hat\lambda_{90}$-pwr&$\lambda_{eq}$&SS Ratio\\
  \cmidrule(r){1-1}\cmidrule(lr){2-2}\cmidrule(lr){3-3}\cmidrule(lr){4-4}\cmidrule(lr){5-5}\cmidrule(lr){6-6}\cmidrule(lr){7-7}\cmidrule(lr){8-8}\cmidrule(lr){9-9}\cmidrule(lr){10-10}\cmidrule(l){11-11}
0.60 & 5 & 70 & 0.15 & 0.704 & 0.426 & 0.524 & 0.216 & 0.249 & 0.570 & 1.500 \\ 
  0.60 & 5 & 80 & 0.15 & 0.795 & 0.584 & 0.681 & 0.308 & 0.396 & 0.622 & 1.500 \\ 
  0.60 & 5 & 90 & 0.15 & 0.860 & 0.738 & 0.797 & 0.409 & 0.538 & 0.673 & 1.467 \\ 
  0.60 & 5 & 100 & 0.15 & 0.906 & 0.867 & 0.875 & 0.518 & 0.657 & 0.721 & 1.380 \\ 
  0.60 & 20 & 50 & 0.15 & 0.664 & 0.265 & 0.307 & 0.042 & 0.028 & 0.607 & 1.500 \\ 
  0.60 & 20 & 60 & 0.15 & 0.781 & 0.610 & 0.630 & 0.139 & 0.157 & 0.695 & 1.500 \\ 
  0.60 & 20 & 70 & 0.15 & 0.859 & 0.894 & 0.877 & 0.320 & 0.378 & 0.765 & 1.343 \\ 
  0.60 & 20 & 80 & 0.15 & 0.911 & 0.991 & 0.961 & 0.563 & 0.599 & 0.819 & 1.225 \\ 
  0.60 & 60 & 40 & 0.15 & 0.706 & 0.280 & 0.315 & 0.005 & 0.002 & 0.666 & 1.500 \\ 
  0.60 & 60 & 50 & 0.15 & 0.823 & 0.903 & 0.891 & 0.088 & 0.099 & 0.771 & 1.360 \\ 
  0.60 & 60 & 60 & 0.15 & 0.895 & 1.000 & 0.995 & 0.453 & 0.492 & 0.842 & 1.183 \\ 
  0.60 & 100 & 30 & 0.15 & 0.632 & 0.037 & 0.031 & 0.000 & 0.000 & 0.609 & 1.500 \\ 
  0.60 & 100 & 40 & 0.15 & 0.788 & 0.786 & 0.789 & 0.011 & 0.006 & 0.750 & 1.450 \\ 
  0.60 & 100 & 50 & 0.15 & 0.879 & 1.000 & 0.999 & 0.278 & 0.270 & 0.838 & 1.200 \\ 
  0.80 & 5 & 40 & 0.15 & 0.697 & 0.417 & 0.496 & 0.212 & 0.252 & 0.566 & 1.500 \\ 
  0.80 & 5 & 50 & 0.15 & 0.844 & 0.696 & 0.784 & 0.380 & 0.502 & 0.659 & 1.500 \\ 
  0.80 & 5 & 60 & 0.15 & 0.924 & 0.915 & 0.895 & 0.574 & 0.704 & 0.742 & 1.333 \\ 
  0.80 & 20 & 30 & 0.15 & 0.692 & 0.330 & 0.352 & 0.057 & 0.036 & 0.626 & 1.500 \\ 
  0.80 & 20 & 40 & 0.15 & 0.859 & 0.892 & 0.893 & 0.319 & 0.357 & 0.764 & 1.350 \\ 
  0.80 & 60 & 20 & 0.15 & 0.616 & 0.062 & 0.060 & 0.001 & 0.000 & 0.590 & 1.500 \\ 
  0.80 & 60 & 30 & 0.15 & 0.845 & 0.963 & 0.952 & 0.148 & 0.147 & 0.791 & 1.333 \\ 
  0.80 & 100 & 20 & 0.15 & 0.717 & 0.286 & 0.284 & 0.001 & 0.003 & 0.684 & 1.500 \\ 
  0.80 & 100 & 30 & 0.15 & 0.896 & 1.000 & 1.000 & 0.453 & 0.506 & 0.855 & 1.167 \\ 
  1.00 & 5 & 30 & 0.15 & 0.790 & 0.574 & 0.680 & 0.304 & 0.392 & 0.617 & 1.500 \\ 
  1.00 & 20 & 20 & 0.15 & 0.699 & 0.350 & 0.431 & 0.064 & 0.045 & 0.630 & 1.500 \\ 
  1.00 & 20 & 30 & 0.15 & 0.913 & 0.992 & 0.966 & 0.579 & 0.614 & 0.821 & 1.200 \\ 
  1.00 & 60 & 20 & 0.15 & 0.853 & 0.978 & 0.959 & 0.183 & 0.225 & 0.799 & 1.300 \\ 
  1.00 & 100 & 20 & 0.15 & 0.904 & 1.000 & 0.999 & 0.549 & 0.586 & 0.863 & 1.150 \\ 
   \bottomrule
\end{tabular}
\endgroup
\caption{Excerpted results from a simulation study modelling biomarker studies, $m=200$. Shown are the $\lambda$-power at$\lambda=75\%$ and at $90\%$ from CLT and from simulations, $\lambda_{_{S/M}}(\pi_{_\mathrm{pi}})$ and samplesize ratio. The IST average power is also shown for comparison.} 
\label{tbl:Lpwr_tbl_Bmkr}
\end{table}
\begin{table}[b]
\centering
\begingroup\tiny
\begin{tabular}{lrrrrrrrr}
  \toprule
 $m$&Eff Sz&Power&$f\tck$&$n_{_{0,0}}$&$n_{_{0,1}}$&$n_{_{1,0}}$&$n_{_{1,1}}$&$\hat\pi_{_{T/J}}(f_0)$\\
  \cmidrule(r){1-1}\cmidrule(lr){2-2}\cmidrule(lr){3-3}\cmidrule(lr){4-4}\cmidrule(lr){5-5}\cmidrule(lr){6-6}\cmidrule(lr){7-7}\cmidrule(lr){8-8}\cmidrule(l){9-9}
1000 & 0.6667 & 0.6 & 0.069 & 51 & 65 & 75 & 89 & 0.0760 \\ 
  1000 & 0.6667 & 0.8 & 0.071 & 66 & 78 & 91 & 102 & 0.0900 \\ 
  1000 & 0.8333 & 0.6 & 0.069 & 33 & 43 & 50 & 59 & 0.0790 \\ 
  1000 & 0.8333 & 0.8 & 0.071 & 43 & 51 & 59 & 67 & 0.0880 \\ 
  1000 & 1.0000 & 0.6 & 0.069 & 24 & 31 & 36 & 42 & 0.0910 \\ 
  1000 & 1.0000 & 0.8 & 0.071 & 31 & 36 & 42 & 47 & 0.0770 \\ 
  2500 & 0.6667 & 0.6 & 0.097 & 51 & 57 & 75 & 83 & 0.1020 \\ 
  2500 & 0.6667 & 0.8 & 0.098 & 66 & 72 & 87 & 94 & 0.0980 \\ 
  2500 & 0.8333 & 0.6 & 0.097 & 33 & 38 & 50 & 54 & 0.0800 \\ 
  2500 & 0.8333 & 0.8 & 0.098 & 43 & 47 & 57 & 61 & 0.1100 \\ 
  2500 & 1.0000 & 0.6 & 0.097 & 24 & 27 & 36 & 39 & 0.1150 \\ 
  2500 & 1.0000 & 0.8 & 0.098 & 31 & 34 & 40 & 43 & 0.1040 \\ 
  5000 & 0.6667 & 0.6 & 0.112 & 51 & 55 & 75 & 80 & 0.1290 \\ 
  5000 & 0.6667 & 0.8 & 0.113 & 66 & 70 & 85 & 90 & 0.0970 \\ 
  5000 & 0.8333 & 0.6 & 0.113 & 33 & 36 & 50 & 53 & 0.1080 \\ 
  5000 & 0.8333 & 0.8 & 0.113 & 43 & 46 & 56 & 58 & 0.1260 \\ 
  5000 & 1.0000 & 0.6 & 0.113 & 24 & 26 & 36 & 38 & 0.1080 \\ 
  5000 & 1.0000 & 0.8 & 0.113 & 31 & 33 & 39 & 41 & 0.1050 \\ 
  7500 & 0.6667 & 0.6 & 0.120 & 51 & 54 & 75 & 80 & 0.1120 \\ 
  7500 & 0.6667 & 0.8 & 0.120 & 66 & 69 & 85 & 88 & 0.1100 \\ 
  7500 & 0.8333 & 0.6 & 0.120 & 33 & 36 & 50 & 53 & 0.1020 \\ 
  7500 & 0.8333 & 0.8 & 0.120 & 43 & 45 & 55 & 57 & 0.1210 \\ 
  7500 & 1.0000 & 0.6 & 0.120 & 24 & 26 & 36 & 38 & 0.1090 \\ 
  7500 & 1.0000 & 0.8 & 0.120 & 31 & 32 & 39 & 41 & 0.1160 \\ 
  10000 & 0.6667 & 0.6 & 0.124 & 51 & 53 & 75 & 78 & 0.1310 \\ 
  10000 & 0.6667 & 0.8 & 0.124 & 66 & 69 & 84 & 87 & 0.1060 \\ 
  10000 & 0.8333 & 0.6 & 0.124 & 33 & 35 & 50 & 51 & 0.1420 \\ 
  10000 & 0.8333 & 0.8 & 0.124 & 43 & 45 & 55 & 57 & 0.1150 \\ 
  10000 & 1.0000 & 0.6 & 0.124 & 24 & 25 & 36 & 38 & 0.1250 \\ 
  10000 & 1.0000 & 0.8 & 0.124 & 31 & 32 & 39 & 40 & 0.1040 \\ 
  20000 & 0.6667 & 0.6 & 0.131 & 51 & 52 & 75 & 78 & 0.1240 \\ 
  20000 & 0.6667 & 0.8 & 0.132 & 66 & 68 & 83 & 85 & 0.1300 \\ 
  20000 & 0.8333 & 0.6 & 0.131 & 33 & 35 & 50 & 51 & 0.1360 \\ 
  20000 & 0.8333 & 0.8 & 0.132 & 43 & 45 & 54 & 55 & 0.1160 \\ 
  20000 & 1.0000 & 0.6 & 0.131 & 24 & 25 & 36 & 36 & 0.1240 \\ 
  20000 & 1.0000 & 0.8 & 0.132 & 31 & 32 & 38 & 39 & 0.1190 \\ 
   \bottomrule
\end{tabular}
\endgroup
\caption{Excerpted results from the simulation study on the use of the CLT for the FDF to bound the FDF with large probability with    
FDR and $r$ fixed at 15\% and 2.5\%, respectively. Displayed are the parameter settigns, $m$, $\theta$, and power,       
followed by the value of the reduced FDR, $f\tck$ required to bound the FDF with probability $1-f_0$, the sample sizes       
$n_{_{0,0}}, n_{_{0,1}}, n_{_{1,0}},$ and $n_{_{1,1}}$, required for specified average power at $\mathrm{BHFDR}(f)$          
specified average power at $\mathrm{BHFDR}(f\tck)$, specified $\lambda_{90}$-power at $\mathrm{BHFDR}(f)$, and specified  
$\lambda_{90}$-power at $\mathrm{BHFDR}(f\tck)$. The last column is the simulated value, $\hat\pi_{_{T/J}}(f_0)$, of the 
tail probability of the FDF under $\mathrm{BHFDR}(f\tck)$.} 
\label{tbl:tbl_FDFincrN}
\end{table}
\begin{table}[b]
\centering
\begingroup\small
\begin{tabular}{lrrrrr}
  \toprule
 $n$&$\rho$&$\pi_{_\mathrm{pi}}$&$\lambda_{75}$-pwr&$\widehat{\mathrm{FDR}}$&$\pi_{_{T/J}}(0.18)$\\
  \cmidrule(r){1-1}\cmidrule(lr){2-2}\cmidrule(lr){3-3}\cmidrule(lr){4-4}\cmidrule(lr){5-5}\cmidrule(l){6-6}
14 & 0.00 & 0.7172 & 0.3000 & 0.1421 & 0.1910 \\ 
  14 & 0.10 & 0.7080 & 0.5020 & 0.1402 & 0.2080 \\ 
  14 & 0.20 & 0.7173 & 0.5260 & 0.1432 & 0.2320 \\ 
  14 & 0.30 & 0.6965 & 0.4780 & 0.1412 & 0.2310 \\ 
  14 & 0.40 & 0.7045 & 0.4930 & 0.1421 & 0.2180 \\ 
  14 & 0.50 & 0.6993 & 0.4860 & 0.1408 & 0.2180 \\ 
  14 & 0.60 & 0.7034 & 0.4830 & 0.1413 & 0.2280 \\ 
  14 & 0.70 & 0.6954 & 0.4910 & 0.1447 & 0.2500 \\ 
  14 & 0.80 & 0.7107 & 0.4900 & 0.1378 & 0.2220 \\ 
  16 & 0.00 & 0.7989 & 0.8680 & 0.1419 & 0.1670 \\ 
  16 & 0.10 & 0.7924 & 0.6950 & 0.1420 & 0.2370 \\ 
  16 & 0.20 & 0.7870 & 0.6880 & 0.1432 & 0.2270 \\ 
  16 & 0.30 & 0.7882 & 0.6940 & 0.1426 & 0.2350 \\ 
  16 & 0.40 & 0.7926 & 0.7020 & 0.1420 & 0.2270 \\ 
  16 & 0.50 & 0.7871 & 0.6940 & 0.1427 & 0.2420 \\ 
  16 & 0.60 & 0.7966 & 0.7050 & 0.1439 & 0.2350 \\ 
  16 & 0.70 & 0.7967 & 0.7060 & 0.1423 & 0.2250 \\ 
  16 & 0.80 & 0.7884 & 0.6780 & 0.1422 & 0.2150 \\ 
   \bottomrule
\end{tabular}
\endgroup
\caption{Average power, $\lambda_{75}$ power, empirical FDR, and $Pr( FDF > 0.18)$ at fixed effect size=1.25, FDR=15\%, r=5\% and 2000simultaneous tests, for sample sizes 14 and 16 and correlation varies over 0 to 80\% in increments of 10\%, with block size 100.} 
\label{tbl:tbl_corr_tests}
\end{table}\fi 

\ifsuppl 
\begin{table}[b]
\centering
\begingroup\tiny
\begin{tabular}{lrrrrrrr}
  \toprule
 Eff Sz&$\E[M_m]$&n&FDR&$\pi^{L}_{\mathrm{av},m}$&$\pi_{_\mathrm{pi}}$&$\pi_o$&$\hat\pi_{_{\mathrm{av},m}}$\\
  \cmidrule(r){1-1}\cmidrule(lr){2-2}\cmidrule(lr){3-3}\cmidrule(lr){4-4}\cmidrule(lr){5-5}\cmidrule(lr){6-6}\cmidrule(lr){7-7}\cmidrule(l){8-8}
0.60 & 100 & 100 & 0.15 & 0.444 & 0.683 & 0.683 & 0.681 \\ 
  0.60 & 100 & 110 & 0.15 & 0.444 & 0.763 & 0.763 & 0.762 \\ 
  0.60 & 100 & 120 & 0.15 & 0.444 & 0.826 & 0.826 & 0.825 \\ 
  0.60 & 100 & 130 & 0.15 & 0.444 & 0.874 & 0.874 & 0.874 \\ 
  0.60 & 100 & 140 & 0.15 & 0.444 & 0.910 & 0.910 & 0.910 \\ 
  0.60 & 1000 & 70 & 0.15 & 0.640 & 0.662 & 0.665 & 0.662 \\ 
  0.60 & 1000 & 80 & 0.15 & 0.743 & 0.761 & 0.764 & 0.761 \\ 
  0.60 & 1000 & 90 & 0.15 & 0.805 & 0.835 & 0.836 & 0.834 \\ 
  0.60 & 1000 & 100 & 0.15 & 0.815 & 0.887 & 0.889 & 0.887 \\ 
  0.60 & 1000 & 110 & 0.15 & 0.816 & 0.924 & 0.925 & 0.924 \\ 
  0.60 & 2000 & 60 & 0.15 & 0.616 & 0.640 & 0.647 & 0.640 \\ 
  0.60 & 2000 & 70 & 0.15 & 0.732 & 0.752 & 0.757 & 0.752 \\ 
  0.60 & 2000 & 80 & 0.15 & 0.817 & 0.832 & 0.836 & 0.832 \\ 
  0.60 & 2000 & 90 & 0.15 & 0.864 & 0.888 & 0.891 & 0.888 \\ 
  0.60 & 2000 & 100 & 0.15 & 0.870 & 0.927 & 0.929 & 0.927 \\ 
  0.80 & 100 & 60 & 0.15 & 0.444 & 0.717 & 0.717 & 0.718 \\ 
  0.80 & 100 & 70 & 0.15 & 0.444 & 0.835 & 0.836 & 0.837 \\ 
  0.80 & 100 & 80 & 0.15 & 0.444 & 0.909 & 0.909 & 0.908 \\ 
  0.80 & 1000 & 40 & 0.15 & 0.630 & 0.653 & 0.656 & 0.654 \\ 
  0.80 & 1000 & 50 & 0.15 & 0.794 & 0.816 & 0.818 & 0.815 \\ 
  0.80 & 1000 & 60 & 0.15 & 0.816 & 0.907 & 0.908 & 0.907 \\ 
  0.80 & 2000 & 40 & 0.15 & 0.727 & 0.747 & 0.753 & 0.747 \\ 
  0.80 & 2000 & 50 & 0.15 & 0.857 & 0.875 & 0.879 & 0.876 \\ 
  1.00 & 100 & 40 & 0.15 & 0.444 & 0.725 & 0.726 & 0.726 \\ 
  1.00 & 100 & 50 & 0.15 & 0.444 & 0.886 & 0.886 & 0.885 \\ 
  1.00 & 1000 & 30 & 0.15 & 0.734 & 0.754 & 0.756 & 0.753 \\ 
  1.00 & 1000 & 40 & 0.15 & 0.816 & 0.915 & 0.916 & 0.914 \\ 
  1.00 & 2000 & 30 & 0.15 & 0.814 & 0.830 & 0.835 & 0.830 \\ 
   \bottomrule
\end{tabular}
\endgroup
\caption{Excerpted results from a simulation study modelling micro-array studies, $m=\numprint{54675}$. IST average power, oracle power, lower bound and simulated average power for a selection of effect sizes, values of $\E[M_m]$, FDR, and $n$.} 
\label{tbl:avgpwr_tbl_Array}
\end{table}
\begin{table}[b]
\centering
\begingroup\tiny
\begin{tabular}{lrrrrrrrrrr}
  \toprule
 Eff Sz&$\E[M_m]$&n&FDR&$\pi_{_\mathrm{pi}}$&$\lambda_{75}$-pwr&$\hat\lambda_{75}$-pwr&$\lambda_{90}$-pwr&$\hat\lambda_{90}$-pwr&$\lambda_{eq}$&SS Ratio\\
  \cmidrule(r){1-1}\cmidrule(lr){2-2}\cmidrule(lr){3-3}\cmidrule(lr){4-4}\cmidrule(lr){5-5}\cmidrule(lr){6-6}\cmidrule(lr){7-7}\cmidrule(lr){8-8}\cmidrule(lr){9-9}\cmidrule(lr){10-10}\cmidrule(l){11-11}
0.60 & 100 & 100 & 0.15 & 0.683 & 0.112 & 0.098 & 0.000 & 0.000 & 0.656 & 1.420 \\ 
  0.60 & 100 & 110 & 0.15 & 0.763 & 0.602 & 0.611 & 0.002 & 0.000 & 0.728 & 1.309 \\ 
  0.60 & 100 & 120 & 0.15 & 0.826 & 0.963 & 0.953 & 0.039 & 0.037 & 0.786 & 1.217 \\ 
  0.60 & 100 & 130 & 0.15 & 0.874 & 1.000 & 0.998 & 0.233 & 0.248 & 0.832 & 1.138 \\ 
  0.60 & 100 & 140 & 0.15 & 0.910 & 1.000 & 1.000 & 0.627 & 0.661 & 0.869 & 1.071 \\ 
  0.60 & 1000 & 70 & 0.15 & 0.662 & 0.000 & 0.000 & 0.000 & 0.000 & 0.654 & 1.500 \\ 
  0.60 & 1000 & 80 & 0.15 & 0.761 & 0.758 & 0.754 & 0.000 & 0.000 & 0.750 & 1.312 \\ 
  0.60 & 1000 & 90 & 0.15 & 0.835 & 1.000 & 1.000 & 0.000 & 0.000 & 0.822 & 1.178 \\ 
  0.60 & 1000 & 100 & 0.15 & 0.887 & 1.000 & 1.000 & 0.121 & 0.110 & 0.874 & 1.070 \\ 
  0.60 & 1000 & 110 & 0.15 & 0.924 & 1.000 & 1.000 & 0.997 & 0.993 & 0.911 & 0.973 \\ 
  0.60 & 2000 & 60 & 0.15 & 0.640 & 0.000 & 0.000 & 0.000 & 0.000 & 0.635 & 1.500 \\ 
  0.60 & 2000 & 70 & 0.15 & 0.752 & 0.557 & 0.549 & 0.000 & 0.000 & 0.744 & 1.343 \\ 
  0.60 & 2000 & 80 & 0.15 & 0.832 & 1.000 & 1.000 & 0.000 & 0.000 & 0.823 & 1.188 \\ 
  0.60 & 2000 & 90 & 0.15 & 0.888 & 1.000 & 1.000 & 0.063 & 0.054 & 0.879 & 1.056 \\ 
  0.60 & 2000 & 100 & 0.15 & 0.927 & 1.000 & 1.000 & 1.000 & 1.000 & 0.918 & 0.960 \\ 
  0.80 & 100 & 60 & 0.15 & 0.717 & 0.266 & 0.311 & 0.000 & 0.000 & 0.687 & 1.367 \\ 
  0.80 & 100 & 70 & 0.15 & 0.835 & 0.981 & 0.984 & 0.058 & 0.055 & 0.795 & 1.200 \\ 
  0.80 & 100 & 80 & 0.15 & 0.909 & 1.000 & 1.000 & 0.612 & 0.625 & 0.868 & 1.075 \\ 
  0.80 & 1000 & 40 & 0.15 & 0.653 & 0.000 & 0.000 & 0.000 & 0.000 & 0.646 & 1.500 \\ 
  0.80 & 1000 & 50 & 0.15 & 0.816 & 1.000 & 1.000 & 0.000 & 0.000 & 0.803 & 1.220 \\ 
  0.80 & 1000 & 60 & 0.15 & 0.907 & 1.000 & 1.000 & 0.762 & 0.780 & 0.894 & 1.017 \\ 
  0.80 & 2000 & 40 & 0.15 & 0.747 & 0.404 & 0.393 & 0.000 & 0.000 & 0.739 & 1.350 \\ 
  0.80 & 2000 & 50 & 0.15 & 0.875 & 1.000 & 1.000 & 0.001 & 0.001 & 0.866 & 1.100 \\ 
  1.00 & 100 & 40 & 0.15 & 0.725 & 0.320 & 0.346 & 0.000 & 0.000 & 0.694 & 1.350 \\ 
  1.00 & 100 & 50 & 0.15 & 0.886 & 1.000 & 0.999 & 0.338 & 0.352 & 0.844 & 1.120 \\ 
  1.00 & 1000 & 30 & 0.15 & 0.754 & 0.592 & 0.593 & 0.000 & 0.000 & 0.743 & 1.333 \\ 
  1.00 & 1000 & 40 & 0.15 & 0.915 & 1.000 & 1.000 & 0.940 & 0.942 & 0.902 & 1.175 \\ 
  1.00 & 2000 & 30 & 0.15 & 0.830 & 1.000 & 1.000 & 0.000 & 0.000 & 0.821 & 1.200 \\ 
   \bottomrule
\end{tabular}
\endgroup
\caption{Excerpted results from a simulation study modelling micro-array studies, $m=\numprint{54675}$, for various values of effect sizes, values of $\E[M_m]$, FDR, and $n$. Shown are the $\lambda$-power at $\lambda=75\%$ and at $90\%$ from CLT and from simulations, $\lambda_{_{S/M}}(\pi_{_\mathrm{pi}})$ and sample size ratio. The IST average power is also shown for comparison.} 
\label{tbl:Lpwr_tbl_Array}
\end{table}
\begin{table}[b]
\centering
\begingroup\tiny
\begin{tabular}{lrrrrrrr}
  \toprule
 Eff Sz&$\E[M_m]$&n&FDR&$\pi^{L}_{\mathrm{av},m}$&$\pi_{_\mathrm{pi}}$&$\pi_o$&$\hat\pi_{_{\mathrm{av},m}}$\\
  \cmidrule(r){1-1}\cmidrule(lr){2-2}\cmidrule(lr){3-3}\cmidrule(lr){4-4}\cmidrule(lr){5-5}\cmidrule(lr){6-6}\cmidrule(lr){7-7}\cmidrule(l){8-8}
0.08 & 400 & 7800 & 0.01 & 0.610 & 0.612 & 0.612 & 0.612 \\ 
  0.08 & 400 & 8400 & 0.01 & 0.679 & 0.690 & 0.691 & 0.691 \\ 
  0.08 & 400 & 9000 & 0.01 & 0.705 & 0.757 & 0.758 & 0.757 \\ 
  0.08 & 400 & 9600 & 0.01 & 0.709 & 0.813 & 0.813 & 0.813 \\ 
  0.08 & 400 & 10200 & 0.01 & 0.709 & 0.858 & 0.858 & 0.857 \\ 
  0.08 & 400 & 10800 & 0.01 & 0.709 & 0.893 & 0.893 & 0.893 \\ 
  0.08 & 400 & 11400 & 0.01 & 0.709 & 0.921 & 0.921 & 0.921 \\ 
  0.28 & 400 & 650 & 0.01 & 0.624 & 0.626 & 0.626 & 0.626 \\ 
  0.28 & 400 & 700 & 0.01 & 0.687 & 0.704 & 0.704 & 0.705 \\ 
  0.28 & 400 & 750 & 0.01 & 0.707 & 0.770 & 0.770 & 0.771 \\ 
  0.28 & 400 & 800 & 0.01 & 0.709 & 0.824 & 0.824 & 0.824 \\ 
  0.28 & 400 & 850 & 0.01 & 0.709 & 0.868 & 0.868 & 0.868 \\ 
  0.28 & 400 & 900 & 0.01 & 0.709 & 0.902 & 0.902 & 0.902 \\ 
  0.28 & 400 & 950 & 0.01 & 0.709 & 0.928 & 0.928 & 0.929 \\ 
  0.68 & 400 & 120 & 0.01 & 0.668 & 0.676 & 0.676 & 0.677 \\ 
  0.68 & 400 & 160 & 0.01 & 0.709 & 0.912 & 0.912 & 0.911 \\ 
  0.08 & 1000 & 7500 & 0.01 & 0.651 & 0.652 & 0.652 & 0.651 \\ 
  0.08 & 1000 & 8000 & 0.01 & 0.715 & 0.716 & 0.716 & 0.716 \\ 
  0.08 & 1000 & 8500 & 0.01 & 0.769 & 0.771 & 0.771 & 0.772 \\ 
  0.08 & 1000 & 9000 & 0.01 & 0.803 & 0.818 & 0.818 & 0.818 \\ 
  0.08 & 1000 & 9500 & 0.01 & 0.813 & 0.856 & 0.856 & 0.856 \\ 
  0.08 & 1000 & 10000 & 0.01 & 0.814 & 0.888 & 0.888 & 0.888 \\ 
  0.08 & 1000 & 10500 & 0.01 & 0.814 & 0.913 & 0.913 & 0.913 \\ 
  0.28 & 1000 & 600 & 0.01 & 0.622 & 0.623 & 0.623 & 0.621 \\ 
  0.28 & 1000 & 640 & 0.01 & 0.689 & 0.689 & 0.690 & 0.689 \\ 
  0.28 & 1000 & 680 & 0.01 & 0.746 & 0.747 & 0.747 & 0.748 \\ 
  0.28 & 1000 & 720 & 0.01 & 0.791 & 0.797 & 0.797 & 0.797 \\ 
  0.28 & 1000 & 760 & 0.01 & 0.810 & 0.838 & 0.838 & 0.838 \\ 
  0.28 & 1000 & 800 & 0.01 & 0.813 & 0.872 & 0.873 & 0.873 \\ 
  0.28 & 1000 & 840 & 0.01 & 0.814 & 0.900 & 0.900 & 0.900 \\ 
  0.28 & 1000 & 880 & 0.01 & 0.814 & 0.923 & 0.923 & 0.923 \\ 
  0.68 & 1000 & 120 & 0.01 & 0.751 & 0.752 & 0.752 & 0.753 \\ 
   \bottomrule
\end{tabular}
\endgroup
\caption{Excerpted results from a simulation study modelling GWA studies, $m=\numprint{1000000}$. IST average power, oracle power, lower bound and simulated average power for a selection of effect sizes, values of $\E[M_m]$, FDR, and $n$.} 
\label{tbl:avgpwr_tbl_GWAS}
\end{table}
\begin{table}[ht]
\centering
\begingroup\tiny
\begin{tabular}{lrrrrrrrrrr}
  \toprule
 Eff Sz&$\E[M_m]$&n&FDR&$\pi_{_\mathrm{pi}}$&$\lambda_{75}$-pwr&$\hat\lambda_{75}$-pwr&$\lambda_{90}$-pwr&$\hat\lambda_{90}$-pwr&$\lambda_{eq}$&SS Ratio\\
  \cmidrule(r){1-1}\cmidrule(lr){2-2}\cmidrule(lr){3-3}\cmidrule(lr){4-4}\cmidrule(lr){5-5}\cmidrule(lr){6-6}\cmidrule(lr){7-7}\cmidrule(lr){8-8}\cmidrule(lr){9-9}\cmidrule(lr){10-10}\cmidrule(l){11-11}
0.08 & 400 & 7800 & 0.01 & 0.612 & 0.000 & 0.000 & 0.000 & 0.000 & 0.604 & 1.413 \\ 
  0.08 & 400 & 8400 & 0.01 & 0.690 & 0.011 & 0.012 & 0.000 & 0.000 & 0.677 & 1.320 \\ 
  0.08 & 400 & 9000 & 0.01 & 0.757 & 0.624 & 0.627 & 0.000 & 0.000 & 0.741 & 1.239 \\ 
  0.08 & 400 & 9600 & 0.01 & 0.813 & 0.999 & 0.999 & 0.000 & 0.000 & 0.794 & 1.168 \\ 
  0.08 & 400 & 10200 & 0.01 & 0.858 & 1.000 & 1.000 & 0.012 & 0.011 & 0.838 & 1.105 \\ 
  0.08 & 400 & 10800 & 0.01 & 0.893 & 1.000 & 1.000 & 0.341 & 0.347 & 0.873 & 1.048 \\ 
  0.08 & 400 & 11400 & 0.01 & 0.921 & 1.000 & 1.000 & 0.933 & 0.927 & 0.901 & 0.998 \\ 
  0.28 & 400 & 650 & 0.01 & 0.626 & 0.000 & 0.000 & 0.000 & 0.000 & 0.617 & 1.394 \\ 
  0.28 & 400 & 700 & 0.01 & 0.704 & 0.037 & 0.040 & 0.000 & 0.000 & 0.690 & 1.303 \\ 
  0.28 & 400 & 750 & 0.01 & 0.770 & 0.807 & 0.827 & 0.000 & 0.000 & 0.753 & 1.223 \\ 
  0.28 & 400 & 800 & 0.01 & 0.824 & 1.000 & 1.000 & 0.000 & 0.000 & 0.805 & 1.153 \\ 
  0.28 & 400 & 850 & 0.01 & 0.868 & 1.000 & 1.000 & 0.036 & 0.031 & 0.848 & 1.089 \\ 
  0.28 & 400 & 900 & 0.01 & 0.902 & 1.000 & 1.000 & 0.543 & 0.548 & 0.882 & 1.034 \\ 
  0.28 & 400 & 950 & 0.01 & 0.928 & 1.000 & 1.000 & 0.981 & 0.972 & 0.908 & 0.984 \\ 
  0.68 & 400 & 120 & 0.01 & 0.676 & 0.003 & 0.001 & 0.000 & 0.000 & 0.663 & 1.325 \\ 
  0.68 & 400 & 160 & 0.01 & 0.912 & 1.000 & 1.000 & 0.784 & 0.785 & 0.892 & 1.019 \\ 
  0.08 & 1000 & 7500 & 0.01 & 0.652 & 0.000 & 0.000 & 0.000 & 0.000 & 0.645 & 1.374 \\ 
  0.08 & 1000 & 8000 & 0.01 & 0.716 & 0.017 & 0.008 & 0.000 & 0.000 & 0.707 & 1.292 \\ 
  0.08 & 1000 & 8500 & 0.01 & 0.771 & 0.926 & 0.938 & 0.000 & 0.000 & 0.760 & 1.220 \\ 
  0.08 & 1000 & 9000 & 0.01 & 0.818 & 1.000 & 1.000 & 0.000 & 0.000 & 0.806 & 1.156 \\ 
  0.08 & 1000 & 9500 & 0.01 & 0.856 & 1.000 & 1.000 & 0.000 & 0.000 & 0.844 & 1.098 \\ 
  0.08 & 1000 & 10000 & 0.01 & 0.888 & 1.000 & 1.000 & 0.123 & 0.130 & 0.875 & 1.046 \\ 
  0.08 & 1000 & 10500 & 0.01 & 0.913 & 1.000 & 1.000 & 0.921 & 0.913 & 0.901 & 1.151 \\ 
  0.28 & 1000 & 600 & 0.01 & 0.623 & 0.000 & 0.000 & 0.000 & 0.000 & 0.617 & 1.408 \\ 
  0.28 & 1000 & 640 & 0.01 & 0.689 & 0.000 & 0.000 & 0.000 & 0.000 & 0.681 & 1.325 \\ 
  0.28 & 1000 & 680 & 0.01 & 0.747 & 0.432 & 0.474 & 0.000 & 0.000 & 0.737 & 1.250 \\ 
  0.28 & 1000 & 720 & 0.01 & 0.797 & 1.000 & 1.000 & 0.000 & 0.000 & 0.785 & 1.185 \\ 
  0.28 & 1000 & 760 & 0.01 & 0.838 & 1.000 & 1.000 & 0.000 & 0.000 & 0.826 & 1.125 \\ 
  0.28 & 1000 & 800 & 0.01 & 0.872 & 1.000 & 1.000 & 0.007 & 0.006 & 0.860 & 1.072 \\ 
  0.28 & 1000 & 840 & 0.01 & 0.900 & 1.000 & 1.000 & 0.516 & 0.506 & 0.888 & 1.151 \\ 
  0.28 & 1000 & 880 & 0.01 & 0.923 & 1.000 & 1.000 & 0.995 & 0.993 & 0.910 & 0.980 \\ 
  0.68 & 1000 & 120 & 0.01 & 0.752 & 0.546 & 0.586 & 0.000 & 0.000 & 0.741 & 1.242 \\ 
   \bottomrule
\end{tabular}
\endgroup
\caption{Excerpted results from a simulation study modelling GWA studies, $m=\numprint{1000000}$. Shown are the $\lambda$-powerat $\lambda=75\%$ and at $90\%$ from CLT and from simulations, $\lambda_{_{S/M}}(\pi_{_\mathrm{pi}})$ and samplesize ratio. The IST average power is also shown for comparison.} 
\label{tbl:Lpwr_tbl_GWAS}
\end{table}\fi 


\begin{thebibliography}{}

\bibitem[\protect\citeauthoryear{Alizadeh, Eisen, Davis, Ma, Lossos, Rosenwald,
  Boldrick, Sabet, Tran, Yu, et~al.}{Alizadeh et~al.}{2000}]{AlizadehA:2000}
Alizadeh, A.~A., M.~B. Eisen, R.~E. Davis, C.~Ma, I.~S. Lossos, A.~Rosenwald,
  J.~C. Boldrick, H.~Sabet, T.~Tran, X.~Yu, et~al. (2000).
\newblock Distinct types of diffuse large b-cell lymphoma identified by gene
  expression profiling.
\newblock {\em Nature\/}~{\em 403\/}(6769), 503--511.

\bibitem[\protect\citeauthoryear{Baggerly and Coombes}{Baggerly and
  Coombes}{2009}]{BaggerlyKA:2009}
Baggerly, K.~A. and K.~R. Coombes ({2009}, {Dec}).
\newblock {Deriving chemosensitivity from cell lines: forensic bioinformatics
  and reproducible research in high-throughput biology}.
\newblock {\em {Ann. of Appl. Statist.}\/}~{\em {3}\/}({4}), {1309--1334}.

\bibitem[\protect\citeauthoryear{{Baldi, Pierre and Long, A. D.}}{{Baldi,
  Pierre and Long, A. D.}}{2001}]{BaldiP:2001}
{Baldi, Pierre and Long, A. D.} ({2001}).
\newblock {A Bayesian Framework for the Analysis of Microarray Expression Data:
  Regularized t-Test and Inference of Gene Changes}.
\newblock {\em {Bioinformatics}\/}~{\em {17}\/}({6}), {509--519}.

\bibitem[\protect\citeauthoryear{Benjamini and Hochberg}{Benjamini and
  Hochberg}{1995}]{BenjaminiY:1995}
Benjamini, Y. and Y.~Hochberg ({1995}).
\newblock {Controlling the false discovery rate - a practical and powerful
  approach to multiple testing}.
\newblock {\em {J. R. Stat. Soc. Ser. B Stat. Methodol.}\/}~{\em {57}\/}({1}),
  {289--300}.

\bibitem[\protect\citeauthoryear{Genovese and Wasserman}{Genovese and
  Wasserman}{2002}]{GenoveseC:2002}
Genovese, C. and L.~Wasserman ({2002}).
\newblock {Operating characteristics and extensions of the false discovery rate
  procedure}.
\newblock {\em {J. R. Stat. Soc. Ser. B Stat. Methodol.}\/}~{\em {64}\/}({3}),
  {499--517}.

\bibitem[\protect\citeauthoryear{Genovese and Wasserman}{Genovese and
  Wasserman}{2004}]{GenoveseC:2004}
Genovese, C. and L.~Wasserman ({2004}, {JUN}).
\newblock {A stochastic process approach to false discovery control}.
\newblock {\em {Ann. Stat.}\/}~{\em {32}\/}({3}), {1035--1061}.

\bibitem[\protect\citeauthoryear{Glueck, Mandel, Karimpour-Fard, Hunter, and
  Muller}{Glueck et~al.}{2008}]{GlueckD:2008}
Glueck, D.~H., J.~Mandel, A.~Karimpour-Fard, L.~Hunter, and K.~E. Muller
  ({2008}).
\newblock {Exact Calculations of Average Power for the Benjamini-Hochberg
  Procedure}.
\newblock {\em {The International Journal of Biostatistics}\/}~{\em
  {4}\/}({1}), {11--28}.

\bibitem[\protect\citeauthoryear{Ibrahim, H., and Gray}{Ibrahim
  et~al.}{2002}]{IbrahimJG:2002}
Ibrahim, J.~G., C.~M. H., and R.~J. Gray ({2002}, {MAR}).
\newblock {Bayesian models for gene expression with DNA microarray data}.
\newblock {\em J. Amer. Statist. Assoc.\/}~{\em {97}\/}({457}), {88--99}.

\bibitem[\protect\citeauthoryear{Ioannidis}{Ioannidis}{2005}]{IonnidisJPA:2005}
Ioannidis, J. P.~A. (2005).
\newblock Why most published research findings are false.
\newblock {\em PLoS Med\/}~{\em 2\/}(8), e124.

\bibitem[\protect\citeauthoryear{Izmirlian}{Izmirlian}{2018}]{IzmirlianG:2018}
Izmirlian, G. (2018).
\newblock {\em pwrFDR: FDR Power}.
\newblock \href{https://CRAN.R-project.org/package=pwrFDR}{R package version
  1.90}.

\bibitem[\protect\citeauthoryear{Jung}{Jung}{2005}]{JungSH:2005}
Jung, S.~H. ({2005}, {Jul 15}).
\newblock {Sample size for FDR-control in microarray data analysis}.
\newblock {\em {Bioinf.}\/}~{\em {21}\/}({14}), {3097--3104}.

\bibitem[\protect\citeauthoryear{Lee and Whitmore}{Lee and
  Whitmore}{2002}]{LeeMLT:2002}
Lee, M. L.~T. and G.~A. Whitmore ({2002}, {Dec 15}).
\newblock {Power and sample size for DNA microarray studies}.
\newblock {\em {Stat. in Med.}\/}~{\em {21}\/}({23}), {3543--3570}.

\bibitem[\protect\citeauthoryear{Liu and Hwang}{Liu and
  Hwang}{2007}]{LiuP:2007}
Liu, P. and J.~T.~G. Hwang ({2007}, {Mar 15}).
\newblock {Quick calculation for sample size while controlling false discovery
  rate with application to microarray analysis}.
\newblock {\em {Bioinf.}\/}~{\em {23}\/}({6}), {739--746}.

\bibitem[\protect\citeauthoryear{{NIH High Performance Computing Staff}}{{NIH
  High Performance Computing Staff}}{2017}]{Biowulf:2017}
{NIH High Performance Computing Staff} ({2017}).
\newblock {The NIH Biowulf cluster}.
\newblock {http://hpc.nih.gov/}.
\newblock {[Online; accessed 22-December-2017]}.

\bibitem[\protect\citeauthoryear{{R Core Team}}{{R Core
  Team}}{2016}]{CiteR:2016}
{R Core Team} (2016).
\newblock {\em R: A Language and Environment for Statistical Computing}.
\newblock Vienna, Austria: R Foundation for Statistical Computing.

\bibitem[\protect\citeauthoryear{Silvestrov}{Silvestrov}{2004}]{SilvestrovD:2004}
Silvestrov, D. ({2004}).
\newblock {\em {Limit Theorems for Randomly Stopped Stochastic Processes}}.
\newblock {London}: {Springer-Verlag}.

\bibitem[\protect\citeauthoryear{Storey}{Storey}{2002}]{StoreyJD:2002}
Storey, J.~D. ({2002}).
\newblock {A direct approach to false discovery rates}.
\newblock {\em {J. R. Stat. Soc. Ser. B Stat. Methodol.}\/}~{\em {64}\/}({3}),
  {479--498}.

\bibitem[\protect\citeauthoryear{Sun and Cai}{Sun and Cai}{2007}]{SunWG:2007}
Sun, W. and T.~T. Cai (2007).
\newblock Oracle and adaptive compound decision rules for false discovery rate
  control.
\newblock {\em J. Amer. Statist. Assoc.\/}~{\em 102\/}(479), 901--912.

\bibitem[\protect\citeauthoryear{Taylor and Patterson}{Taylor and
  Patterson}{1985}]{TaylorRL:1985}
Taylor, R.~L. and R.~F. Patterson (1985).
\newblock Strong laws of large number for arrays of row-wise exchangeable
  random elements.
\newblock {\em Internat. J. Math. \& Math. Sci.\/}~{\em 8\/}(1), 135--144.

\end{thebibliography}

\begin{thebibliography}{}

\bibitem[\protect\citeauthoryear{Genovese and Wasserman}{Genovese and
  Wasserman}{2004}]{GenoveseC:2004}
Genovese, C. and L.~Wasserman ({2004}, {JUN}).
\newblock {A stochastic process approach to false discovery control}.
\newblock {\em {Ann. Stat.}\/}~{\em {32}\/}({3}), {1035--1061}.

\bibitem[\protect\citeauthoryear{Kurtz}{Kurtz}{2016}]{KurtzT:2016}
Kurtz, T.~G. ({2016}, {Nov}).
\newblock {Personal communication}.

\bibitem[\protect\citeauthoryear{{Shorack, Galen R. and Wellner, Jon
  A.}}{{Shorack, Galen R. and Wellner, Jon A.}}{1984}]{ShorackWellner:1986}
{Shorack, Galen R. and Wellner, Jon A.} ({1984}).
\newblock {\em {Empirical processes with applicationis to statistics}}.
\newblock {New York}: {John Wiley and Sons}.

\bibitem[\protect\citeauthoryear{Silvestrov}{Silvestrov}{2004}]{SilvestrovD:2004}
Silvestrov, D. ({2004}).
\newblock {\em {Limit Theorems for Randomly Stopped Stochastic Processes}}.
\newblock {London}: {Springer-Verlag}.

\bibitem[\protect\citeauthoryear{Taylor and Patterson}{Taylor and
  Patterson}{1985}]{TaylorRL:1985}
Taylor, R.~L. and R.~F. Patterson (1985).
\newblock Strong laws of large number for arrays of row-wise exchangeable
  random elements.
\newblock {\em Internat. J. Math. \& Math. Sci.\/}~{\em 8\/}(1), 135--144.

\end{thebibliography}
\end{document}